# Dynamic Mode Decomposition Accelerated Forecast and Optimization of Geological $CO_2$ Storage in Deep Saline Aquifers

Dimitrios Voulanas[1,2*], Eduardo Gildin[1]

[1] Harold Vance Department of Petroleum Engineering, Texas A&M University, College Station, TX, USA

[2] Texas A&M Energy Institute, Texas A&M University, College Station, TX, USA

dvoulanas@tamu.edu, egildin@tamu.edu

*corresponding author

keywords: data-driven, dynamic mode decomposition, optimization, $CO_2$ sequestration, high-fidelity reservoir modeling and simulation, sparsity-promoting

Abstract

Data-driven and non-intrusive DMDc and DMDspc models successfully expedite $CO_2$ fluid flow forecast and optimization, aiding in the acceleration of risk assessment and overall decision-making for geological $CO_2$ storage, while requiring fewer computational resources than traditional high-fidelity reservoir simulators, machine learning, and reduced physics proxy models. This aims to assist in the broader effort of climate change mitigation. DMDc and DMDspc models were trained independently with weekly, monthly, and yearly commercial simulator pressure and $CO_2$ saturation fields. The domain of interest is a large-scale, offshore, highly heterogeneous reservoir model with over 130,000 cells. DMDc/DMDspc snapshot reconstruction came with significantly reduced runtimes, from several hours to minutes. Two operation cases were considered: 1. $CO_2$ injection and 2. $CO_2$ injection and water production for pressure maintenance. Injection/production occurs for the first 15 years while the plume is monitored for the next 85 years. The two DMD model variants' forecast performance was cross compared with independent and individual simulator snapshots sets generated with different well rates and evaluated using percent change error, mean absolute error and Pearson's R correlation coefficient metrics. Training and forecasting were conducted on a 3 GHz 12 core CPU with 32 GB RAM workstation. Regarding pressure, DMDspc achieved a slightly higher than DMDc average error by discarding several modes. On the other hand, DMDspc showed limited success in discarding modes for $CO_2$ saturation. Almost all DMD pressure models managed to successfully forecast pressure fields, while a smaller number of DMD models managed to forecast $CO_2$ saturation. Only DMD models with errors below 5% PCE for pressure and 0.01 MAE for saturation were considered acceptable for geological $CO_2$ storage optimization. Optimized $CO_2$ injection and water production amounts were consistent across selected DMD models and all time scales. Regarding optimization we propose the reconstruction of monitored-during-optimization cells only as it reduces even further optimization time while providing consistent results with the optimization that used full snapshot reconstruction. Our contributions are two-fold: 1. we propose a practical algorithm for optimization using a proxy model based on DMD that can be applied to any reservoir operation, 2. we analyzed metrics used for incorporating the proxy model into an optimization framework, especially for optimization of geological $CO_2$ storage. To the best of our knowledge, this is the first application of DMD, particularly DMDspc, for forecast and optimization of geological $CO_2$ storage.

1. Introduction
1.1 Aim of Study

This study introduces dynamic mode decomposition with control (DMDc) as an alternative to neural network (NN) machine learning (ML) and reduced physics-based (RP) proxy modeling to reduce the computational time of geological $CO_2$ storage forecast and optimization. DMD's non-intrusive nature doesn't require to know a given simulator's internal structure. While the DMD and NN approaches are data-driven and have on-par accuracy, ML requires much more training data and effort, being a highly iterative procedure to design the neural network's dimensions in contrast to DMDc, which requires a few single set snapshots, as a bare minimum, to train and validate. DMDc uses several central processing unit (CPU) cores to train while ML uses at least one graphics processing unit (GPU), as a minimum. Our implementation is equally applicable to GPUs. RP proxy modeling, on the other hand, implicitly solves governing equations across the model domain while including assumptions in the physics that simplify the problem, losing little to no accuracy but sacrificing generality, unlike the two other approaches.

We also explore the benefits of applying sparsity-promoting DMD with control (DMDspc), which further decreases computational time with minimal added accuracy loss. The DMDspc implementation we use enables the reconstruction of specific reservoir cells' values, e.g., top-layer cells, which are the monitored cells during optimization significantly lowering computational costs even further. This approach contributes to the broader effort of developing sustainable and efficient methods for modeling geological $CO_2$ storage, a critical solution against climate change. We also propose an efficient and straightforward workflow that establishes the basis of a simple but robust framework for any reservoir operation. Ultimately, this framework provides users with an easy-to-use tool to enable rapid forecast and optimization across reservoirs of different sizes and a variety of operations.

1.2 Background

The advent of climate change and its associated challenges necessitate innovative approaches to mitigate the increase of atmospheric $CO_2$ [1]. Risk assessment of $CO_2$ sequestration primarily focuses on the probability and consequences of $CO_2$ leakage from a geologic storage site over time and the potential adverse impact on health, safety, the environment, and public policy [2], [3], [4], [5]. Injecting captured $CO_2$ from industrial sources into saline aquifers that are sealed with an impermeable cap rock provides a secure environment for geological $CO_2$ storage [1], [6]. Currently, more than 50 $CO_2$ storage sites around the globe require a long-term integrity assessment to avoid leakages. Gholami et al. (2021) reviewed different mechanisms that could lead to $CO_2$ leakage. They attempted to provide a risk assessment scheme that may improve the safety of injection and storage operations [7]. Various geological factors affect geological storage, with the two most important factors being the integrity of the cap rock, which, if low, can leak stored $CO_2$, and the brine's impact on the formations. Good knowledge of such factors leads to decreased uncertainty and accurate risk assessment [2], [5], [6], [7], vital components of successful geological $CO_2$ storage [6], leading to robust overall decision-making.

After characterizing the target formation's reservoir and before initiating $CO_2$ injection, the physics that define the flow and overall behavior of injected $CO_2$ must be hypothesized and validated. Over the past ten years, this approach has undergone considerable study regarding risk assessment [8], [9]. The mechanisms that control flow behavior need to be studied and understood via physics-based modeling, which takes into account the fluid and chemical processes active in the subsurface, as shown by the aforementioned hypotheses, and history match by the available injection/production and monitoring data [10].

Simulations are also used to systematically develop a monitoring program for a $CO_2$ sequestration project [8], [9], [10]. Designing and running simulations to serve the aforementioned assessments are often expensive since accurate modeling must include complex interactions between $CO_2$, brine, and complex geological formations [5]. These simulations must be run over extensive spatial and temporal scales, often spanning decades or even centuries, with many time steps to ensure accurate results. High-fidelity models such as these demand substantial computational resources, both in terms of processing power and data storage.

Moridis et al. (2023) analyzed the impact of salinity during long-term $CO_2$ sequestration and investigated possible solutions to mitigate salt precipitation. In that study, they accounted for all multiphase flow and transport processes of $CO_2$ sequestration. They used a high-definition mesh, as small as 0.01 m, to correctly capture the important effect of halite precipitation. Considering $H_2O$, $CO_2$, and salt as fluid components and thermal evolution in the simulations, this discretization resulted in over 414,000 simultaneous equations. All cases used ~50,000 to more than 100,000 time steps. The simulator used, TOUGH+RGB, required seven to twelve months of continuous computations using a fully implicit scheme [5]. Despite such costs, these simulations are mandatory for informing decision-making and ensuring successful geological $CO_2$ storage projects, as mentioned earlier. This inherent cost necessitates using surrogate models that reduce computational time while preserving the key aspects of high-fidelity simulations (HFS). Much research has been conducted recently that includes different combinations of physics [11], [12], [13], [14].

Surrogate or proxy modeling presents a viable alternative to overcome these computational barriers. It simplifies complex models by significantly reducing the number of computations needed, thus accelerating the overall process. However, surrogate modeling endeavours to preserve, as mentioned earlier, the critical dynamics of the studied system as a geological $CO_2$ storage or subsurface flow system, ensuring that key aspects of the flow and interaction dynamics are accurately captured. This approach aims to provide a balance between computational efficiency and the retention of essential system dynamics. Therefore, extensive risk analyses, forecasts, and optimization of geological $CO_2$ storage strategies, along with other integrated reservoir studies, become more feasible, especially in scenarios where computational resources are constrained.

Proxy models can be based on the mathematical formulation, e.g., physics-based or data-driven methods and combinations thereof. The physics-based approaches typically implement simplifications by neglecting and/or simplifying physics or numerical aspects of the problem. Some robust and popular examples are (1) reduced-physics modeling, (2) machine learning-based surrogate modeling, or (3) projection-based reduced-order modeling (ROM). Common methods include proper orthogonal decomposition (POD), balanced truncation (BT), and DMD.

Reduced-physics modeling refers to approaches that simplify complex physical systems by incorporating certain assumptions, which reduces the number of governing equations or variables. One such example is the Capacitance-resistance model (CRM), which is an analytical model and is widely used in reservoir engineering to model reservoir fluid flow. CRM assumes that the reservoir behaves like a network of interconnected tanks (capacitors) and flow resistances, ignoring the detailed physics of fluid flow, such as Darcy's law or multiphase flow. By focusing on the relationship between fluid storage (capacitance) and flow resistance, CRM allows for faster computation and real-time optimization of reservoir management, though it comes with limitations in accurately capturing the full dynamics of complex reservoirs and fluids [15]. This approach has been applied mainly to primary production and waterflooding operations [16], [17]. Another RP approach is the Fast Marching Method (FMM). FMM is a computational algorithm that is used in reservoir engineering to model fluid front propagation in heterogeneous media [18]. It is an efficient way to solve the Eikonal equation, which describes the evolution of a wavefront over time. In the context of reservoirs, FMM is employed to estimate the arrival times of fluid fronts, such as water or gas, during secondary recovery processes like waterflooding. This method leverages a grid-based approach to trace the movement of the front based on the speed of fluid propagation, determined by reservoir properties like permeability and porosity [18], [19].

Iino et al. (2017) introduce a novel and practical approach using the FMM for rapid multi-phase simulation of shale reservoirs with multi-continua heterogeneity. Accurate modeling of unconventional reservoirs is essential due to complex interactions between reservoir rocks, microfractures, and hydraulic fractures. The proposed FMM-based method recasts the 3D flow equation into a 1D equation along the "diffusive time of flight" (DTOF) coordinate, which captures the 3D spatial heterogeneity. This transformation leads to orders of magnitude faster computation than traditional 3D finite-difference simulations. Additionally, their results showed that the higher the number of grid blocks, the higher the computational efficiency. The computation time of one million grid blocks drops from 6766 to 177 seconds and from 279 to 28 seconds for one hundred thousand grid blocks [20]. Holanda et al. (2018) report that typical CRM simulations for reservoirs often take between 1 to 2 hours or more, depending on factors like the complexity of the reservoir and the number of wells involved. For field-scale applications, this relatively quick run time makes CRM models particularly useful in settings that require iterative simulations for tasks like history matching or optimizing waterflood operations [16].

Neural networks (NN) serve as proxy models for various complex processes. In our case, reservoir engineering is the area where ML has been most widely used in the petroleum industry, achieving great results in complex subsurface processes (e.g., shale gas production prediction, well test interpretation, and history matching) [21], [22]. NNs are often employed to create surrogate models that approximate the behavior of complex reservoir simulations in whole or in part. For instance, regarding the former, they can be used to predict fluid flow and pressure distribution based on historical data and simulation results, or regarding the latter, they can predict liquid production using only injection rates during waterflooding [14], [22], [23], [24].

To capture such complex pattern NNs utilize multiple layers of interconnected nodes or "neurons": an input layer, one or more hidden layers, and an output layer [25], [26]. The architecture of a neural network is given by the number of layers and neurons, which influence its ability to learn complex patterns. Designing an effective NN is a highly iterative and usually very time-consuming procedure, as a trade-off between model complexity and generalization must be balanced [26]. Too few layers or neurons may easily lead to underfitting, while too many suffer from overfitting that degrades its performance on unseen data [25], [26]. Coupling multiple neural operators and architectures naturally increases model complexity; hence, it is challenging to tune and optimize with new hyperparameters [25], [27], [28].

The amount of data needed to train a neural network also plays a crucial role in its performance. The larger the network, the greater number of parameters and size of a dataset to avoid overfitting. Training neural networks in their most basic form involves the running of optimization algorithms, such as stochastic gradient descent or Adam, to minimize the prediction error. The time taken for training is highly variable depending on network size, dataset size, and difficulty of the problem-from minutes for small networks to days or even weeks for deep networks with very large datasets using modern GPU/s [25], [26].

Ng et al. (2021) implemented Smart Proxy Modeling (SPM), an ML approach requiring a spatio-temporal database from numerical simulations. The study used artificial neural networks and applied SPM to a dual-porosity, dual-permeability fractured reservoir model, focusing on production optimization. Two Smart Proxy Models (SPMs) were developed to predict oil and water production rates at a given injection rate. Each model underwent separate neural network training using three algorithms: Stochastic Gradient Descent (SGD), Particle Swarm Optimization (PSO), and Adam. As a result, a total of six highly accurate SPMs were constructed ($R^2$>0.99). One 30-year-long simulation scenario with 361 timesteps took ~160 seconds to complete while running five scenarios simultaneously took about 290 seconds. In comparison, SPM computation times were reduced: SGD took ~110 seconds, PSO ~50 seconds, and Adam took approximately 120 seconds [29].

Bi et al. (2024) proposed a physics-informed spatial-temporal neural network (PI-STNN) that integrates flow theory into its loss function and combines a deep convolutional encoder-decoder (DCED) with a convolutional long short-term memory (ConvLSTM) network. The model is designed to capture spatial-temporal dependencies in reservoir simulations while leveraging physical laws. Its performance was compared with a purely data-driven model using the same neural architecture and the Fourier Neural Operator (FNO). The results showed that PI-STNN outperforms the data-driven approach and demonstrates competitive accuracy, even surpassing the FNO in some reservoir simulation tasks. This highlights its robustness and generalization capabilities for complex fluid flow prediction [30].

Several ROM methods exist in which the state variables or snapshots and/or the system of equations are projected into low-dimensional space and then computed and/or solved, have been applied for a range of subsurface flow problems [11], [12], [31], [32], [33], [34], [35]. While ROMs can be highly effective, their accuracy typically hinges on how closely new (test) runs resemble the training runs. In numerous applications, significant increases in speed ranging from two- to three-fold (100-1000 times faster) or more are anticipated. One such method, DMD, our study focus, is particularly robust at capturing the dynamics of complex systems by decomposing the temporal evolution of high-dimensional data into a set of spatial modes and corresponding frequencies.

Jin and Durlofsky (2018) introduced a POD-TPWL reduced-order modeling framework to simulate and optimize the injection stage of geological $CO_2$ storage. This framework uses the combination of trajectory piecewise linearization (TPWL) and low-dimensional subspace projection via POD to produce new solutions with new sets of well controls based on the linearization of previously simulated (training) solutions. The resulting representation is low-dimensional and linear, in contrast to the original nonlinear full-order simulations. Several new POD-TPWL models were devised using multiple derivatives and rate-controlled injection wells. These models can provide accurate estimates of $CO_2$ molar fraction at given domain locations. The POD-TPWL model was then incorporated into a mesh adaptive direct search optimization scheme where the objective is to minimize the amount of $CO_2$ reaching a target layer at the end of the injection period. The POD-TPWL model results were similar to those provided by the full-order simulations. The preprocessing computations needed to construct the POD-TPWL models entail a (serial) time equivalent of about 6.7 full-order simulations, though the resulting runtime speedups, relative to full-order simulation, are about 100–150 for the cases considered [36].

Dall'Aqua et al. (2023) suggested a new multi-reservoir framework to obtain ROMs for two-phase fluid flow for optimal well-control design, which is equipped with input-output tracking capabilities (well response). This work proposed four methods; methods 1 and 2 rely only on Peaceman's equation, whereas methods 3 and 4 perform a Koopman linearization to the state equations. The first method uses the state predictions at the well location from the state proxy and then uses Peaceman's equation to make an output prediction. The second method differs from the first in the training process, which is slightly different. This method combines the state snapshots (pressure and saturation) and the output term in the training process. The third method performs a Koopman linearization of the output equation, and the fourth method uses the Balanced Truncation (BT) concept and attempts to achieve state and the best correlated input-output matching simultaneously. Simulations were run via the Matlab Reservoir Simulation Toolbox (MRST) and (CMG IMEX). A single run MRST and CMG IMEX took 45 secs and 2.39 secs, respectively, while the corresponding ROMs took 0.7 secs [35].

2. Materials and Methods

The materials and methods section includes six sub-sections and is structured as follows. Sub-section 1 shows the overall proposed workflow introduced via this study. Sub-section 2 provides a brief overview of porous media flow governing equations used to produce primary variables snapshots. Sub-section 3 presents information about the model domain and well placement. Sub-section 4 shows an overview of DMDc/DMDspc theory and its formulations. In sub-section 5, information on DMD forecast and temporal analysis of the results is presented and analysed. Finally, sub-section 6 shows the optimization procedure and the objective function we used for optimizing geological $CO_2$ storage.

2.1 Overview

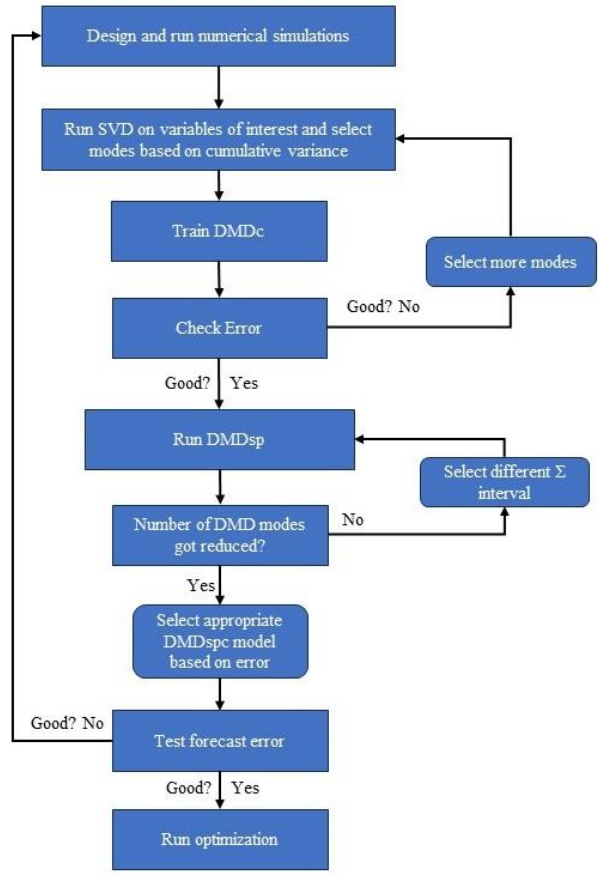

Fig. 1 – Flowchart of DMD accelerated forecast and optimization of geological $CO_2$ storage.

The DMD accelerated procedure that forecasts and optimizes geological $CO_2$ storage (Fig. 1) we propose is briefly described here:

1. Design and run numerical reservoir simulations with the simulator and well rates of choice. All simulations must show all the required physics since DMD will only learn from the physics it can see from the snapshots it's fed with. The number and size of grid blocks in these numerical simulations are such that best approximate the domain's geology while they remain at a minimum number to minimize computation time without reducing accuracy. Therefore, out methodology relies on high-fidelity reservoir domain and numerical reservoir simulations that best balance accuracy and simulation speed.
2. Run SVD on the variables of interest and select modes based on the normalized cumulative variance. The threshold of such choice is recommended to be 0.995 and above. However, the user could select a different threshold if it is deemed necessary.
3. Train DMDc model and check error. If the error is acceptable, continue; if not, then include more SVD modes.
4. Run DMDsp using the chosen Σ value range. Σ value range can be determined by trial and error. If the Σ value range is acceptable, then the number of modes will reduce. If not, repeat step 4. with a different Σ value range.
5. Select DMDspc model based on the error.
6. Test forecast error. If it is acceptable, then continue. If not, then start again from step 1., and then proceed to step 7.
7. Run optimization

Simulations and DMD related computations were run on a workstation with 12 cores 3.0GHz CPU and 32GB RAM.

2.2 Governing Equations

In a compositional model, the composition of reservoir fluids consists of a finite number of hydrocarbon or non-hydrocarbon components. These components are associated as phases in a reservoir. Our numerical simulation assumes that the flow process is isothermal (i.e., at constant temperature), and the components considered in our project are water, $CO_2$, and sodium chloride dissolved in the aquifer's water. Some important governing equations are shown below for completeness' sake.

$$\frac{\partial(\phi \xi_w S_w)}{\partial t} + \nabla(\xi_w u_w) = q_w, \text{ Aquifer} \tag{1}$$

$$\frac{\partial(\phi[x_{ig}\xi_g S_g])}{\partial t} + \nabla(x_{ig}\xi_g u_g) + \nabla(d_{ig}) = q_i, \ i = 1,2,\dots,N_c, \text{ CO}_2 \text{ gas} \tag{2}$$

$$u_\alpha = -\frac{1}{\mu_\alpha} K_\alpha (\nabla p_\alpha - \rho_\alpha \wp \nabla_z), \alpha = w, g \tag{3}$$

where $\alpha = w, g$, $\nabla = \left(\frac{\partial}{\partial x} + \frac{\partial}{\partial y} + \frac{\partial}{\partial z}\right)$, $\nabla p_\alpha = \left(\frac{\partial p_\alpha}{\partial x}, \frac{\partial p_\alpha}{\partial y}, \frac{\partial p_\alpha}{\partial z}\right)$, and $\sum_{i=1}^{N_c} x_{ig} = 1$.

$\xi$ is the molar density of water (w) or gas (g), $q_w$ and $q_i$ are the molar flow rates of water and ith component, respectively, $d_{ig}$ is the diffusive flux, and $u_\alpha$ is the volumetric velocity. $S_w$ and $S_g$ are saturations of water and $CO_2$ gas, $\wp$ is the gravitational constant, $\nabla$ is the gradient operator, and $p_\alpha$ is the pressure of any phase in the system.

Relative permeability models [37] are integrated into the simulator's governing equations to describe fluid flow within the reservoir. Relative permeability values for each fluid (e.g., water, $CO_2$) are incorporated into Darcy's law in Eq. (3) above. These values adjust the permeability of the reservoir rock for each fluid, accounting for the presence and interaction of multiple fluids in the pores. In multiphase flow (like water and $CO_2$ in aquifer storage), the relative permeability of each phase impacts the flow rate according to its saturation level. As $CO_2$ is injected and its saturation increases, the relative permeability to $CO_2$ increases, affecting the flow patterns of both $CO_2$ and water. It predicts the movement of the $CO_2$ plume within the aquifer. Different relative permeabilities of $CO_2$ and water lead to different plume behaviors and spreading patterns. The commercial reservoir simulator (e.g., ECLIPSE 300) encapsulates all governing equations, including robust fluid modeling where the equations of state (EOS), like the Peng-Robinson EOS [38], are used for aquifer - $CO_2$ solubility in brine modeling [39].

$$P_c = -\frac{1}{\alpha^\gamma}\left[\left(\frac{S_l - S_{l\,min}}{1 - S_{gr}^\Delta - S_{l\,min}}\right)^{-\left(\frac{n^\gamma}{n^\gamma - 1}\right)} - 1\right]^{\left(\frac{1}{n^\gamma}\right)} \tag{4}$$

$$k_{rg} = k_{rg(slr)}\left(1 - (\bar{S}_l + \bar{S}_{gt})\right)^\gamma \left(1 - (\bar{S}_l + \bar{S}_{gt})^{\frac{1}{m}}\right)^{2m} \tag{5}$$

$$k_{rl} = \sqrt{\bar{S}_l}\left[1 - \left(1 - \frac{\bar{S}_{gt}}{1 - \bar{S}_l^\Delta}\right)\left(1 - (\bar{S}_l + \bar{S}_{gt})^{\frac{1}{m}}\right)^m - \left(\frac{\bar{S}_{gt}}{1 - \bar{S}_l^\Delta}\right)\left(1 - (\bar{S}_l^\Delta)^{\frac{1}{m}}\right)^m\right]^2 \tag{6}$$

where $P_c$ is the capillary pressure, $S_l$ is the liquid or brine saturation, $S_{l\,min}$ is the minimum liquid saturation, $S_{gr}^\Delta$ is residual gas saturation, $\gamma$ denotes the branch (d for drainage, w for imbibition), $\alpha^\gamma$, $n^\gamma$ and, $S_{l\,min}$ are fitting parameters, $S_l^\Delta$ is the saturation at the drainage-to-imbibition turning point, $k_{rl}$ is the liquid phase or brine relative permeability, $\bar{S}_l, \bar{S}_l^\Delta$ & $\bar{S}_{gt}$ are effective liquid, turning point liquid, and trapped gas-phase saturations, respectively. Readers are referred to [40] for a full description of ECLIPSE 300.

2.3 Model Domain

The model domain is extracted from a larger proprietary dataset. It has a 3050 m average depth a.m.s.l (Fig. 2). The sea depth at its location is about 1700 m. The reservoir's geological environment is highly heterogeneous; it consists of deltaic deposits with a noncontinuous shaly barrier with channels connected both vertically and horizontally. This necessitated the development of a high-fidelity geological model that will enable proper tracking of $CO_2$ fluid flow with respect to reservoir heterogeneity. The model's mesh cells are thin along the z-axis since $CO_2$ plume migration is more vertical than horizontal.

In summary, the simulation configuration is:
- Cell dimension is 75 m by 75 m
- The maximum i, j, and k indices are 52, 28, and 101, respectively
- The total active cells are 135,340
- The reservoir area is 4.2 km by 2.2 km, and the reservoir thickness is 720 m
- Two cases: 1. One well for injecting $CO_2$ (Fig. 2), 2. Two wells, one for injecting $CO_2$ and one for pressure maintenance by producing brine (Fig. 2)
- Initial reservoir temperature and pressure of 56°C and 241 bars, respectively
- 5200 seven days (weekly) $\Delta t$ time-steps
- 780 injection/production time-steps

In the geological $CO_2$ storage simulation, we placed a $CO_2$ injection well at the reservoir's deepest area of the lower part, where permeable-less permeable-impermeable formation alternations are found, while the production well was placed at the bottom of the reservoir's shallowest part. The idea behind such well placement is: 1. the injected $CO_2$ will take a longer path to reach the reservoir's cap rock, which will allow more $CO_2$ to dissolve into the brine; 2. the brine production well will not produce the injected $CO_2$ as it rises to the caprock.

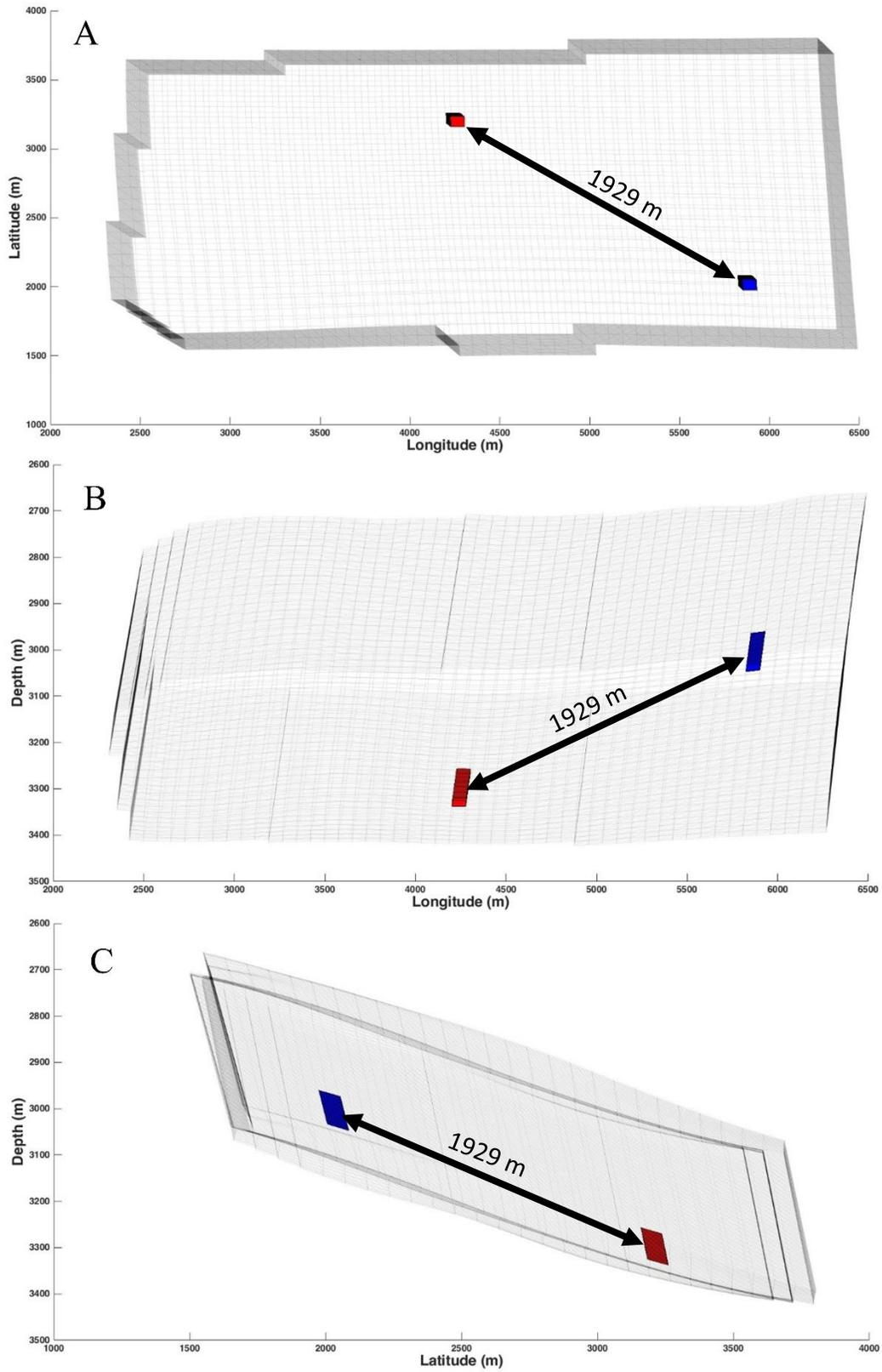

Fig. 2 – Reservoir shape and outline (A: XY, B: XZ, C: YZ planes) along with the injection well (red) and production well (blue) and the overall distance between them.

## 2.4 Dynamic Mode Decomposition

Geological $CO_2$ storage simulations' computational expense arises from the need to resolve complex multiphase flow and transport processes over large meshes and time scales. DMD serves as a mitigation to this computational burden that leverages the speed of data-driven algorithms to approximate the high-dimensional dynamical systems associated with $CO_2$ storage simulations.

DMD stands out due to its ability to extract dynamic features from complex systems efficiently, offering a significant reduction in computational time without substantially compromising accuracy. Unlike other methods, such as POD, which primarily focus on capturing the most energetic modes, DMD provides an eigenvalue decomposition that identifies spatiotemporal coherent structures and captures the flow dynamics' essence. Methods like POD are adept at data compression and feature extraction; they often require a large amount of data to achieve a suitable level of accuracy. Furthermore, machine learning approaches, though powerful, often require extensive training data and can be opaque in their decision-making process.

DMD's distinction lies in its algorithmic simplicity and robustness in handling nonlinear, high-dimensional data. It excels in providing a predictive framework that is not only accurate but also interpretable, enabling the projection of future states of the $CO_2$ plume with reduced computational resources. This characteristic is particularly advantageous for geological $CO_2$ storage simulations, where forecasting the behavior of the injected $CO_2$ over extended periods is paramount. Readers are referred to the following series of papers that comprehensively describe the basics behind DMD, several of its variants, and some of its successful applications [41], [42], [43], [44], [45], [46], [47]. In this study, we will only briefly explain DMD basics and the two DMD variants we used.

### 2.4.1 Sparsity-Promoting DMD with Control

At its basis, DMD analyzes the relationship between measurements from a dynamical system. The stacked snapshot vectors are represented by the two following matrices (7) and (8):

$$X = [x_1 \ x_2 \ x_3, \dots, x_{m-1}] \in \mathbb{R}^{n \times (m-1)} \tag{7}$$
$$X' = [x_2 \ x_3 \ x_4, \dots, x_m] \in \mathbb{R}^{n \times (m-1)} \tag{8}$$

$X$ holds the first time step snapshot vectors from the first to the next-to-last snapshot vector column-wise. $X'$ holds the second time step snapshot vectors from the second to the last column vector. The total number of snapshot vectors is $m$. The vectors are collected at regular time intervals $\Delta t$. DMD generates a linear operator $A$ that maps $X$ matrix to $X'$ matrix. $n$ is the number of measurements or reservoir cells for each vector, and $m$ is the number of the dynamical system's time steps.

$$X' \approx AX \tag{9}$$

However, standard DMD may yield inaccurate dynamics when external system control exists. The DMDc method [48] enables the discovery of underlying physics dynamics without knowing explicit external control. It also quantifies the effect of control inputs on the system's evolution. In this case, Eq. (9) became Eq. (10), which has the added term $BY$, $B \in \mathbb{R}^{n \times l}$ is the control mapping and $Y \in \mathbb{R}^{l \times (m-1)}$ is the control input snapshots, where $l$ is the number of the distinct controls.

$$X' \approx AX + BY \tag{10}$$

DMDc finds the best-fit mappings $A$ and $B$ approximations. The matrix $B$ can be known or accurately estimated. This assumption is ideal for most complex dynamical systems and implies a significant amount of knowledge about how control inputs affect the system. The unknown $B$ and $A$ matrices are estimated from the control input snapshots and state snapshots. In addition, DMDc, like standard DMD, greatly reduces dimensionality and therefore computational effort by truncating non-important singular values while incorporating control. The number of singular values kept after truncations is $r$. Therefore, if $r \ll m$, lead to a smaller, more compact model is created via a linear subspace of dimension $r$ onto which $X$ can be projected.

The sparsity-promoting DMD (DMDsp), another DMD variant, can further truncate modes in a more systematic manner while usually adding a little more error in computations. DMDsp was first introduced by Jovanovic et al. (2014) [43] and later extended to account for inputs [44]. The DMDsp is an enhancement to the traditional DMD method, which adds a penalty to a convex optimization problem whose performance index includes a sparsity-promoting term (the $\mathcal{L}_1$ norm of the DMD mode). This is useful in situations where we want to retain only the most significant DMD modes while discarding the rest, which may represent less important dynamics or noise. In this study, we use the work of Tsolovikos et al. (2021) [45], because it combines the two aforementioned DMD variants into one workflow, the DMDspc.

To achieve the dimension reduction mentioned earlier, the $X$ and $X'$ is projected using $\widetilde{U}^T$ as $\widetilde{U}^T X$ and $\widetilde{U}^T X'$, respectively, where $T$ is the complex conjugate transpose. The $\widetilde{U}$ comes from an singular value decomposition (SVD) on $X$ which yields $X = U\Sigma V^T \approx \widetilde{U}\widetilde{\Sigma}\widetilde{V}^T$, where $U \in \mathbb{R}^{n\times n}$, $\Sigma \in \mathbb{R}^{n\times(m-1)}$, $V^T \in \mathbb{R}^{(m-1)\times(m-1)}$, $\widetilde{U} \in \mathbb{R}^{n\times r}$, $\widetilde{\Sigma} \in \mathbb{R}^{r\times r}$, $\widetilde{V}^T \in \mathbb{R}^{r\times(m-1)}$. This yields $H \in \mathbb{R}^{r\times(m-1)}$ and $H' \in \mathbb{R}^{r\times(m-1)}$ which are the backward and forward projected snapshots, respectively.

A second SVD is required to find the reduced-order subspace of the output space. This SVD is applied to the following augmented matrix: $\begin{bmatrix} H \\ \Upsilon \end{bmatrix}$ and yields $U_c \Sigma_c V_c^T \approx \widetilde{U}_c \widetilde{\Sigma}_c \widetilde{V}_c^T$, where $U_c \in \mathbb{R}^{(r+l)\times(r+l)}$, $\Sigma_c \in \mathbb{R}^{(r+l)\times(r+l)}$, and $V_c^T \in \mathbb{R}^{(r+l)\times(r+l)}$ and the truncated matrices are $\widetilde{U}_c = \begin{bmatrix} \widetilde{U}_{c,F} \\ \widetilde{U}_{c,G} \end{bmatrix} \in \mathbb{R}^{(r+l)\times r_c}$, $\widetilde{\Sigma}_c \in \mathbb{R}^{r_c \times r_c}$, and $\widetilde{V}_c^T \in \mathbb{R}^{r_c \times (m-1)}$. Then $F$ and $G$ are calculated with $F = H' \widetilde{V}_c^T \widetilde{\Sigma}_c^{-1} \widetilde{U}_{c,F}^T$ and $G = H' \widetilde{V}_c^T \widetilde{\Sigma}_c^{-1} \widetilde{U}_{c,G}^T$, which are the projected $\tilde{A}$ and $\tilde{B}$, respectively [49].

Therefore, Eq. (10) becomes, in projected space:

$$H' = FH + GU \qquad (11)$$

$F$ is eigendecomposed as:

$$FW = W\Lambda \qquad (12)$$

where $W = [w_1, w_2, \ldots, w_{r_c-1}, w_{r_c}] \in \mathbb{C}^{r_c \times r_c}$ is a non-singular matrix containing $F$'s right eigenvectors, $\Lambda = diag\{\lambda\} \in \mathbb{C}^{r_c \times r_c}$ and $\lambda = [\lambda_1, \lambda_2, \ldots, \lambda_{r_c-1}, \lambda_{r_c}] \in \mathbb{C}^{r_c}$ is a vector that contains $F$'s eigenvalues. The transformation $H = W\Psi$ yields the modal form of Eq. (11):

$$\Psi' = \Lambda\Psi + \Gamma U \qquad (13)$$

where $\Psi \in \mathbb{C}^{r_c}$ is the DMD mode amplitude vector and $\Gamma = W^{-1}G$. The snapshot output is then approximated by:

$$X \approx \Phi\Psi \qquad (14)$$

with $\Phi = \widetilde{U}W = [\varphi_1, \varphi_2, \ldots, \varphi_{r_c-1}, \varphi_{r_c}]$.

At this point, before sparsity promotion, DMDc reconstruction equations (shown below) can be used to obtain the DMDc generated snapshots. $\Phi^\dagger$ is used to project snapshots to reduced space and $\Phi$ to unprojected them back to the original space.

$$\eta_{k+1} = F\eta_k + Gu_k = F(\Phi^\dagger x_k) + Gu_k \qquad (15)$$
$$x_{k+1} = \Phi\eta_{k+1} \qquad (16)$$

DMD does not offer objective ranking, which makes discarding the least important DMD modes a non-trivial task. Below follows a similar procedure, by [45], to sparsity-promoting DMD [43] that selects the most important DMDc modes.

Each spectral amplitude, $\psi_i \in \mathbb{C}$, is decoupled and corresponds to individual eigenvalues $\lambda_i$, each mode $\varphi_i$ oscillates with a frequency and a growth/decay rate characterized by generally complex eigenvalue $\lambda_i$ and amplitude $\psi_{i,k}$ at each time k. In addition, the relative importance of each $\varphi_i$ must be determined via a weighting factor $a_i \in [0,1]$ for each mode $i$. Thus, the snapshot $y_{k+1}$ can be approximated by:

$$x_{k+1} \approx \sum_{i=1}^{r} a_i \varphi_i (\lambda_i \psi_i + \Gamma_{i,:} u_k) = \Phi diag\{a\}(\Lambda \psi_k + \Gamma u_k) \qquad (17)$$

where $\Gamma_{i,:} \in \mathbb{R}^{r_c}$ is the i-th row of $\Gamma$, $a = [a_1, a_2, \ldots, a_{r_c-1}, a_{r_c}]^T \in \mathbb{R}^{r_c}$ and $k$ is the current step. The training data's mode amplitudes $\psi_k$ can be approximated as $\Psi = \Phi^\dagger X$, where † is the Moore–Penrose pseudoinverse [50]. The snapshot matrix $X'$ expressed with DMD modes and weights is:

$$X' \approx \Phi diag\{\alpha\} R \qquad (18)$$

where $R = \Lambda\Psi + \Gamma U$. The objective function to be minimized, based on the above, is the following:

$$J_{LS}(a) = \|X' - \Phi diag\{a\}R\|_F^2 \quad (19)$$

This can be rewritten in quadratic form while balancing the minimization of $J_{LS}(a)$ and $\|a\|_0 = card\{a\}$. However, a non-convex problem will be generated, which is reformulated to a convex problem by approximating the $\mathcal{L}_0$ norm with the reweighted $\mathcal{L}_1$:

$$\min_{a} \quad J_1(a) = a^T P a - a^T d - d^T a + s + \Sigma \|S\alpha\|_1 \quad (20)$$

where $\mathcal{E}$ is a non-negative scalar term that balances sparsity against approximation error, $P = (\Phi^T \Phi) \circ \overline{(RR^T)}$, $d = \overline{vdiag\{R(X')^T \Phi\}}$, $s = trace(X'^T X)$, $\|x\|_1 = \Sigma_1^\mu |x_i|$, if $x \in \mathbb{R}^\mu$, and $S = diag\{s_1, s_2, \ldots, s_{r_c-1}, s_{r_c}\}$ such that $\|Sa\|_1 = \Sigma_{i=1}^{r_c} |s_i a_i| = \|a\|_0$. The symbol $\circ$ denotes elementwise multiplication, $\overline{M}$ is the conjugate of $M$, $M^T$ is the conjugate transpose of $M$, and $vdiag\{M\}$ is the vector consisting of the diagonal elements of the square matrix $M$. $S$ is not known initially, so it needs to be solved for. The weights' initial values are after each iteration $f$ using the solution $a$ as:

$$s_i^{(f+1)} = 1 \Big/ \left( \left| \alpha_i^{(l)} \right| + e \right), i = 1, \ldots, r_c \quad (21)$$

The $e$'s value is $10^{-6}$ to avoid division by zero. Eq. (21) is solved iteratively. After each iteration, $S$ is updated. This approximates the non-convex optimization problems' solution while it yields a sparse solution for (20).

Next, the least important DMD modes, where corresponding $a$ values are zeros, are truncated. The now sparse linear model of the DMD mode amplitudes is obtained from $\psi_{k+1} = \Lambda \psi_k + \Gamma u_k$ and the transformation $\tilde{\psi} = E\psi_k$, where $E \in \mathbb{R}^{r_s \times r_c}$. $E$ is derived from the identity matrix $I \in \mathbb{R}^{r_c \times r_c}$ by deleting rows $i \in \{i: a_i = 0\}$ and $r_s = \|a\|_0 \leq r_c$.

The truncated reduced-order model and measurement output $x$ are:

$$\tilde{\psi}_{k+1} = \tilde{\Lambda}\tilde{\psi}_k + \tilde{\Gamma} u_k \quad (22)$$
$$x_k = \tilde{\Phi}\tilde{\psi}_{k+1} \quad (23)$$

where $\tilde{\Lambda} = E\Lambda E^T$, $\tilde{\Gamma} = E\Gamma$, and $\tilde{\Phi} = \Phi diag\{a\}E^T$. The linear state-space model above contains only the $r_s$ most important nodes of (13). This complex modal state-space model corresponds to the following real modal state-space model, which uses $\Theta^\dagger$ is used to project snapshots to reduced space and $\Theta$ to unprojected them back to the original space.

$$\xi_{k+1} = \tilde{A}\xi_k + \tilde{B}u_k = \tilde{A}(\Theta^\dagger x_k) + \tilde{B}u_k \quad (24)$$
$$x_{k+1} = \Theta \xi_{k+1} \quad (25)$$

2.5 DMDspc Training, Validation, and Temporal Analysis

In this study, we trained the DMDspc on single high-fidelity ECLIPSE 300 simulation run results. Well rates of runs 1 and 6 (not shown here) were ultimately discarded because their rates were reduced substantially due to injectivity. Fig. 2 shows the selected injection schemes that are applied to each simulation individually, with weekly time steps used for training and/or validating the DMDspc models. We did not keep runs 4 and 9 since the low rates applied did not show any significant difference from runs 5 and 10, respectively, which have half the rates of runs 4 and 9. Each simulation run includes 780 time steps or 15 years of injection time. For the remainder of the 100 years of simulation time, the wells are shut in to allow for pressure and $CO_2$ saturation monitoring.

With regards to training DMDc and DMDspc, we used ECLIPSE 300 runs 2, 5, 7, and 10 individually. Regarding validation, the reconstructed snapshots of the DMD models were compared against the original snapshots of the runs that the DMD models were not trained on. We did this to quantify a model's ability to train with high rates to forecast a case with low rates and vice versa. During temporal analysis, we used models trained with coarser resolutions (e.g., monthly and yearly) to forecast the corresponding snapshots from the weekly HFS runs.

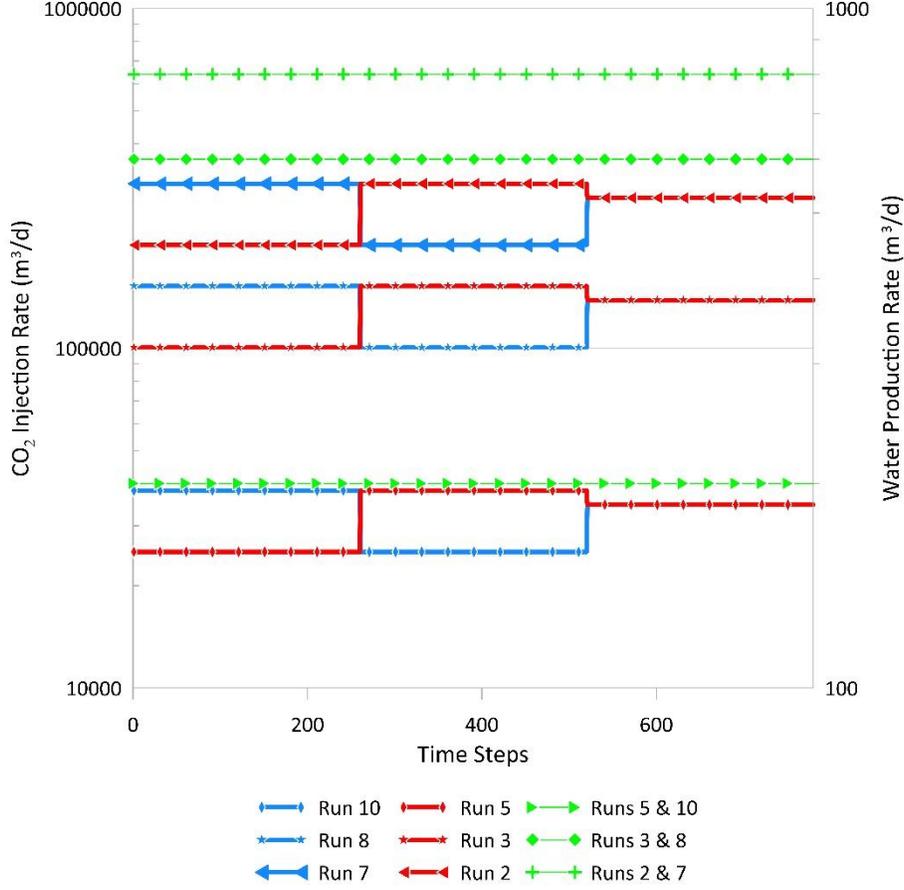

Fig. 3 - $CO_2$ injection and water production rates for DMD training and validation ECLIPSE 300 runs.

2.6 Geological $CO_2$ storage optimization

The optimization procedure we used in this study closely follows the one found in MRST's co2lab [51], [52], [53]. $CO_2$ sequestration optimization requires multiple sequential HFS runs, as mentioned earlier. DMD replaces the HFS with matrix computations that decrease computation time when compared to HFS. We used Pytorch's optimizer [54] and the L-BFGS method [55], [56] to minimize the following objective functions, which we simplified accordingly to fit our problem. Moreover, the saturation leakage in the original co2lab objective function uses spill point dynamics that apply only when multiple $CO_2$ traps are present. In our case, there is only one trap present, so the trapped $CO_2$ does not leak to another trap, but it still leaks via the caprock itself. Therefore, in our case, $Leakage$ equals $\sum_{i=1}^{N} C_L \circ Sat_i \circ PV$ at last time step. Both of the following objective functions $J_1$ and $J_2$ are scalar, as all terms (e.g., leakage, pressure penalty, injected $CO_2$, and produced water volumes) are summed over time and monitored cells.

$$\min J_1 = -\sum_{t=1}^{Q} CO_{2,t} + \sum_{t=1}^{Q} Water_t + Pressure\ Penalty \qquad (26)$$

$$\min J_2 = (1 - C_L)\sum_{t=1}^{Q} (CO_{2,t} \circ \Delta t) + \sum_{t=1}^{Q} (Water_t \circ \Delta t) + Pressure\ Penalty + Leakage \qquad (27)$$

where Q is the total number of time steps, $N$ is the total number of monitored cells, $CO_{2,t} \circ \Delta t$ is the injected $CO_2$ volume with $CO_{2,t}$ being the injected $CO_2$ rate at a given time step $t$, $Water \circ \Delta t$ is the produced Water volume with $Water_t \circ \Delta t$ with $Water_t$ being the produced Water rate at a given time step $t$, $Pressure\ Penalty = \sum_{t=1}^{Q}[\max(0, \text{sign}(p_{i=1,\dots,N} - p_{max})) \circ PPF \circ \sum_{i=1}^{N}(p_i - p_{max})^2]$ and $C_L$ is the leakage factor. $p_{i=1,\dots,N}$ (pressure) and $Sat_{i=1,\dots,N}$ ($CO_2$ saturation) represent the monitored cells during optimization which are found at the reservoir's top layer. In addition, $p_{max}$ is the maximum allowed pressure, which is 90% of the overburden pressure, and PV is the pore volume. The pressure penalty factor, $PPF$, receives one of the following values: $10^{-9}, 10^{-8}, 10^{-7}, 10^{-6}, 10^{-5}, 10^{-4}, 10^{-3}, 10^{-2}, 10^{-1}$. $\Delta t$, in this case, is 7, 30, or 365 days for weekly, monthly, and yearly time scales, respectively.

The inner optimization iterations, using a given pressure penalty factor (starting from the first value of $10^{-9}$), terminate when the gradient norm or the objective function value drops below the set threshold of $10^{-3}$ or when the number of iterations exceeds 15. Following that, as part of the outer optimization iteration, the top layer pressure is compared against a 2% margin above the maximum allowed pressure. If the top layer's pressure is within that margin, then the nearly optimized rates are used as an initial guess for the following optimization run, which uses the next pressure penalty factor from the values given above. If not, the "nearly" optimized rates are considered optimal. The objective function monitors specific cells regarding maximum allowed pressure and leakage, and then the rates reduce progressively along each outer optimization iteration from the maximum values. In case B, all rates were kept within the operational range, with a minimum of 0 m³/d and a maximum water production of 1,000 m³/d and 350,000 m³/d of $CO_2$ injection. Case A was left unconstrained because of high-pressure buildup during injection.

We perform this optimization procedure using DMDc and DMDspc, which reconstruct entire reservoir snapshots (all mesh cells). However, the DMDspc implementation reconstructs specific user-defined cells like those monitored-during-optimization top-layer cells. This approach reduces optimization time even further. All computational times recorded are presented in the Results and Discussion section and the Appendix (Table A2 and Table A3). Finally, we perform a temporal analysis using the DMD models trained with coarser time scale snapshot, e.g., monthly and yearly, and compare the optimization results to those received from the DMD models trained on weekly snapshots. This comparison further evaluates the overall DMD model performance.

4. Results and Discussion

The results and discussion section includes five sub-sections and is structured as follows. Sub-section 1 provides a brief introduction to the SVD rank selection required for applying DMD successfully and efficiently. Sub-section 2 presents the runtime and memory requirements of DMD models. Sub-section 3 presents and discusses the results of DMDc/DMDspc training and validation with its training data. Sub-section 4, presents and discusses DMDc/DMDspc forecast performance under different input controls with individual and independent snapshot data and a temporal analysis across the three different time scales. Sub-section 5 presents and discusses the $CO_2$ storage optimization in both cases (A: $CO_2$ injection case and B: $CO_2$ injection and water production case).

4.1 SVD Rank Selection

We performed full SVD on the ECLIPSE 300 simulation results at three timescales: weekly, monthly, and yearly. SVD produced different results when applied to pressure and $CO_2$ saturation data, as shown in Fig. 4. Normalized cumulative variance (NCV) of each variable was similar across cases A and B. Therefore, we only present run 3 of case B in Fig. 4.

The lowest pressure SVD weekly and monthly snapshots NCV (at rank 1) is ~0.99972. While for the yearly time scale, the lowest value (at rank 1) is 0.99965. The 0.99999 NCV corresponds to 2 modes for all time scales. This shows that with a few SVD ranks, reservoir pressure can be accurately reproduced. We ultimately chose 15 modes, which represent more than 0.99999 of NCV. This was done so the sparsity-promoting algorithm could work.

In contrast, the smallest $CO_2$ saturation SVD weekly and monthly snapshots NCV (at rank 1) is much lower than that of pressure, at ~0.7257, while for yearly snapshots is at ~0.7269. The 0.9999999 NCV corresponds to 887, 561, and 79, respectively. These values negate the usage of a uniform number of modes across time scales. Therefore, we used 900, 600, and 100 modes for weekly, monthly, and yearly time scales, respectively, that ensure capturing at least 0.9999999 of cumulative variance.

POD reconstruction errors are minimal in all cases. Calculation time ranges from ~500 to ~700 secs for the SVD for either pressure or saturation snapshots. The sparsity-promoting algorithm time ranges from

~10 to ~30 secs for pressure and ~100 to ~250 secs for saturation. POD reconstruction errors are presented in detail in the appendix's Table A1.

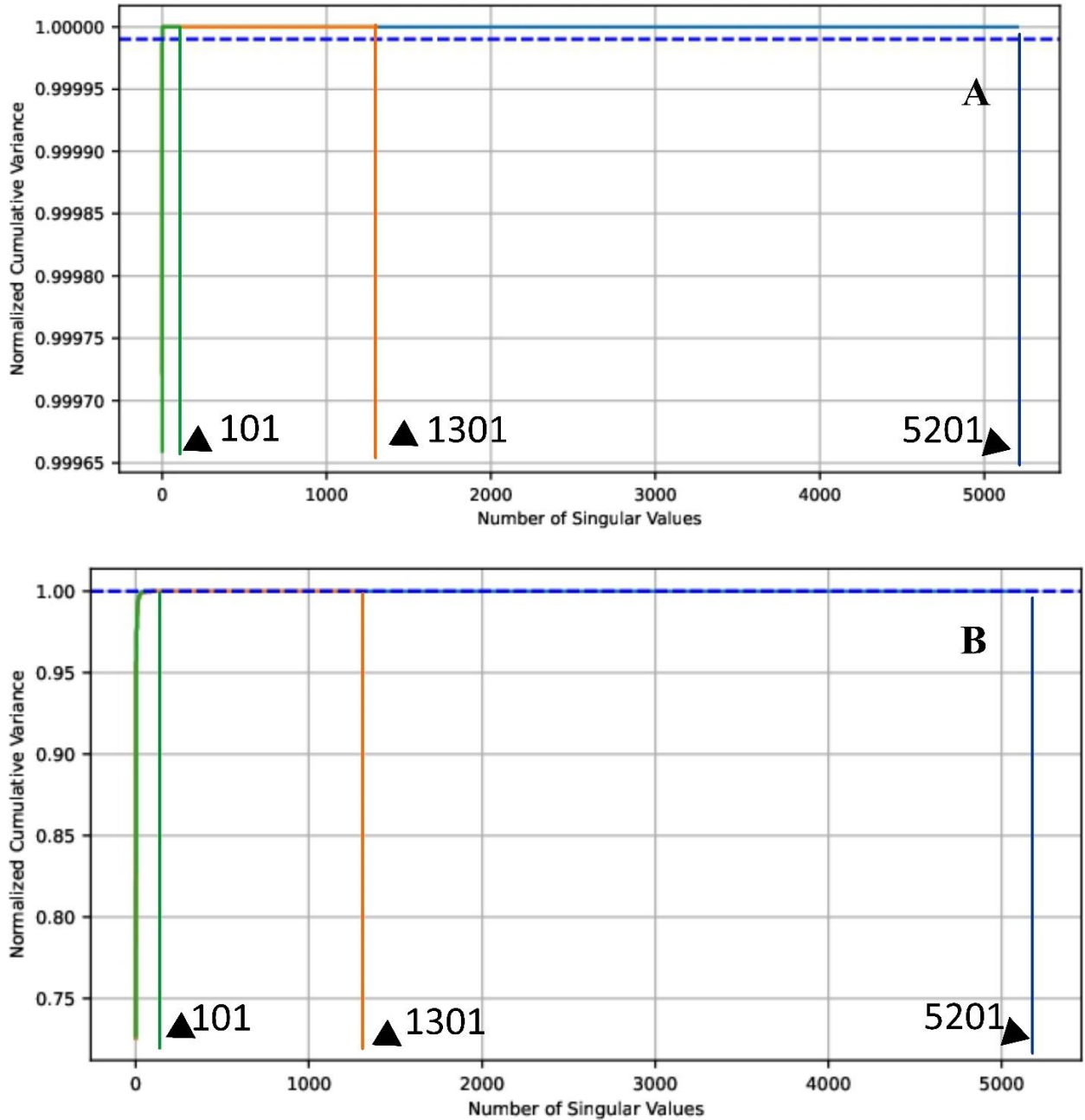

Fig. 4 – Normalized cumulative variance of pressure (A) and $CO_2$ saturation (B) singular values for yearly, monthly, and weekly snapshots of pressure (green, orange, and light blue, respectively). The dashed dark blue line denotes the 0.99999 and 0.9999999 variance thresholds.

4.2 DMDc/DMDspc Reconstruction Runtime and Memory Requirements

The forecast models' computational times are shown in Table A2 and Table A3 across each scale, case, variable, and period. Both DMDc and DMDspc are faster than the ECLIPSE 300 in every case, even when the two variables are computed sequentially, as shown in Table 1. ECLIPSE 300 average simulation time is ~223 mins for case A and ~140 mins for case B. In contrast, for the entire simulation period, the largest pressure DMDc time is ~1.28 mins and ~1.4 mins for case A and case B, respectively, while the largest saturation time is ~33 mins and ~31 mins for case A and case B, respectively. Both DMDc times are less by far, so they are the largest well-active DMDc times (less than 0.5 mins) from their respective ECLIPSE 300 times of ~45 mins and ~33 mins for case A and case B, respectively. DMDspc times yield smaller times than DMDc, as mentioned earlier in this section. On average, as shown by Table A2 and Table A3, case A and B times are similar since they have a similar number of computations. The overall DMD time reduction performance illustrates the speedup that data-driven model order reduction offers.

Regarding memory usage, ECLIPSE 300 simulations require 564 MB memory. DMDc memory requirements at any given time step, shown at Eq. (15) and (16) or Eq. (24) and (25), are the previous and next snapshot, the $\tilde{A}$, $\tilde{B}$, $\Phi$ or $\Theta$, $\Phi^{\dagger}$ or $\Theta^{\dagger}$, and a column of $Y$ that advances the current snapshot to the next state. This translates to the following generalized formula, respectively:

$$Total\ Memory = \frac{\left((2\ x\ n + r_c\ x\ r_c + r_c\ x\ k + n\ x\ r_c + r_c\ x\ n + l\ x\ 1)\ x\ 8\ Bytes\right)}{1024^2}\ MB \tag{28}$$

For example, run 10 case A weekly $CO_2$ saturation DMDc memory requirements are: $2\ x\ 135,340 + 900\ x\ 900 + 900\ x\ 1 + 135,340\ x\ 900 + 900\ x\ 135,340 + 1\ x\ 1 = 244,698,780$ double precision values or 1,957,590,240 Bytes or ~1,867 MB. Regarding run 7 case A weekly pressure for the same case, $2\ x\ 135,340 + 15\ x\ 15 + 15\ x\ 1 + 135,340\ x\ 15 + 15\ x\ 135,340 + 1\ x\ 1 = 4,331,121$ double precision values or 34,648,968 Bytes or ~33 MB. DMDspc memory requirements are significantly lower than those of DMDc, as $r_c$ is replaced by $r_s$ which can be substantially smaller. Similarly, as above, run 10 case A weekly $CO_2$ saturation DMDspc memory requirements are: $2\ x\ 135,340 + 493\ x\ 493 + 493\ x\ 1 + 135,340\ x\ 493 + 493\ x\ 135,340 + 1\ x\ 1 = 133,959,463$ double precision values or 1,071,675,704 Bytes or ~1,022 MB while run 5 case A weekly pressure memory requirements are: $2\ x\ 135,340 + 5\ x\ 5 + 5\ x\ 1 + 135,340\ x\ 5 + 5\ x\ 135,340 + 1\ x\ 1 = 1,624,111$ double precision values or 12,992,888 Bytes or ~12.39 MB

Table 1 – Average reconstruction times across forecast DMDc and DMDspc models for pressure and $CO_2$ saturation at weekly, monthly, and yearly time scales for case A: $CO_2$ injection and case B: $CO_2$ injection and water production

| | $CO_2$ Injection | | | |
|---|---|---|---|---|
| | Pressure | | $CO_2$ saturation | |
| | Simulation period | | | |
| | DMDc (mins) | DMDspc (mins) | DMDc (mins) | DMDspc (mins) |
| Weekly | 1.2844 | 1.1027 | 32.6102 | 18.6320 |
| Monthly | 0.2973 | 0.1733 | 6.7161 | 6.3609 |
| Yearly | 0.0196 | 0.0088 | 0.0897 | 0.0830 |
| | Wells Active | | | |
| Weekly | 0.1633 | 0.0556 | 5.6367 | 2.5913 |
| Monthly | 0.0624 | 0.0249 | 1.4815 | 1.4501 |
| Yearly | 0.0165 | 0.0054 | 0.0580 | 0.0527 |
| | $CO_2$ Injection and Water Production | | | |
| | Pressure | | $CO_2$ saturation | |
| | Simulation period | | | |
| | DMDc (mins) | DMDspc (mins) | DMDc (mins) | DMDspc (mins) |
| Weekly | 1.3959 | 1.0087 | 30.6388 | 22.1182 |
| Monthly | 0.2844 | 0.1458 | 7.0348 | 6.4489 |
| Yearly | 0.0271 | 0.0138 | 0.0689 | 0.0735 |

| | Wells Active | | | |
|---|---|---|---|---|
| Weekly | 0.1417 | 0.0676 | 5.7418 | 4.2037 |
| Monthly | 0.0207 | 0.0089 | 1.5765 | 1.6080 |
| Yearly | 0.0172 | 0.0069 | 0.0467 | 0.0369 |

4.3 DMDc/DMDspc Training

Training starts with SVD rank selection (see subsection 4.1), then with DMDc training, and finally with the DMDsp algorithm, which determines the most important DMD modes, as mentioned earlier. We validate DMD results by comparing the reconstructed results with the original training data. For saturation, the yearly DMDc error is between 1.44% and 3.7%, while monthly and weekly errors range from 0.075% to 0.119% and 0.078% to 0.16%, respectively. Yearly pressure DMDc error is between 0.0352% to 0.4738%, while monthly and weekly errors range from 0.0013% to 0.0144% and 0.0003% to 0.0062%. Table 2 summarises the percent change error (PCE) for runs 2, 5, 7, and 10 trained DMDc models.

Table 2 – DMDc PCE for pressure and $CO_2$ saturation at weekly, monthly, and yearly time scales

| $CO_2$ Injection | | | |
|---|---|---|---|
| Pressure | Weekly | Monthly | Yearly |
| Run 2 | 0.0009 | 0.0056 | 0.1793 |
| Run 5 | 0.0003 | 0.0013 | 0.0352 |
| Run 7 | 0.0009 | 0.0057 | 0.2690 |
| Run 10 | 0.0003 | 0.0014 | 0.0530 |
| $CO_2$ Saturation | Weekly | Monthly | Yearly |
| Run 2 | 0.078 | 0.105 | 1.679 |
| Run 5 | 0.139 | 0.080 | 2.831 |
| Run 7 | 0.079 | 0.119 | 2.041 |
| Run 10 | 0.142 | 0.087 | 3.442 |
| $CO_2$ Injection and Water Production | | | |
| Pressure | Weekly | Monthly | Yearly |
| Run 2 | 0.0059 | 0.0142 | 0.4738 |
| Run 5 | 0.0007 | 0.0030 | 0.1481 |
| Run 7 | 0.0062 | 0.0144 | 0.2260 |
| Run 10 | 0.0007 | 0.0031 | 0.1238 |
| $CO_2$ Saturation | Weekly | Monthly | Yearly |
| Run 2 | 0.157 | 0.075 | 1.440 |
| Run 5 | 0.157 | 0.093 | 3.038 |
| Run 7 | 0.101 | 0.075 | 2.041 |
| Run 10 | 0.160 | 0.090 | 3.694 |

The DMDsp keeps the most important DMD modes and discards the rest. This leads to added error and computational time reduction in general, as mentioned earlier. However, there are cases where the algorithm removes "noisy" DMD modes, which improves accuracy, either during the reconstruction of training data and/or during forecast while maintaining computational time reduction (Fig. 5 A1 and see subsection 4.2.2). This is observed only for pressure in this study. The PCE stays the same as DMDc, where the DMDsp was not able to remove any DMD modes and increased when DMD modes were suppressed. Fig. 5 and Fig. 6 show representative percent loss versus reduced model order for DMDsp models trained individually with runs 2, 5, 7, and 10. Fig. 5 A1 shows accuracy improvement in reconstructing pressure by removing "noisy" DMD modes. All other cases in that figure show a slight reduction in accuracy by removing modes. Table 3 presents the total PCE when reconstructing DMDspc training data, which includes the sequentially added error caused by SVD, DMDc, and DMDsp for both variables of interest across all time scales.

DMDspc managed to reduce the number of DMD modes regarding pressure across almost all trained models (see Table 3). Although the accuracy decreased by several orders of magnitude for weekly and monthly time scales, this new accuracy is still good given the number of modes used, the resulting speed increase, and that the simulated and reconstructed snapshots are rough estimates due to data unavailability in reservoirs. DMDspc yearly trained models showed a slight reduction in accuracy within the same order of magnitude. The monthly run 7 trained model showed a drop of two magnitude of order, while the rest of the models showed approximately three to four orders drop for both cases A and B. Case A has a 0.27 average PCE for removing an average of nine modes, while case B showed an average PCE of 0.35 with the removal of six modes on average.

Regarding $CO_2$ saturation, DMDspc did not manage to decrease DMD modes without introducing significant errors in general. Only 5 out of 24 trained models maintained accuracy with fewer DMD modes (see Table 3). Fig. 6 B1 and B2 show a slight reduction in $CO_2$ saturation accuracy by removing a considerable number of DMD modes. Weekly and monthly errors are like those of pressure except for the weekly run 7 trained model, which yields the same two orders of magnitude higher errors as the yearly trained models. Case B has a 0.16 weekly average PCE for removing 59 modes on average and a 0.09 monthly average PCE for removing 105 modes on average.

Table 3 – DMDspc PCE for pressure and $CO_2$ saturation at weekly, monthly, and yearly time scales

| Case A - $CO_2$ Injection | | | |
| --- | --- | --- | --- |
| Pressure (PCE) / Modes | Weekly | Monthly | Yearly |
| Run 2 | 0.51 / 8 | 0.25 / 3 | 0.51 / 8 |
| Run 5 | 0.2 / 5 | 0.15 / 4 | 0.24 / 5 |
| Run 7 | 0.0009 / 15 | 0.06 / 2 | 0.36 / 10 |
| Run 10 | 0.23 / 2 | 0.2 / 3 | 0.2 / 6 |
| $CO_2$ saturation (PCE) / Modes | Weekly | Monthly | Yearly |
| Run 2 | 0.08 / 900 | 0.11 / 600 | 1.68 / 100 |
| Run 5 | 0.14 / 900 | 0.08 / 600 | 2.83 / 100 |
| Run 7 | 1 / 493 | 0.12 / 600 | 2.04 / 100 |
| Run 10 | 0.14 / 900 | 0.08 / 600 | 3.44 / 98 |
| Case B - $CO_2$ Injection and Water Production | | | |
| Pressure (PCE) / Modes | Weekly | Monthly | Yearly |
| Run 2 | 0.35 / 12 | 0.33 / 9 | 0.5 / 12 |
| Run 5 | 0.55 / 5 | 0.29 / 3 | 0.245 / 8 |
| Run 7 | 0.26 / 10 | 0.45 / 11 | 0.4 / 9 |
| Run 10 | 0.5 / 4 | 0.14 / 4 | 0.28 / 13 |
| $CO_2$ saturation (PCE) / Modes | Weekly | Monthly | Yearly |
| Run 2 | 0.2 / 710 | 0.08 / 498 | 1.44 / 100 |
| Run 5 | 0.16 / 900 | 0.09 / 600 | 3.04 / 100 |
| Run 7 | 0.1 / 801 | 0.1 / 496 | 2.04 / 99 |
| Run 10 | 0.16 / 900 | 0.09 / 600 | 3.69 / 100 |

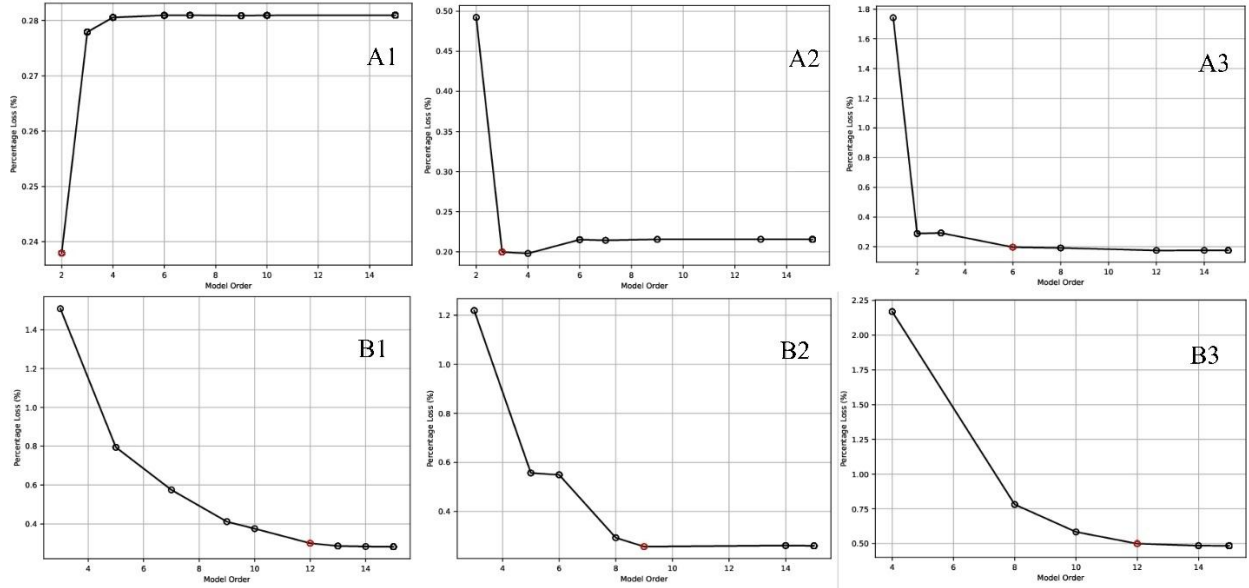

Fig. 5 – Pressure PCE by DMDspc for case A – Run 10 ($CO_2$ injection) and case B – Run 2 ($CO_2$ injection and water production). 1 denotes weekly, 2 monthly, and 3 yearly time scales.

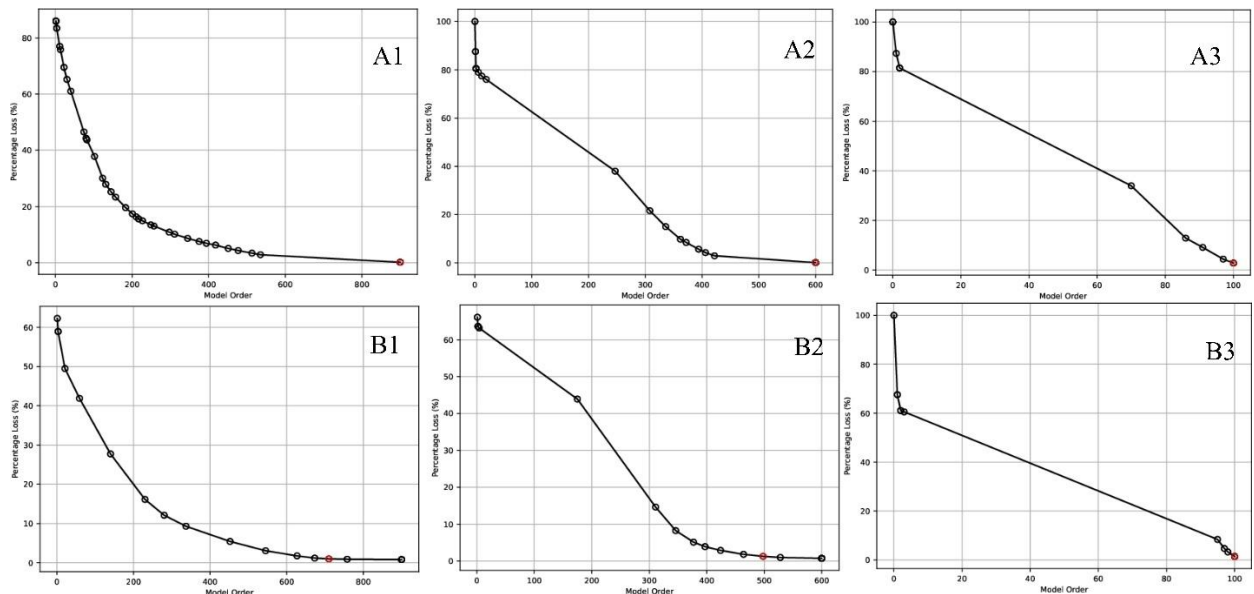

Fig. 6 – $CO_2$ saturation PCE by DMDspc for case A – Run 5 ($CO_2$ injection) and case B - Run 2 ($CO_2$ injection and water production). 1 denotes weekly, 2 monthly, and 3 yearly time scales

4.4 DMDc/DMDspc Forecast and Temporal Analysis

Control input optimization—in this case, well rates—may result in optimized values that differ significantly from those used to train surrogate models. This necessitates an evaluation of a given surrogate model's forecast capabilities, which is the subject of this subsection. Here, we discuss the global average forecast performance across the different DMDc/DMDspc models for all three time scales, which comes with a cross-comparison against the rest of the ECLIPSE 300 simulated snapshots. Models with exceedingly high error are excluded from the analysis. Complete tables are presented in the Appendix (Table A4 - Table A7) This assists in selecting the most robust DMDc/DMDspc models that can both serve in optimization and long term plume monitoring.

Regarding the well active period case A pressure, DMDc PCE averages 1.8%, 12%, and 3.2% while DMDspc PCE is 2.4%, 9.8%, and 2.3% for weekly, monthly, and yearly time scales, respectively. Only at the weekly time scale, some slight error is added by the DMDspc, while at monthly and yearly time scales, the error decreased due to noisy mode removal. Here, almost all models are capable of forecasting different pressure states with less than 5% error. In contrast, simulation period case A pressure DMDc PCE averages 8% and 5.5%, excluding weekly run 10 and yearly run 5, 10 models, while DMDspc PCE averages 8.1% and 12%, excluding weekly run 2 and yearly run 5, 10 models, for weekly and yearly time scales, respectively. Monthly DMD models are excluded due to high error. Errors for this period are larger than those for the well's active period. Weekly and yearly run 2 DMDc and weekly run 7 and yearly run 2 DMDspc models reliably forecast long-term pressure states.

Regarding the active well period case B pressure, we noticed smaller accuracy errors than case A. DMDc PCE averages at 3.35%, 7%, and 5%, excluding weekly run 10 model, while DMDspc PCE averages at 49%, 10.2%, and 7.32%, excluding monthly run 5, 10 models, for weekly, monthly, and yearly time scales, respectively. In contrast, the entire simulation period case B pressure DMDc PCE averages 13.4%, 21.1%, and 13.3%, excluding weekly 7,10, monthly 2,10, and yearly 5,10 models, while DMDspc PCE averages at 15.7%, 22.5%, and 11.7%, excluding weekly run 7,10, monthly run 2 and yearly run 5,10 models, for weekly, monthly, yearly time scales, respectively.

Regarding case A $CO_2$ saturation, DMDc and DMDspc give the same errors since DMDsp did not manage to discard any DMD modes, as mentioned earlier. For the well's active period, only the weekly run 2 trained DMDc and DMDspc models were not able to short-range forecast. DMDc and DMDspc MAE averages at 0.0075 across all time scales. All DMD models are capable of forecasting different saturation states, except for weekly run 2 trained DMD models. Similarly, for the simulation period weekly time scale, only run 7 trained DMDc and DMDspc models managed to forecast different saturation states with 0.0052 average MAE. At the monthly time scale, DMDc and DMDspc MAE averages at 0.0052 and 0.0146 for run 7 and 10 trained DMDspc models, respectively. For the yearly time scale, all DMD models managed to forecast different saturation states with a 0.0114 MAE across all model forecasts. Run 7 trained DMD models are the best at long-term forecast across all time scales.

Regarding active wells period case B $CO_2$ saturation, weekly DMDc and DMDspc MAE averages at 0.0063, excluding run 5 and 10 trained DMD models, while monthly and yearly trained DMDc and DMDspc MAE averages at 0.0126 and 0.0109, respectively. For the entire simulation period, weekly DMDc and DMDspc MAE averages at 0.0083 for the run 2 and 7 trained DMD models. At the monthly time scale, DMDc MAE averages at 0.0094 while DMDspc MAE averages at 0.0082, excluding run 5 and 10 trained DMDspc models. At the yearly time scale, DMDc and DMDspc MAE averages at 0.012 with no DMD model excluded.

Finally, based on the DMD models' forecasting behavior, it is shown that DMD forecasts pressure with greater accuracy when it is trained on snapshots that correspond to relatively low well rates in contrast to $CO_2$ saturation forecasts, which show greater accuracy when trained on snapshots that correspond to relatively high well rates. Also, the DMDsp manages to discard modes for pressure with a high-value $\Sigma$ interval, $10^4 - 10^6$, and for $CO_2$ saturation with a low-value $\Sigma$ interval, $10^{-1}$-$10^4$.

### 4.4.1 Temporal Evolution of Primary Variables

The global (the entire reservoir) temporal forecast evolution, which shows how reconstructed DMD snapshots capture temporal fluid dynamics within the reservoir. Similarly, as before this forecast evolution comes with a cross-comparison and figures of select and representative DMDc/DMDspc models against the rest of the ECLIPSE 300 simulated snapshots at weekly, monthly, and yearly time scales (e.g., run 2 with run 7 trained DMDc model). Pearson's correlation coefficient (R), in addition to PCE and MAE, was selected as a global temporal error evolution metric to show how close reconstructed values are to those produced by ECLIPSE 300 simulations. Below, some representative DMD models are presented to discuss the temporal evolution of two global variables. Case A includes: 1. run 7 simulated pressure with run 5 trained DMD models snapshots and 2. run 5 simulated $CO_2$ saturation with run 7 trained DMD models snapshots. Case B includes: 1. run 10 weekly simulated pressure 10 with run 5 trained DMD models snapshots, 2. run 2 monthly and yearly pressure with run 7 trained DMD models snapshots, and 3. run 10 weekly and monthly $CO_2$ saturation with run 2 trained DMD models snapshots.

Regarding case A, weekly and yearly pressure R is at least 0.999 across the entire simulation period, while MAE progressively increases to 100 bars as time passes. DMDc and DMDspc show the same trend on average (Fig. 7 and Fig. 9). Monthly pressure R is at least above 0.98 for the well active period in contrast for the rest of the simulation period where both the DMD models deteriorate to very low R and very high MAE (Fig. 8). Weekly, monthly, and yearly $CO_2$ saturation MAE and R show similar curve shapes. MAE increases up to 0.004 while R stabilizes at around 0.4, which shows a poor correlation between reconstructed and original snapshots (Fig. 10 – Fig. 12).

Regarding case B, monthly and yearly pressure R is always above 0.998, while weekly is always above 0.999. In addition, it improves with time after its drop at the well active period, unlike R at forecasting run 7 with run 5 trained DMD models case (Fig. 7 - Fig. 9). Weekly DMDc MAE below 1 bar while DMDspc is below 10 bars. Monthly and yearly MAE average on ~5 bars and have similar temporal evolution at both DMDc and DMDspc (Fig. 13 - Fig. 15). $CO_2$ saturation MAE is always below 0.01 across all time scales. Wells active period MAE keeps increasing while late simulation MAE stabilizes at around 0.006. Weekly and monthly R DMDc decrease from 0.8 and ~0.75 to 0.1 while DMDspc loses a little accuracy at the simulation period's beginning, but it increases to the same value as that of DMDc (Fig. 16 and Fig. 17).

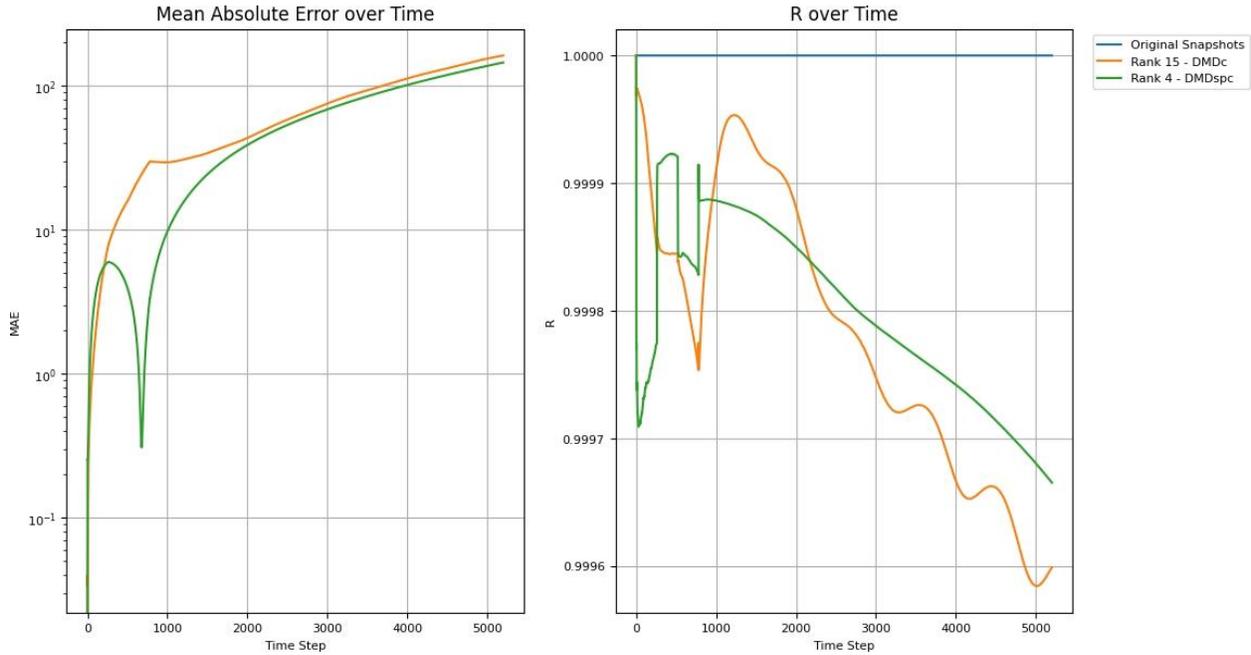

Fig. 7 – Simulation period weekly pressure MAE and R over time metrics for forecasting run 7 with run 5 trained DMD models. Case A: $CO_2$ injection

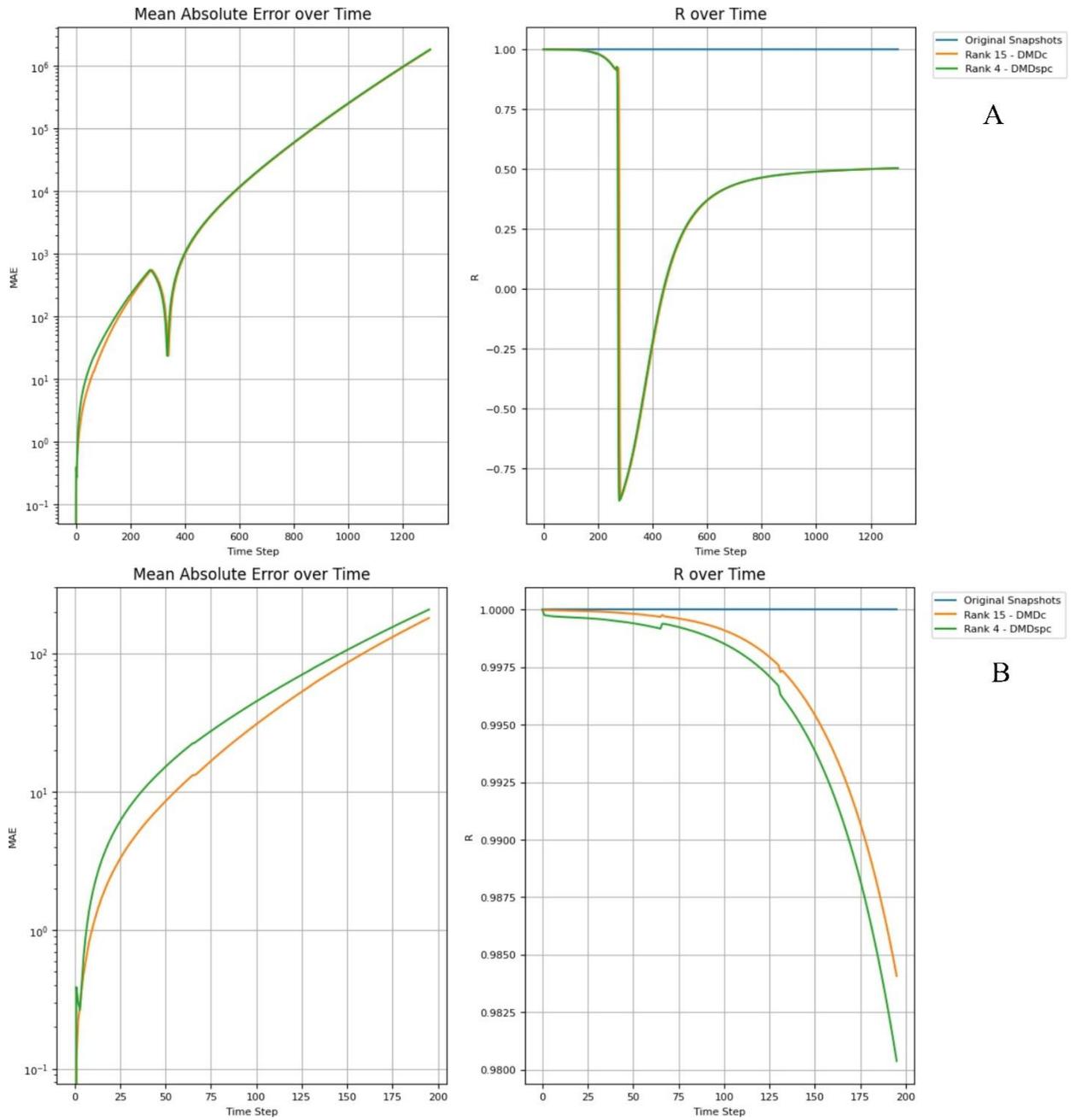

Fig. 8 – Simulations period monthly pressure MAE and R over time metrics for forecasting run 7 with run 5 trained DMD models. Case A: $CO_2$ injection. A: Entire simulation period, B: Well active

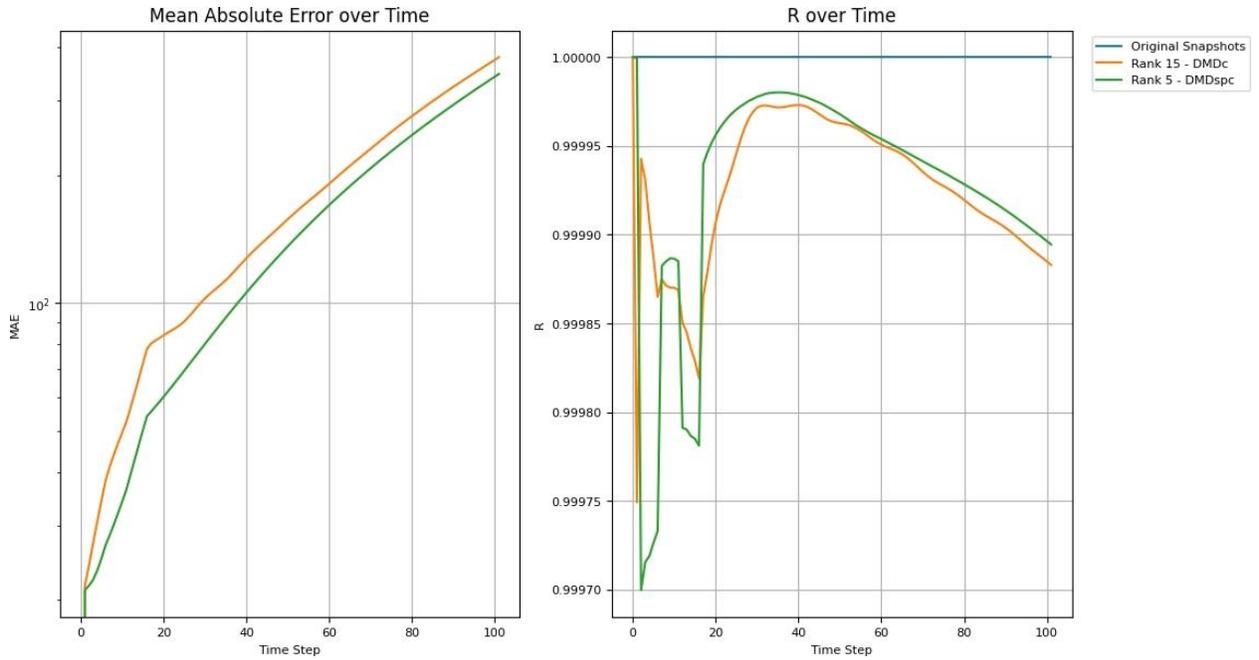

Fig. 9 – Simulation period yearly pressure MAE and R over time metrics for forecasting run 7 with run 5 trained DMD models. Case A: $CO_2$ injection

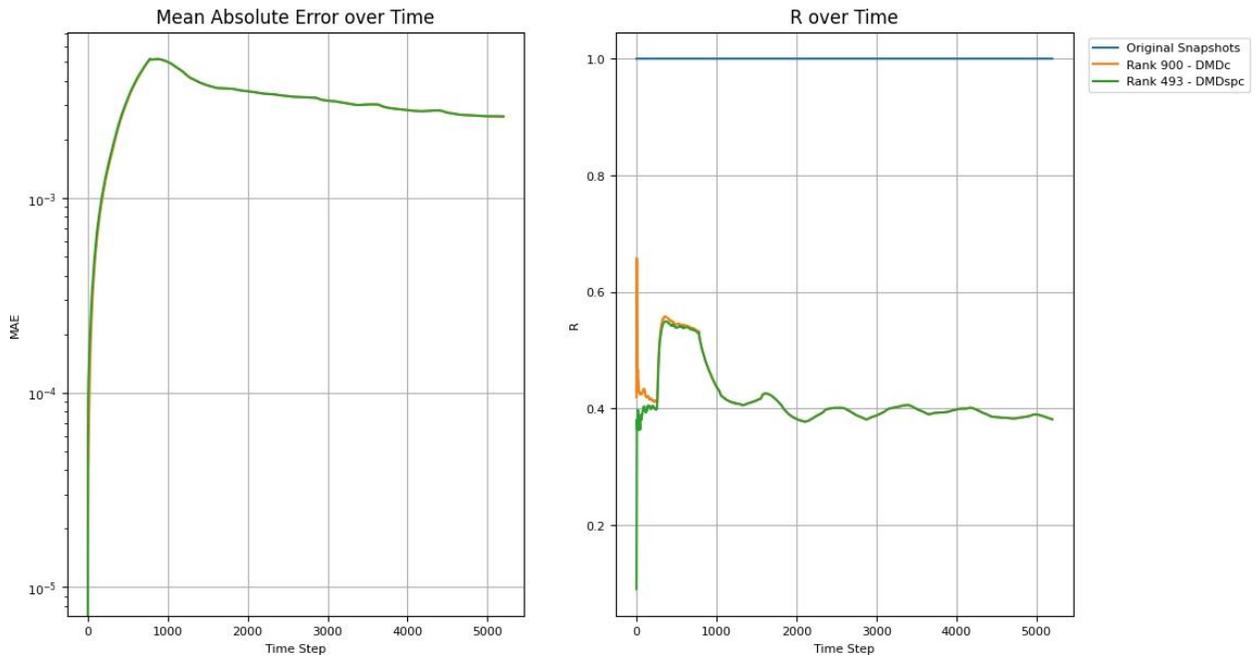

Fig. 10 – Simulation period weekly $CO_2$ saturation MAE and R over time metrics for forecasting run 5 with run 7 trained DMD models. Case A: $CO_2$ injection

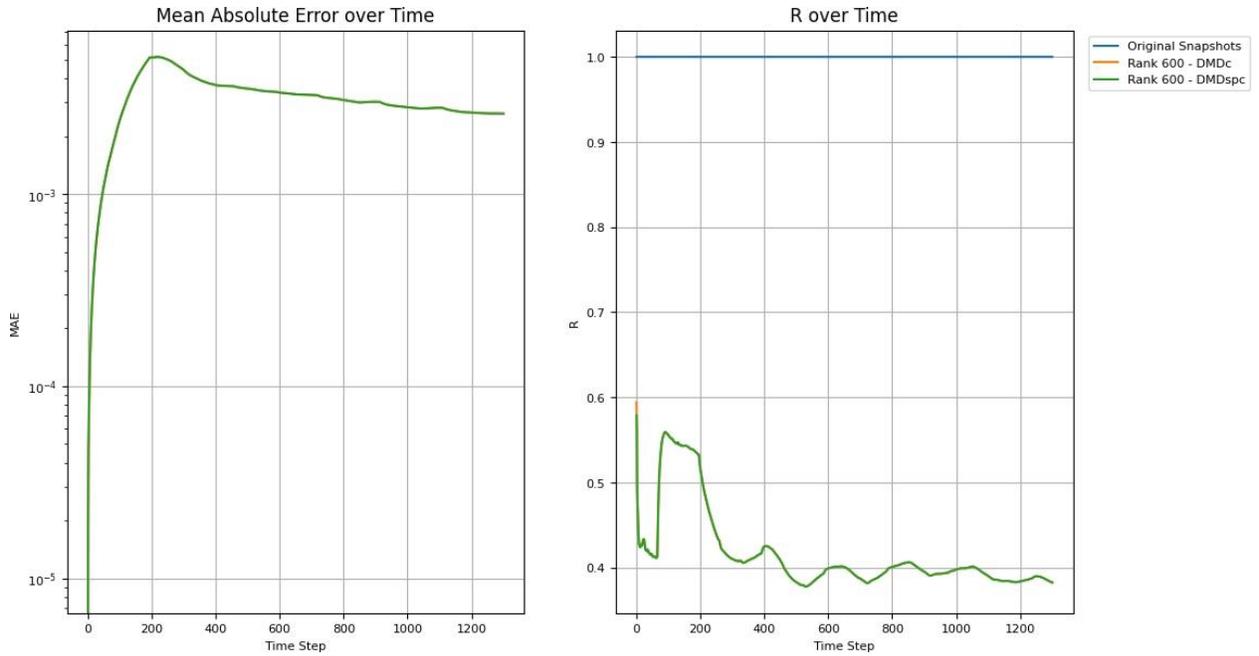

Fig. 11 – Simulation period monthly $CO_2$ saturation MAE and R over time metrics for forecasting run 5 with run 7 trained DMD models. Case A: $CO_2$ injection

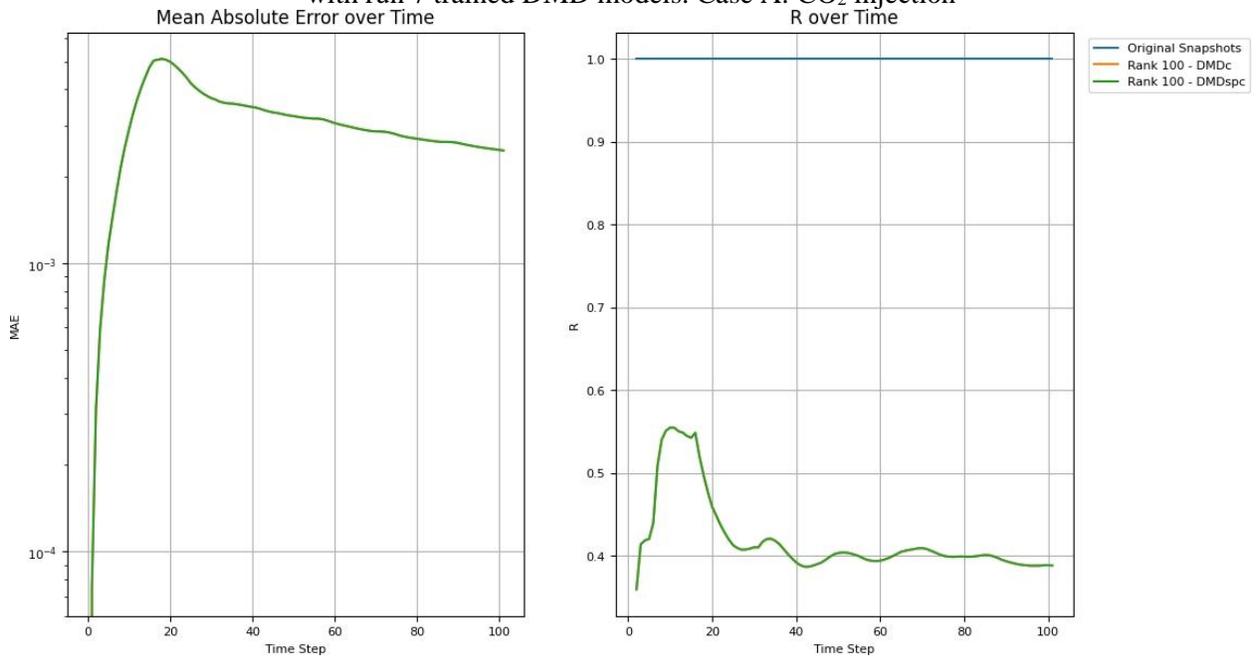

Fig. 12 – Simulation period yearly MAE and R over time metrics for forecasting run 5 $CO_2$ saturation with run 7 trained DMD models. Case A: $CO_2$ injection

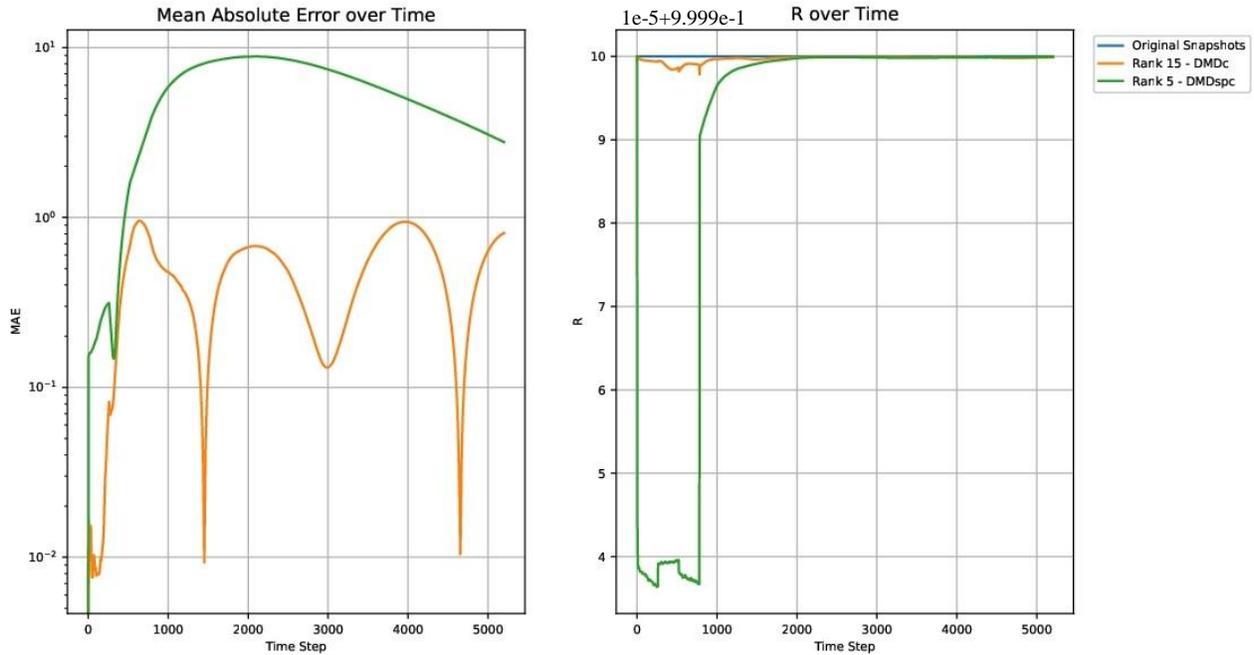
Fig. 13 – Simulation period weekly pressure MAE and R over time metrics for forecasting run 10 with run 5 trained DMD models. Case B: $CO_2$ injection and water production

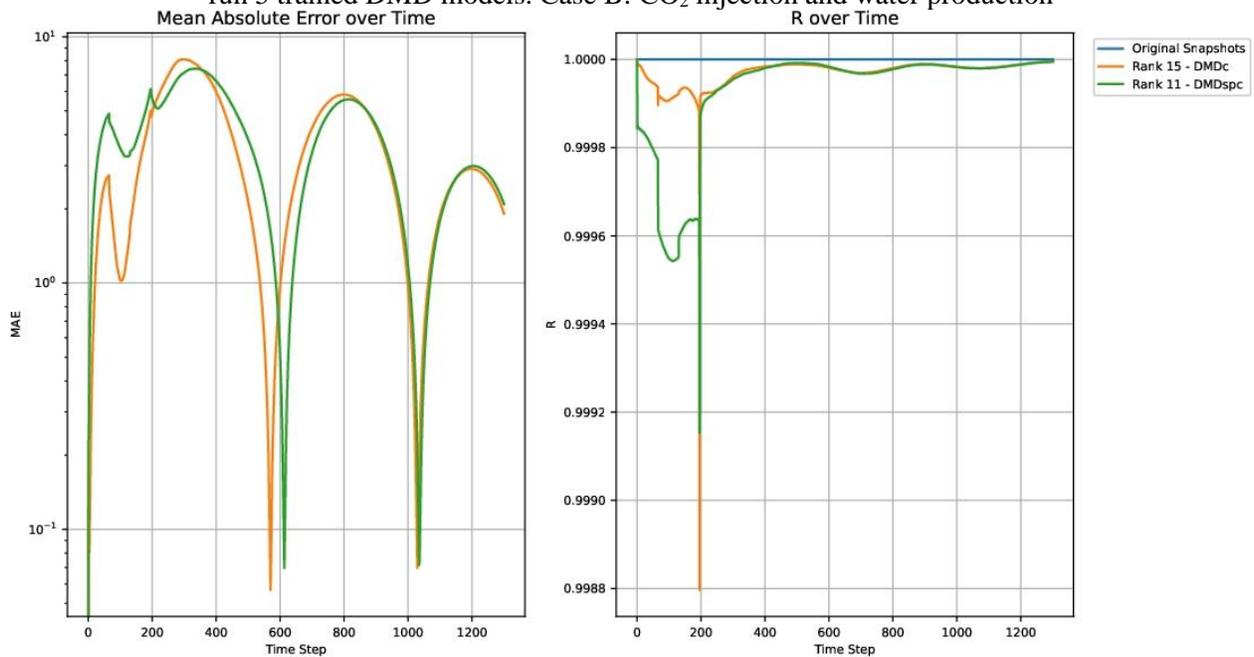
Fig. 14 – Simulations period monthly pressure MAE and R over time metrics for forecasting run 2 with run 7 trained DMD models. Case B: $CO_2$ injection and water production

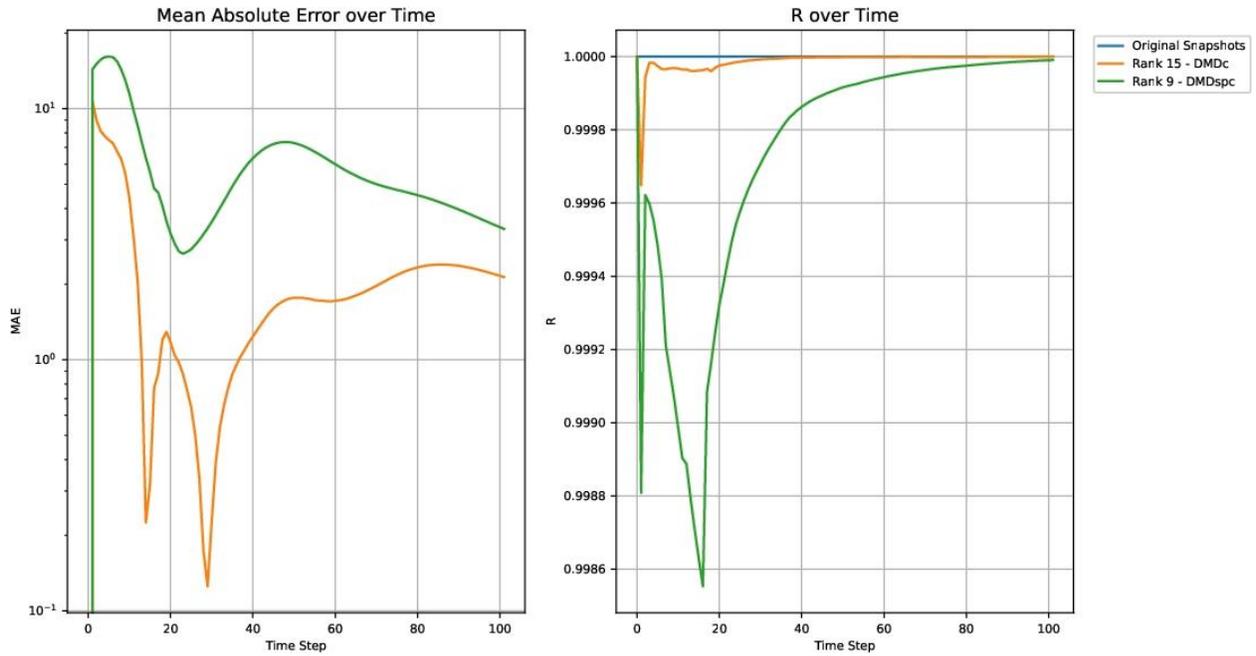

Fig. 15 – Simulation period yearly pressure MAE and R over time metrics for forecasting run 2 with run 7 trained DMD models. Case B: $CO_2$ injection and water production

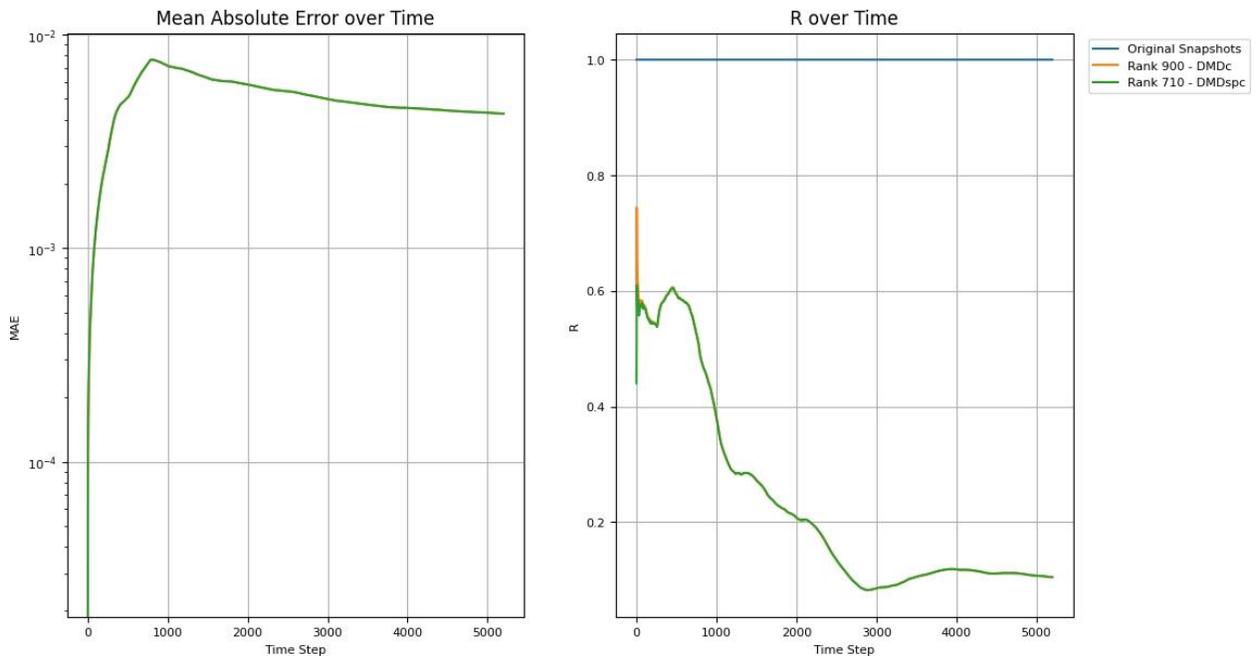

Fig. 16 – Simulation period weekly $CO_2$ saturation MAE and R over time metrics for forecasting run 10 with run 2 trained DMD models. Case B: $CO_2$ injection and water production

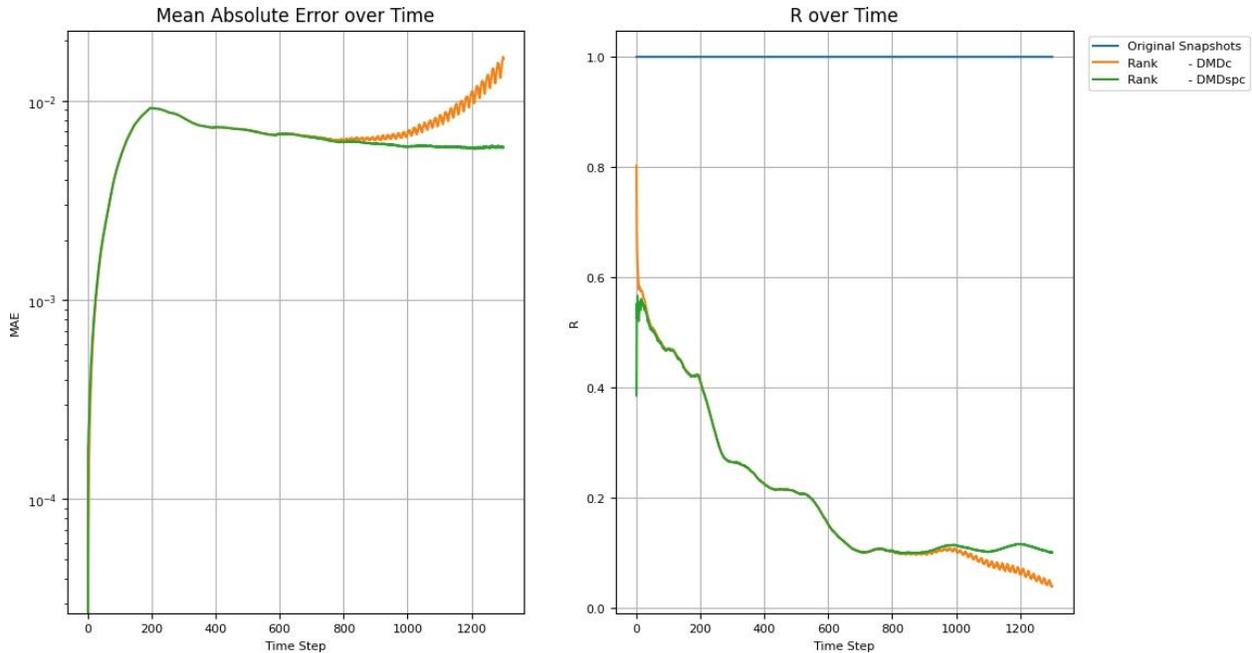

Fig. 17 – Simulation period monthly $CO_2$ saturation MAE and R over time metrics for forecasting run 10 with run 2 trained DMD models. Case B: $CO_2$ injection and water production

### 4.4.2 Temporal Evolution of Select Secondary Variables

While the temporal evolution of primary variables demonstrates the ability of reconstructed DMD snapshots to capture fluid dynamics within the reservoir, the requirements of risk assessment necessitate the inclusion of additional secondary variables to ensure a comprehensive assessment of storage performance. BHP management ensures maximum injection rate while maintaining operation safety. Dissolved $CO_2$ are one of the main trapping mechanisms of geological $CO_2$ storage and is vital in assessing storage efficiency. Below we discuss some representative cases. BHP cases includes: 1. Case A run 7 $CO_2$ injection well BHP run 5 trained DMD models, and 2. Case B monthly and yearly run 2 $CO_2$ injection well and water production well BHP with run 7 trained DMD models. Dissolved $CO_2$ cases includes: 1. Case A run 5 dissolved $CO_2$ with run 7 trained DMD models, and case B run 10 dissolved $CO_2$ with run 2 trained DMD models.

Regarding case A run 7 $CO_2$ injection well BHP run 5 trained DMD models weekly and yearly BHP DMD reconstructed values have less than 5% overestimation difference from the original values. Weekly BHP DMDc PCE increases up to ~4.5% while DMDspc PCE is at 2% at the beginning of the active well period and afterward drops close to zero (Fig. 18). Monthly DMDc and DMDspc PCE values both overestimate BHP by ~32% and ~36%, respectively (Fig. 19). Yearly DMDc PCE increases to ~12% and DMDspc increases to ~8%, an improvement of ~4% (Fig. 20). Case B weekly DMDc slightly underestimates the water production well and $CO_2$ injection well BHP by 0.3% and 0.5% PCE peak values, respectively. DMDspc, on the other hand, does the opposite and at higher peak values of ~1.4% and 1.75%, respectively (Fig. 21). Monthly DMDc and DMDspc slightly overestimate the two wells' BHP by 2.8% average peak values. The influence of the well rate change is reflected by the PCE curves (Fig. 22). Yearly DMD models underestimate the BHP by a higher peak average of 4.75% (Fig. 23).

Forecasted dissolved $CO_2$ in the reservoir closely matches original values across all time scales and DMD models presented here. Dissolved $CO_2$ amounts averages to ~0.000175 Mtons for all three time scales. Yearly DMDc and DMDspc errors show similar behavior monthly DMDc and DMDspc errors (Fig. 24 - Fig. 26). Weekly and yearly dissolved $CO_2$ amount to ~0.00017 Mtons while monthly dissolved $CO_2$ amounts to ~0.00018 Mtons same as case A. (Fig. 27 -Fig. 29).

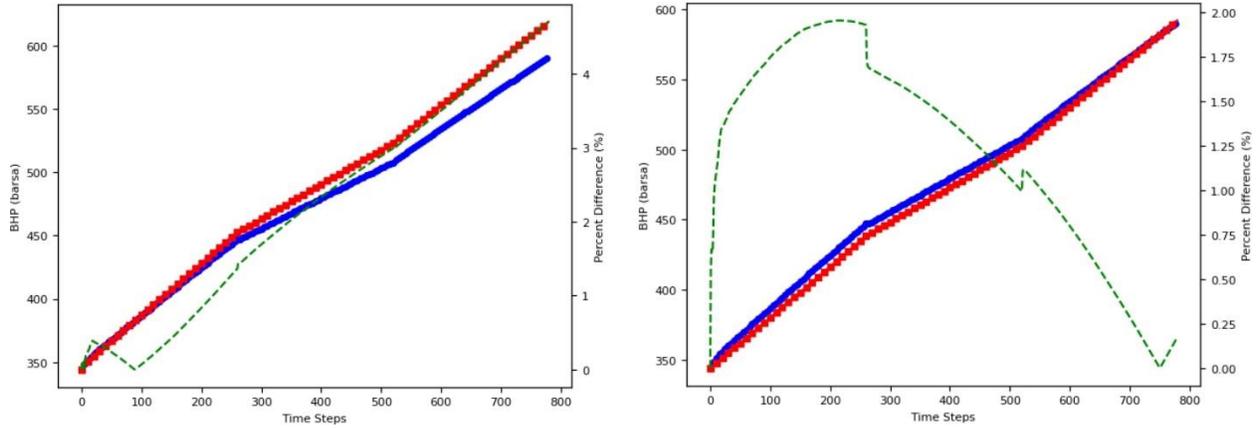

Fig. 18 – Simulation period weekly run 7 $CO_2$ injection well BHP (blue line – original values) forecasted with run 5 trained DMD models (red line – forecasted values) by DMDc - 15 modes (left) and DMDspc – 4 modes (right). Case A: $CO_2$ injection. Green dashed line denotes the percent difference between the two.

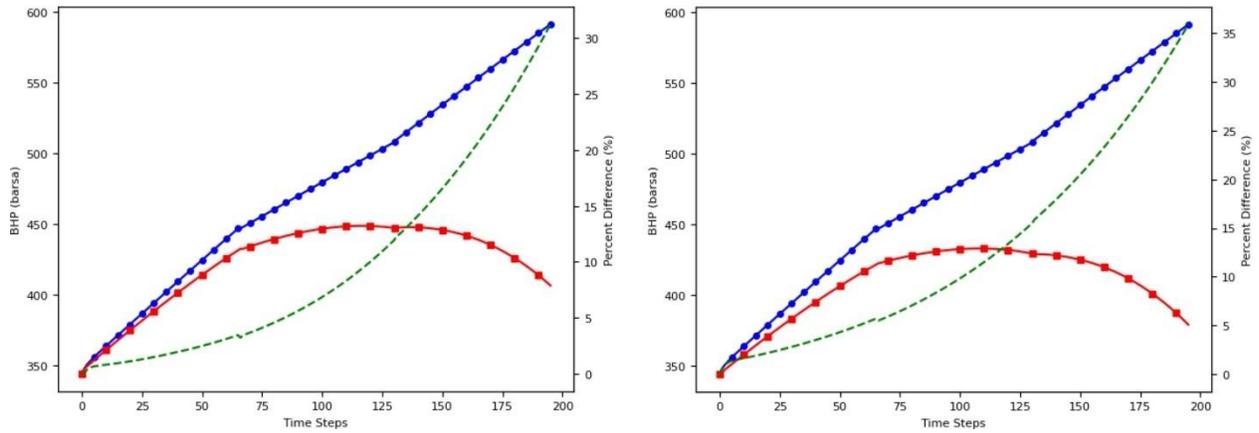

Fig. 19 – Simulation period monthly run 7 $CO_2$ injection well BHP (blue line – original values) forecasted with run 5 trained DMD models (red line – forecasted values) by DMDc - 15 modes (left) and DMDspc – 4 modes (right). Case A: $CO_2$ injection.

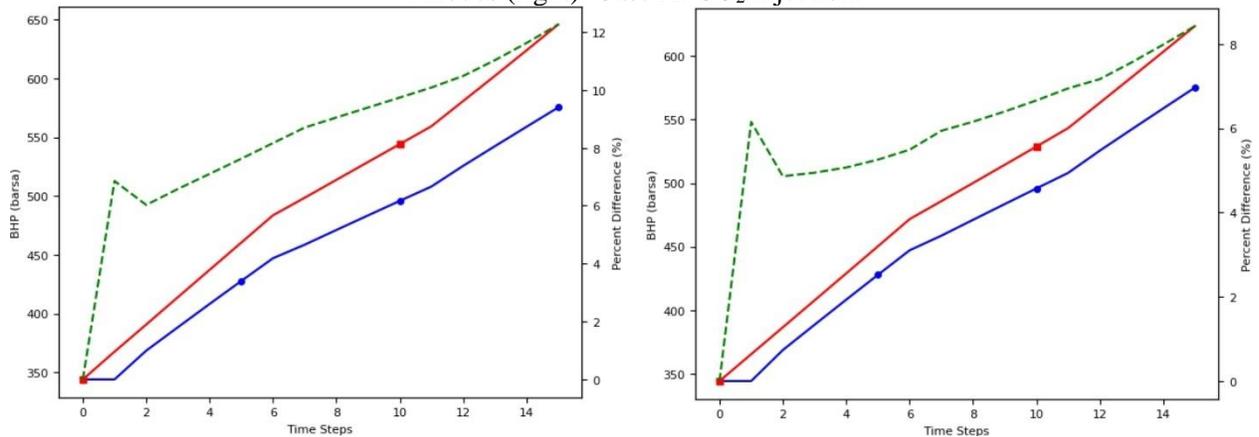

Fig. 20 – Simulation period yearly run 7 $CO_2$ injection well BHP (blue line – original snapshots) forecasted with run 5 trained DMD (red line – forecasted snapshots) by DMDc - 15 modes (left) and DMDspc – 5 modes (right). Case A: $CO_2$ injection.

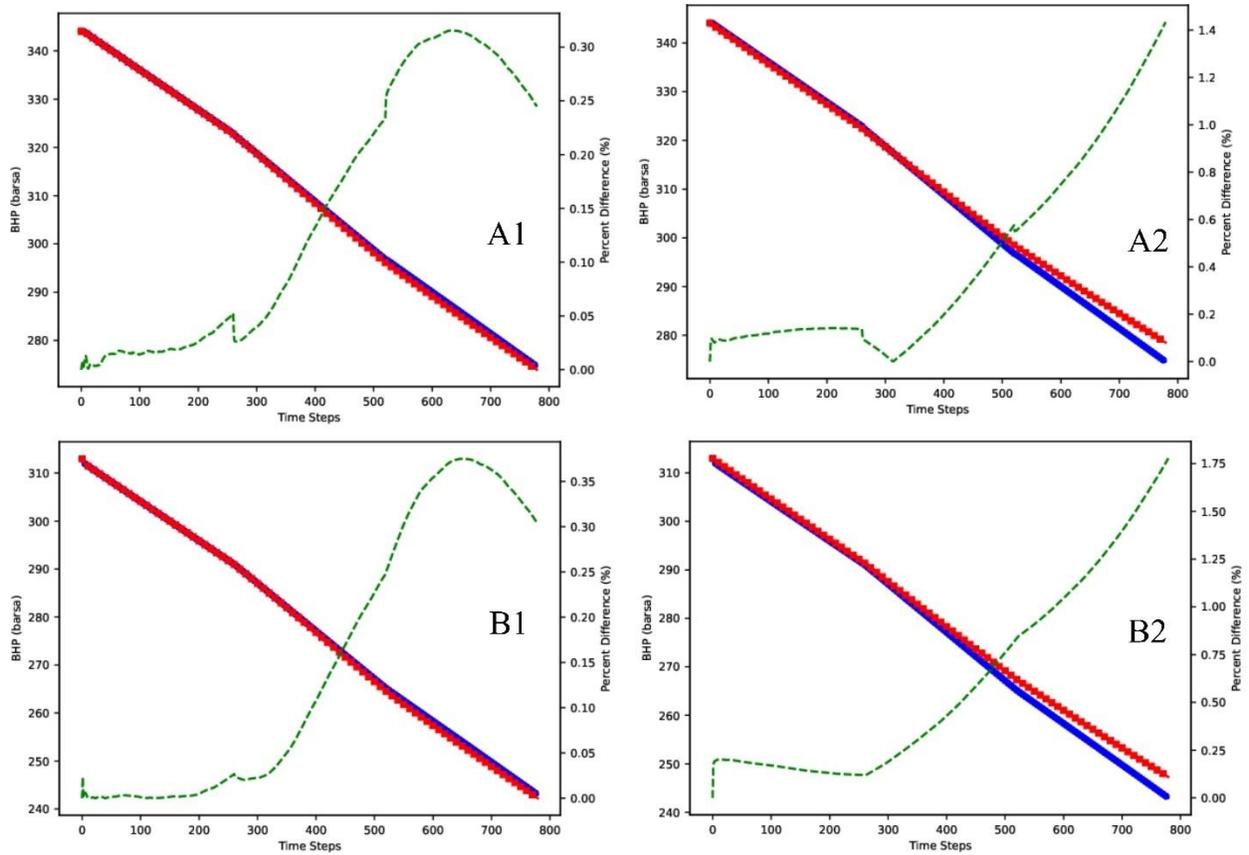

Fig. 21 – Simulation period weekly run 10 BHP (blue line – original values) forecasted with run 5 trained DMD (red line – forecasted values) by DMDc - 15 modes (left column) and DMDspc – 5 modes (right column). A: $CO_2$ injection well BHP, B: Water production well BHP. Case B: $CO_2$ injection and water production.

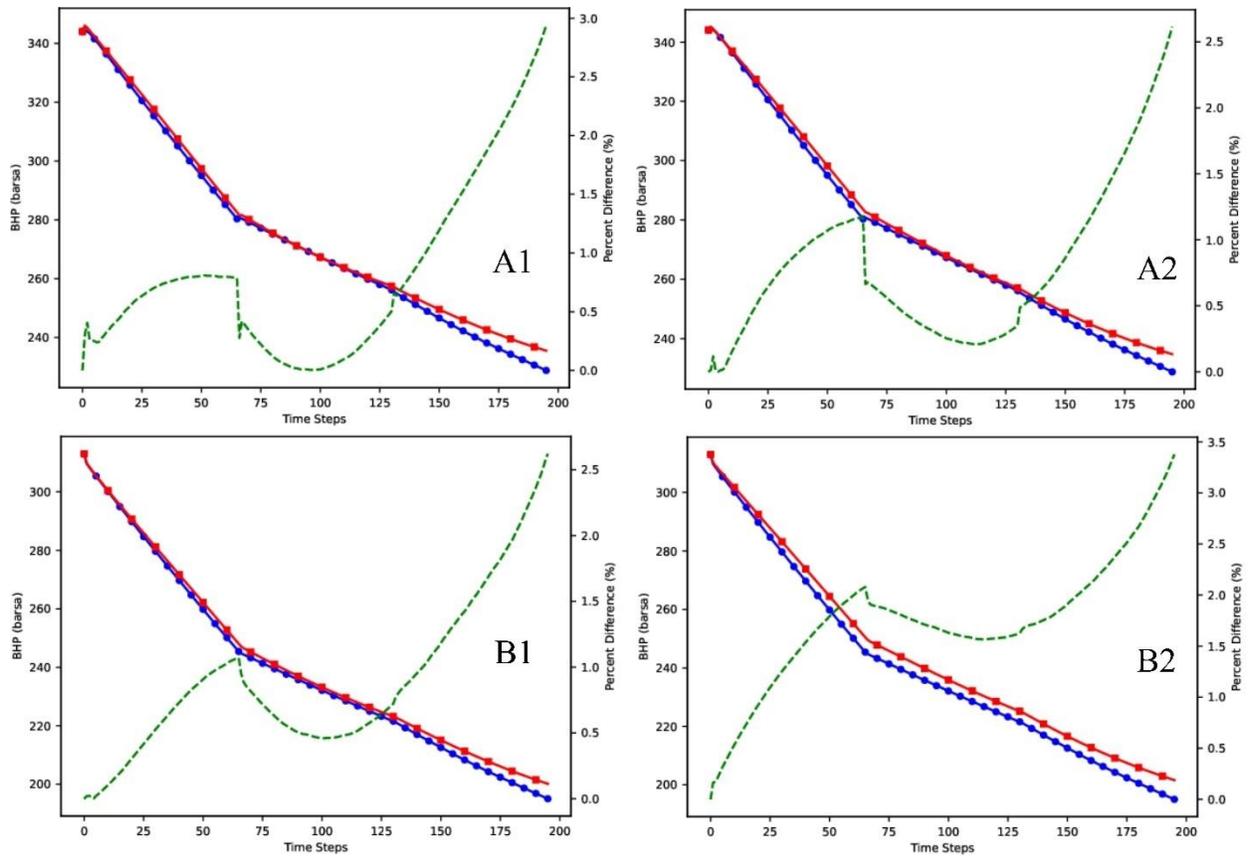

Fig. 22 – Simulation period monthly run 2 BHP (blue line – original values) forecasted with run 7 trained DMD (red line – forecasted values) by DMDc - 15 modes (left column) and DMDspc – 11 modes (right column). A: $CO_2$ injection well BHP, B: Water production well BHP, 1: entire simulation period, 2: wells active period. Case B: $CO_2$ injection and water production.

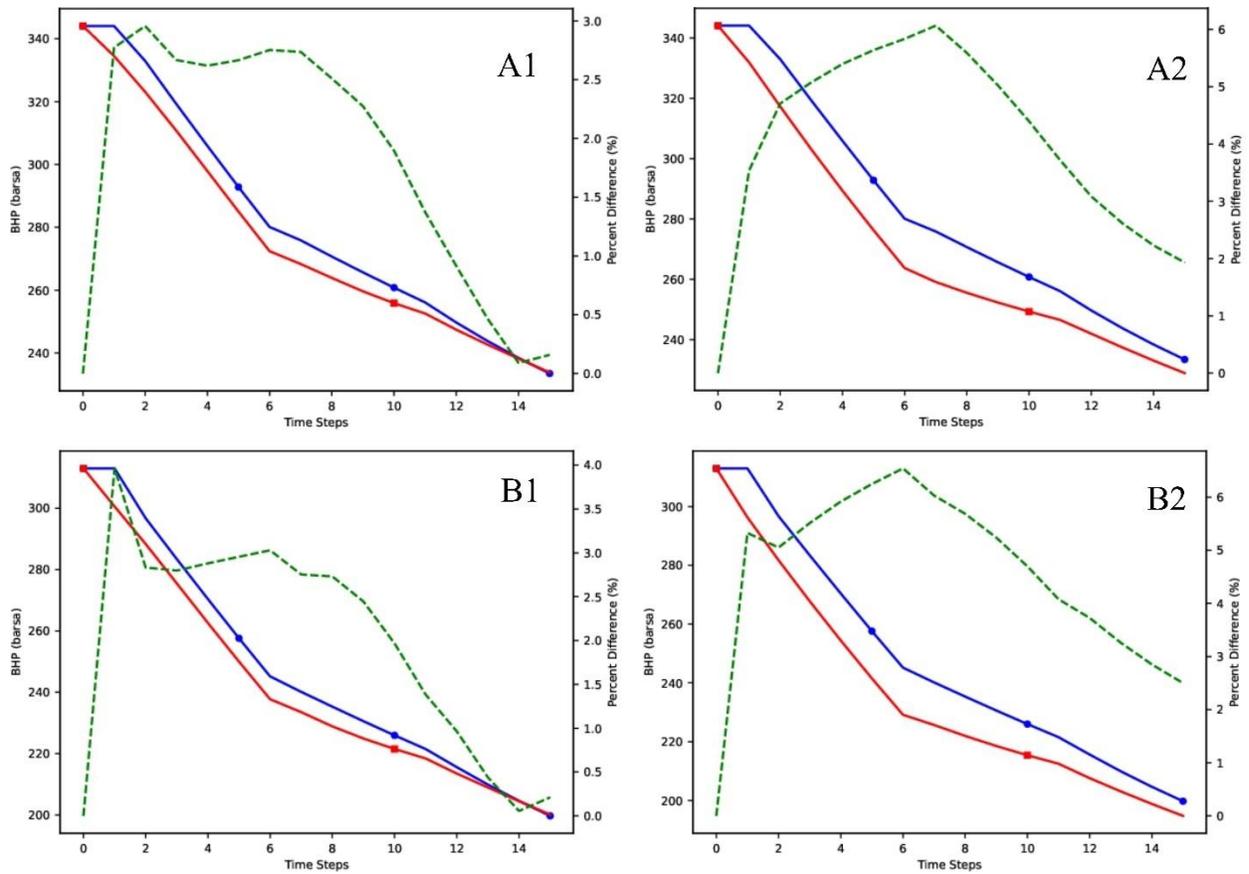

Fig. 23 – Simulation period yearly run 2 BHP (blue line – original values) forecasted with run 7 trained DMD (red line – forecasted values) by DMDc - 15 modes (left column) and DMDspc – 9 modes (right column). A: $CO_2$ injection well BHP, B: Water production well BHP, 1: entire simulation period, 2: wells active period. Case B: $CO_2$ injection and water production.

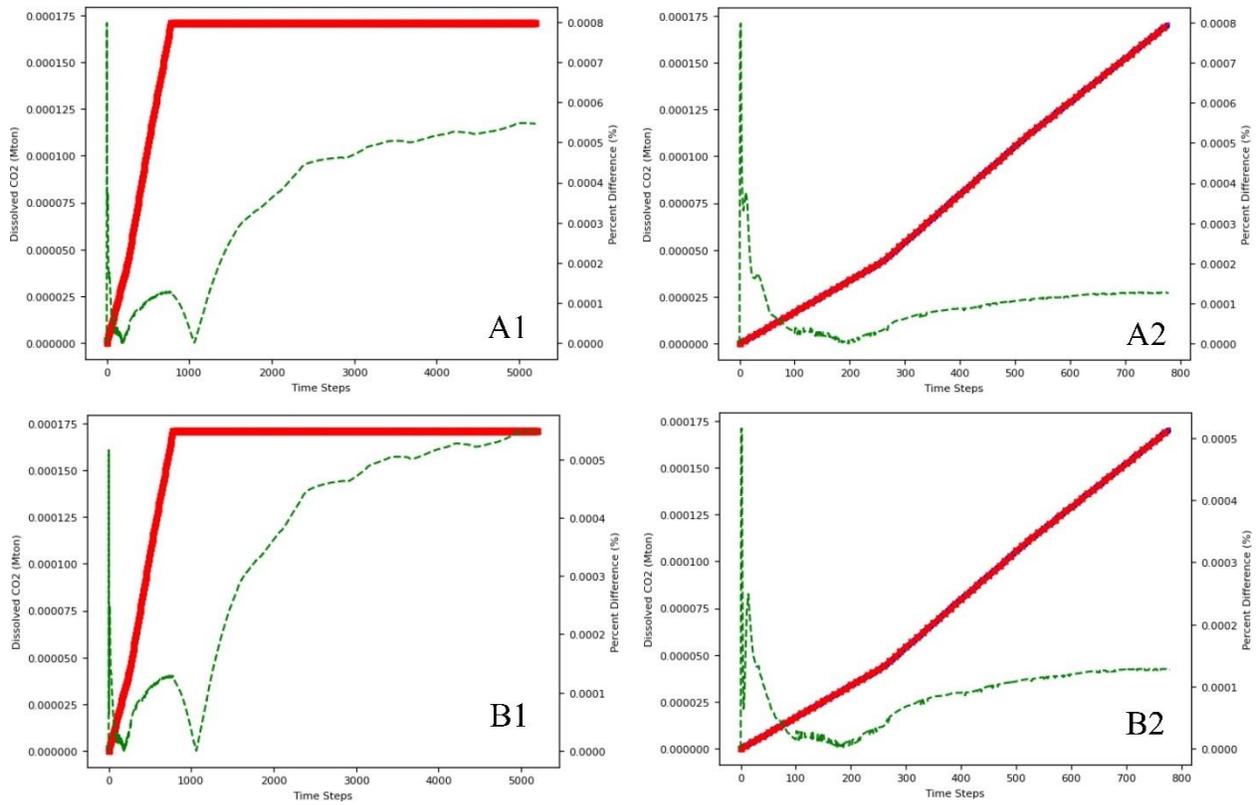

Fig. 24 – Simulation period weekly run 5 dissolved $CO_2$ (blue line – original values) forecasted with run 7 trained DMD (red line – forecasted values) by DMDc - 900 modes (A) and DMDspc – 493 modes (B). 1: entire simulation period, 2: wells active period. Case A: $CO_2$ injection.

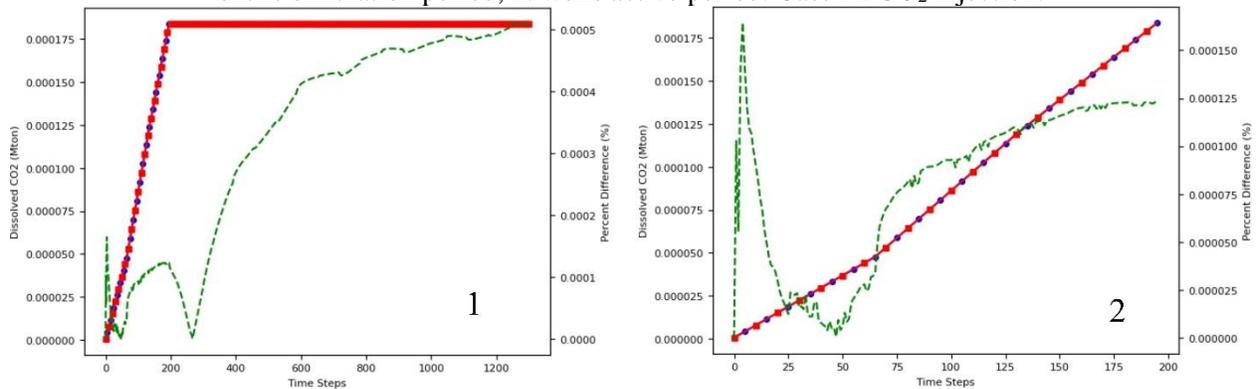

Fig. 25 – Simulation period monthly run 5 dissolved $CO_2$ (blue line – original snapshots) forecasted with run 7 trained DMD (red line – forecasted snapshots) by DMD models – 600 modes. 1: entire simulation period, 2: wells active period. Case A: $CO_2$ injection.

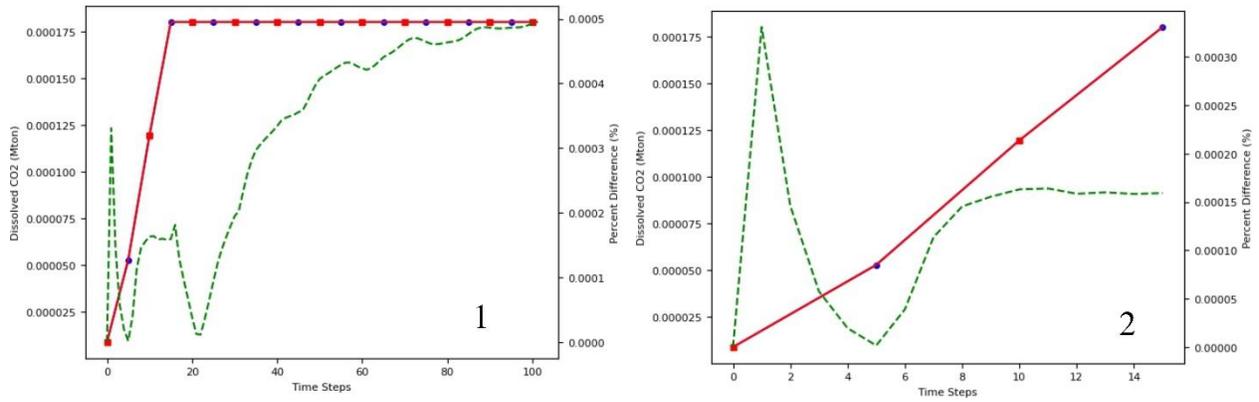

Fig. 26 – Simulation period yearly run 5 dissolved $CO_2$ (blue line – original values) forecasted with run 7 trained DMD (red line – forecasted values) by DMD - 100 modes. 1: entire simulation period, 2: well active period. Case A: $CO_2$ injection.

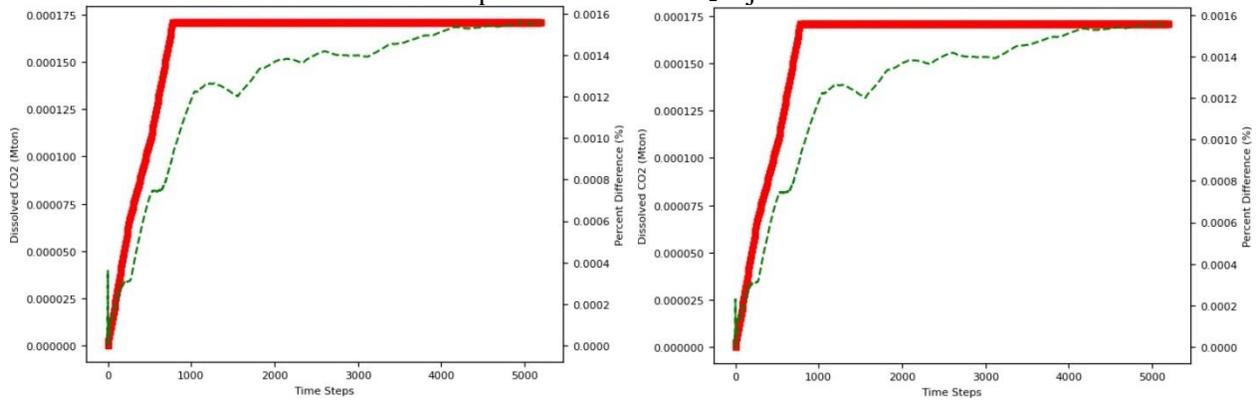

Fig. 27 – Simulation period weekly run 10 dissolved $CO_2$ (blue line – original values) forecasted with run 2 trained DMD (red line – forecasted values) by DMDc - 900 modes (left) and DMDspc – 710 modes (right). Case B: $CO_2$ injection and water production.

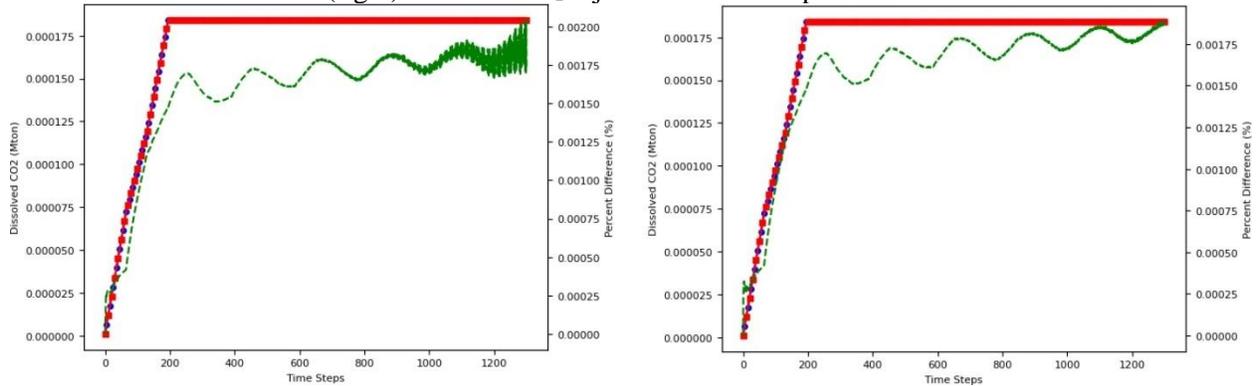

Fig. 28 – Simulation period monthly run 10 dissolved $CO_2$ (blue line – original values) forecasted with run 2 trained DMD (red line – forecasted values) by DMDc - 600 modes (left) and DMDspc – 498 modes (right). Case B: $CO_2$ injection and water production.

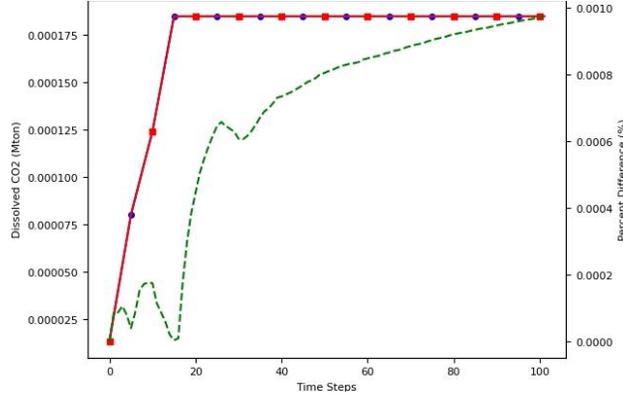

Fig. 29 – Simulation period yearly run 10 dissolved $CO_2$ (blue line – original values) forecasted with run 2 trained DMD models (red line – forecasted values) by DMD - 100 modes. Case B: $CO_2$ injection and water production.

4.5 Geological $CO_2$ Storage Optimization

Accurate geological $CO_2$ storage optimization requires accurate calculations to ensure reliable conclusions. In this case, most DMD models have relatively small errors <5% PCE for pressure or <0.01 MAE for saturation that are generally acceptable to be used for optimization and/or geological $CO_2$ storage forecast. Pressure-constrained optimization trials show that even a little error produced by the DMD models can have a substantial impact on the optimization results.

The DMD models we choose for the optimization procedure have the least amount of forecasting error and are the following, according to time scale and variable. For pressure-constrained case A, we choose weekly and monthly run 5 trained DMDc, weekly run 5, monthly run 2 trained DMDspc as it has less error than run 5 DMDc and yearly run 7 pressure models while for case B, we choose weekly run 5, monthly run 7, and yearly run 5 trained pressure models. Regarding the pressure-leakage-constrained cases, we choose the $CO_2$ saturation DMD models in addition to the pressure DMD models selected above with the same criteria as the pressure-constrained cases. For case A, we selected weekly, monthly, and yearly run 7 trained $CO_2$ saturation models, while for case B, we selected weekly, monthly, and yearly run 2 trained models.

Optimization time varied between different DMD models because of different time scales and different pressure penalty factors. Table 4 presents the total optimization time for pressure-constrained and pressure-leakage-constrained optimization regarding case A and case B. Most optimization runs used a $10^{-9}$ pressure penalty factor as it satisfied pressure constraints. Those cases were faster, while the rest had to iterate through each element of the pressure penalty factor vector until pressure constraints were satisfied. DMDc accelerated weekly pressure–leakage constrained optimizations were not run since they are infeasible because of the similar run times with ECLIPSE 300 run times.

Reconstructing only the reservoir cells monitored by the optimizer showed that this approach can even further reduce computational time by a huge amount (see DMDspc – Monitored Cells in Table 4). The case with the largest recorded time is weekly pressure-leakage constrained $CO_2$ injection and water production with ~60 mins execution time. This occurred because of the penalty vector element iteration mentioned earlier. However, this did not impact the overall speed as the monitored cells are 1,340, which is ~0.99% of the 135,340 reservoir cells. The time needed to do the same optimization with ECLIPSE 300 instead of the DMD models or the DMDspc-monitored cells approach would have been extremely large and therefore infeasible, given the number of function calls and the time of a single ECLIPSE 300 run.

Table 4 - Pressure-constrained and pressure-leakage-constrained optimization total time (mins) for case A: $CO_2$ injection and case B: $CO_2$ injection and water production

| Pressure Constrained | | | |
| --- | --- | --- | --- |
| $CO_2$ Injection | | | |
| Total Optimization Time (mins) | DMDc | DMDspc | DMDspc - Monitored Cells |
| Weekly | 1124 | 405 | 10 |
| Monthly | 446 | 253 | 4.2 |
| Yearly | 0.8 | 0.8 | 0.015 |
| $CO_2$ Injection and Water Production | | | |
| Total Optimization Time (mins) | DMDc | DMDspc | DMDspc - Monitored Cells |
| Weekly | 1364 | 935 | 6.4 |
| Monthly | 76 | 75 | 1.7 |
| Yearly | 0.8 | 0.8 | 0.006 |
| Pressure - Leakage Constrained | | | |
| $CO_2$ Injection | | | |
| Total Optimization Time (mins) | DMDc | DMDspc | DMDspc - Monitored Cells |
| Weekly | - | 3660 | 8 |
| Monthly | - | 588 | 3 |
| Yearly | 10.5 | 9.5 | 0.29 |
| $CO_2$ Injection and Water Production | | | |
| Total Optimization Time (mins) | DMDc | DMDspc | DMDspc - Monitored Cells |
| Weekly | - | 2756 | 60.4 |
| Monthly | - | 160 | 1.3 |
| Yearly | 1 | 1.5 | 0.4 |

The optimized injected $CO_2$ amounts, in addition to the produced water, when it comes to case B, are similar across models and time scales. Table 5 presents the optimized injected $CO_2$ and produced water amounts for pressure-constrained and pressure-leakage-constrained cases A and B in Mtons. Monthly pressure-constrained case A DMDc is an exception to that because the DMDc model used has 11.84% PCE. This indicates that a small change in the reconstructed snapshots can lead to large differences in the optimization results, as mentioned earlier.

Table 5 – Optimized total input amounts for pressure-constrained and pressure–leakage–constrained case A: $CO_2$ injection and case B: $CO_2$ injection and water production

| | | | |
|---|---|---|---|
| Pressure Constrained | | | |
| $CO_2$ Injection | | | |
| $CO_2$ Injected (Mton) | DMDc | DMDspc | DMDspc - Monitored Cells |
| Weekly | 2.36E-05 | 1.72E-05 | 2.14E-05 |
| Monthly | 4.65E-04 | 2.36E-05 | 2.14E-05 |
| Yearly | 2.36E-05 | 2.57E-05 | 2.57E-05 |
| $CO_2$ Injection and Water Production | | | |
| $CO_2$ Injected (Mton) / Water Produced (Mton) | DMDc | DMDspc | DMDspc - Monitored Cells |
| Weekly | 4.10E-03 / 5.47E-03 | 4.10E-03 / 5.46E-03 | 4.10E-03 / 5.43E-03 |
| Monthly | 4.41E-03 / 5.88E-03 | 4.41E-03 / 5.90E-03 | 4.41E-03 / 3.50E-03 |
| Yearly | 4.38E-03 / 5.84E-03 | 4.38E-03 / 5.84E-03 | 4.38E-03 / 5.48E-03 |
| Pressure and Leakage-constrained | | | |
| $CO_2$ Injection | | | |
| $CO_2$ Injected (Mton) | DMDc | DMDspc | DMDspc - Monitored Cells |
| Weekly | - | 2.87E-04 | 2.85E-04 |
| Monthly | - | 4.01E-04 | 3.97E-04 |
| Yearly | 2.32E-04 | 2.96E-04 | 2.92E-04 |
| $CO_2$ Injection and Water Production | | | |
| $CO_2$ Injected (Mton) / Water Produced (Mton) | DMDc | DMDspc | DMDspc - Monitored Cells |
| Weekly | - | 4.07E-03 / 4.78E-03 | 4.10E-03 / 4.82E-03 |
| Monthly | - | 4.40E-03 / 4.88E-03 | 4.41E-03 / 4.83E-03 |
| Yearly | 4.38E-03 / 4.56E-03 | 4.38E-03 / 4.92E-03 | 4.38E-03 / 4.56E-03 |

5. Conclusions, Limitations, and Future Work

In this study, we applied data-driven and non-intrusive DMDc and DMDspc to pressure and $CO_2$ saturation fields to accelerate forecast and optimization of geological $CO_2$ storage, aiding in the acceleration of risk assessment, overall decision-making, and the broader effort of climate change mitigation. This work demonstrates that rapid and cheap forecast and optimization is achieved using the simple but robust framework of DMD-based proxy models, that require fewer computational resources than traditional high-fidelity reservoir simulators, machine learning, and reduced physics proxy models. Moreover, the reconstruction of specific reservoir cells' values which are the monitored cells during optimization lowers computational costs even further (e.g., less memory and fewer computations that DMDspc). DMD requires much less data than NN to train with a single non-iterative DMD application using a few cores of CPU as a minimum (e.g. a single set of snapshots and no need to figure NN dimensions). Since DMD relies on linear algebra our methodology is equally applicable to GPUs. Regarding RP proxy models, our approach with DMD enables generalization to many reservoir problems while it doesn't require flash calculations and Jacobian matrix, residual assembly, and linear/non-linear system solving.

DMD models were fed with weekly, monthly, and yearly ECLIPSE 300 simulations with variable well rates. Well rates were kept constant for long periods of time because of our generalized study. However, having restrictions during optimization, e.g., surface facilities and injectivity might lead to differences between consecutive well rates' values. This may cause DMD to have lower accuracy and, therefore, needs to be investigated further. The large-scale offshore reservoir model used includes more than 130,000 cells mesh and is highly heterogeneous. However, to make the procedure we suggest in this study more reliable, it needs to be applied to multiple reservoirs. Two cases were considered: case A, which uses a single well that injects $CO_2$, and case B, which uses one $CO_2$ injection well and one water production well for pressure maintenance.

DMDc and DMDspc models are both able to accurately capture the underlying fluid flow physics within the reservoir. Snapshot reconstruction showed a significant increase in speed by reducing several hours of ECLIPSE 300 simulation time to mere minutes for pressure and to several minutes for saturation. Average reconstruction computation times among case A and case B, since they have similar times, for the entire simulation period DMDc (DMDspc) average pressure reconstruction times are weekly 1.34 (1.05) mins, monthly 0.29 (0.16) mins, and yearly 0.02 (0.01) mins. Average saturation times are 31.61 (20.37) mins, monthly 6.87 (6.41) mins, and yearly 0.08 (0.08) mins. The active well period DMDc (DMDspc) average pressure reconstruction times are weekly 0.15 (0.06) mins, monthly 0.05 (0.02) mins, and yearly 0.017 (0.006) mins. Average saturation times are 5.69 (3.39) mins, monthly 1.52 (1.52) mins, and yearly 0.05 (0.04) mins.

DMDspc effectively reduced the number of DMD modes related to pressure across almost all trained models, resulting in a good balance of accuracy and speed improvement. For Case A, an average error of 0.27 was achieved by removing an average of 9 modes. Case B showed an average error of 0.35 with an average removal of 6 modes. However, in terms of $CO_2$ saturation, DMDspc had limited success in decreasing DMD modes. Only 5 out of 24 trained models retained accuracy while removing a significant number of modes. No modes were removed for Case A. In Case B, the weekly average error was 0.16 with an average removal of 59 modes, and the monthly average error was 0.09 with an average removal of 105 modes. This limitation must be further investigated because the $CO_2$ saturation computational times are still high when compared with those of pressure.

Optimizing control inputs, such as well rates, may result in optimized values that differ from the original surrogate model training inputs. This requires an evaluation of the surrogate model's forecasting capabilities. Therefore, one of the key components to successful accelerated optimization of $CO_2$ sequestration is to evaluate the performance of DMDc and DMDspc models in forecasting pressure and $CO_2$ saturation fields for different well rates. PCE, MAE, and R metrics were used to evaluate the DMD models' performance for Case A ($CO_2$ injection) and Case B ($CO_2$ injection and water production) across different time scales. DMD models with large errors were discarded.

For the active well period in Case A, the average PCE values for DMDc are 1.8% (weekly), 12% (monthly), and 3.2% (yearly), while for DMDspc, they are 2.4% (weekly), 9.8% (monthly), and 2.3% (yearly). Both models can forecast pressure states, but DMDspc generally shows lower error at larger time scales, e.g., weekly. Simulation period DMDc exhibits a high PCE for monthly forecasts, indicating poor performance, whereas DMDspc shows lower but still significant errors. Yearly models perform better overall. For the active well period in Case B, the average PCE values for DMDc are 3.35% (weekly), 7% (monthly), and 7% (yearly), while for DMDspc, they are 49% (weekly), 10.2 % (monthly), and 7.32% (yearly). Errors are like those in Case A, with yearly models performing reliably. For the simulation period, errors are higher than in Case A, with only specific runs of DMDc and DMDspc models showing relatively low forecast errors.

Regarding the $CO_2$ saturation, both models exhibit similar errors, since they have the same number of DMD modes. For the active well period in Case A, the average MAE is 0.0085 for weekly and monthly scales and 0.0057 for yearly. Errors slightly increase over the entire simulation period, with yearly models showing the best forecast performance. In Case B, the MAE remains below 0.01 across all time scales, with similar error trends.

Overall, DMD models can capture temporal fluid dynamics, with DMDspc generally performing better at larger time scales, e.g., weekly. The performance varies significantly across different runs and time scales, highlighting the importance of model selection and forecasting validation. Our analysis also shows that DMD forecasts pressure more accurately when trained on snapshots corresponding to relatively low well rates, whereas $CO_2$ saturation forecasts are more accurate when trained on snapshots corresponding to relatively high well rates. Optimized $CO_2$ injection and water production amounts are similar across models and time scales, except for monthly pressure-constrained DMDc case A, which has a high PCE. This indicates that low PCE is necessary for reliable optimization results and that small differences in reconstructed snapshots can lead to significant changes in optimization outcomes. DMD models with errors below 5% PCE or 0.01 MAE are deemed acceptable for geological $CO_2$ storage optimization.

Optimization times vary due to computational demands and pressure penalty factors. Pressure-leakage-constrained optimizations for weekly time scale are infeasible due to DMDc models' relatively long run times. The monitored cells approach significantly reduces optimization time. For instance, weekly pressure-leakage-constrained $CO_2$ injection take about 8 minutes with the monitored cells approach. The optimization process using ECLIPSE 300 would take significantly longer due to the long runtime. The simple optimization approach of this study could be augmented with different optimizers and optimization techniques to further reduce computation times.

Appendix

Table A1 – POD PCE for pressure and $CO_2$ saturation at weekly, monthly, and yearly time scales

| | $CO_2$ Injection | | |
|---|---|---|---|
| Pressure | Weekly | Monthly | Yearly |
| Run 2 | 0.0008 | 0.0008 | 0.0007 |
| Run 5 | 0.0003 | 0.0003 | 0.0003 |
| Run 7 | 0.0008 | 0.0008 | 0.0006 |

|                | Run 10       | 0.0003  | 0.0003  | 0.0003   |
|----------------|--------------|---------|---------|----------|
|                | CO$_2$ saturation | Weekly  | Monthly | Yearly   |
|                | Run 2        | 0.0262  | 0.0399  | 1.28E-13 |
|                | Run 5        | 0.0274  | 0.0294  | 1.42E-13 |
|                | Run 7        | 0.0251  | 0.0388  | 1.35E-13 |
|                | Run 10       | 0.0280  | 0.0294  | 1.74E-13 |
| CO$_2$ Injection and Water Production |  |  |  |  |
|                | Pressure     | Weekly  | Monthly | Yearly   |
|                | Run 2        | 0.0054  | 0.0054  | 0.0050   |
|                | Run 5        | 0.0007  | 0.0007  | 0.0006   |
|                | Run 7        | 0.0055  | 0.0055  | 0.0050   |
|                | Run 10       | 0.0007  | 0.0007  | 0.0006   |
|                | CO$_2$ saturation | Weekly  | Monthly | Yearly   |
|                | Run 2        | 0.0406  | 0.0390  | 1.57E-13 |
|                | Run 5        | 0.0406  | 0.0354  | 1.33E-13 |
|                | Run 7        | 0.0477  | 0.0393  | 1.35E-13 |
|                | Run 10       | 0.0409  | 0.0349  | 1.58E-13 |

Table A2 – DMDc and DMDspc forecast reconstruction times for pressure and CO$_2$ saturation at weekly, monthly, and yearly time scales for case A: CO$_2$ injection.

| Pressure |  |  |  |  |  |  |  |
|---|---|---|---|---|---|---|---|
| Weekly |  |  | Monthly |  |  | Yearly |  |
| Simulation period |  |  |  |  |  |  |  |
| Run | DMDc (mins) | DMDspc (mins) | Run | DMDc (mins) | DMDspc (mins) | Run | DMDc (mins) | DMDspc (mins) |
| 2    | 1.1374  | 1.5750  | 2    | 0.1926 | 0.0969 | 2    | 0.0193 | 0.0085 |
| 5    | 1.2981  | 1.1421  | 5    | 0.2904 | 0.0707 | 5    | 0.0197 | 0.0070 |
| 7    | 1.8512  | 1.1014  | 7    | 0.3855 | 0.3024 | 7    | 0.0193 | 0.0116 |
| 10   | 0.8508  | 0.5924  | 10   | 0.3205 | 0.2232 | 10   | 0.0200 | 0.0081 |
| Mean | 1.2844  | 1.1027  | Mean | 0.2973 | 0.1733 | Mean | 0.0196 | 0.0088 |
| Wells active |  |  |  |  |  |  |  |
| Run | DMDc (mins) | DMDspc (mins) | Run | DMDc (mins) | DMDspc (mins) | Run | DMDc (mins) | DMDspc (mins) |
| 2    | 0.1599  | 0.0965  | 2    | 0.0443 | 0.0091 | 2    | 0.0060 | 0.0017 |
| 5    | 0.1348  | 0.0430  | 5    | 0.0609 | 0.0312 | 5    | 0.0064 | 0.0015 |
| 7    | 0.1774  | 0.0476  | 7    | 0.0628 | 0.0206 | 7    | 0.0299 | 0.0105 |
| 10   | 0.1812  | 0.0353  | 10   | 0.0815 | 0.0387 | 10   | 0.0235 | 0.0080 |
| Mean | 0.1633  | 0.0556  | Mean | 0.0624 | 0.0249 | Mean | 0.0165 | 0.0054 |
| CO$_2$ saturation |  |  |  |  |  |  |  |
| Weekly |  |  | Monthly |  |  | Yearly |  |
| Simulation period |  |  |  |  |  |  |  |
| Run | DMDc (mins) | DMDspc (mins) | Run | DMDc (mins) | DMDspc (mins) | Run | DMDc (mins) | DMDspc (mins) |
| 2    | 30.5143 | 9.9991  | 2    | 5.7064 | 5.1269 | 2    | 0.0741 | 0.0821 |
| 5    | 29.9165 | 27.4394 | 5    | 7.5542 | 6.7960 | 5    | 0.1429 | 0.0907 |
| 7   | 38.9799 | 7.6741  | 7    | 7.7988 | 5.3133 | 7    | 0.0715 | 0.0798 |
| 10  | 31.0301 | 29.4156 | 10   | 5.8051 | 8.2075 | 10   | 0.0702 | 0.0793 |
| Mean| 32.6102 | 18.6320 | Mean | 6.7161 | 6.3609 | Mean | 0.0897 | 0.0830 |

| Wells active | | | | | | | | |
| --- | --- | --- | --- | --- | --- | --- | --- | --- |
| Run | DMDc (mins) | DMDspc (mins) | Run | DMDc (mins) | DMDspc (mins) | Run | DMDc (mins) | DMDspc (mins) |
| 2 | 5.6391 | 1.4931 | 2 | 1.4787 | 1.4505 | 2 | 0.0680 | 0.0783 |
| 5 | 5.6137 | 5.6846 | 5 | 1.4844 | 1.4454 | 5 | 0.0680 | 0.0783 |
| 7 | 5.6403 | 1.1391 | 7 | 1.4798 | 1.4507 | 7 | 0.0548 | 0.0306 |
| 10 | 4.2027 | 2.0483 | 10 | 1.4831 | 1.4537 | 10 | 0.0411 | 0.0235 |
| Mean | 5.2739 | 2.5913 | Mean | 1.4815 | 1.4501 | Mean | 0.0580 | 0.0527 |

Table A3 – DMDc and DMDspc forecast reconstruction times for pressure and $CO_2$ saturation at weekly, monthly, and yearly time scales for case B: $CO_2$ injection and water production

| Pressure | | | | | | | | |
| --- | --- | --- | --- | --- | --- | --- | --- | --- |
| Weekly | | | Monthly | | | Yearly | | |
| Simulation period | | | | | | | | |
| Run | DMDc (mins) | DMDspc (mins) | Run | DMDc (mins) | DMDspc (mins) | Run | DMDc (mins) | DMDspc (mins) |
| 2 | 1.8052 | 1.6578 | 2 | 0.3011 | 0.1871 | 2 | 0.0205 | 0.0123 |
| 5 | 1.2887 | 0.7114 | 5 | 0.2427 | 0.0756 | 5 | 0.0209 | 0.0092 |
| 7 | 0.9159 | 0.6000 | 7 | 0.3129 | 0.2424 | 7 | 0.0463 | 0.0184 |
| 10 | 1.5740 | 1.0655 | 10 | 0.2811 | 0.0782 | 10 | 0.0208 | 0.0153 |
| Mean | 1.3959 | 1.0087 | Mean | 0.2844 | 0.1458 | Mean | 0.0271 | 0.0138 |

| Wells active | | | | | | | | |
| --- | --- | --- | --- | --- | --- | --- | --- | --- |
| Run | DMDc (mins) | DMDspc (mins) | Run | DMDc (mins) | DMDspc (mins) | Run | DMDc (mins) | DMDspc (mins) |
| 2 | 0.1284 | 0.0867 | 2 | 0.0475 | 0.0266 | 2 | 0.0281 | 0.0138 |
| 5 | 0.1336 | 0.0528 | 5 | 0.0351 | 0.0089 | 5 | 0.0074 | 0.0024 |
| 7 | 0.1468 | 0.0660 | 7 | 0.0684 | 0.0123 | 7 | 0.0067 | 0.0026 |
| 10 | 0.1582 | 0.0648 | 10 | 0.0518 | 0.0115 | 10 | 0.0265 | 0.0087 |
| Mean | 0.1417 | 0.0676 | Mean | 0.0207 | 0.0089 | Mean | 0.0172 | 0.0069 |

| $CO_2$ saturation | | | | | | | | |
| --- | --- | --- | --- | --- | --- | --- | --- | --- |
| Weekly | | | Monthly | | | Yearly | | |
| Simulation period | | | | | | | | |
| Run | DMDc (mins) | DMDspc (mins) | Run | DMDc (mins) | DMDspc (mins) | Run | DMDc (mins) | DMDspc (mins) |
| 2 | 30.3463 | 10.9301 | 2 | 7.6383 | 7.6009 | 2 | 0.0630 | 0.0555 |
| 5 | 30.8238 | 30.4159 | 5 | 7.8359 | 7.5582 | 5 | 0.0683 | 0.0799 |
| 7 | 30.8932 | 17.1103 | 7 | 6.5252 | 5.5564 | 7 | 0.0739 | 0.0776 |
| 10 | 30.4919 | 30.0163 | 10 | 6.1396 | 5.0801 | 10 | 0.0702 | 0.0808 |
| Mean | 30.6388 | 22.1182 | Mean | 7.0348 | 6.4489 | Mean | 0.0689 | 0.0735 |

| Wells active | | | | | | | | |
| --- | --- | --- | --- | --- | --- | --- | --- | --- |
| Run | DMDc (mins) | DMDspc (mins) | Run | DMDc (mins) | DMDspc (mins) | Run | DMDc (mins) | DMDspc (mins) |
| 2 | 5.4184 | 1.6209 | 2 | 1.3880 | 1.4073 | 2 | 0.0354 | 0.0385 |
| 5 | 5.6775 | 5.8482 | 5 | 1.8264 | 1.9010 | 5 | 0.0434 | 0.0250 |
| 7 | 5.7250 | 3.4978 | 7 | 1.6814 | 1.7025 | 7 | 0.0479 | 0.0417 |
| 10 | 6.1463 | 5.8480 | 10 | 1.4103 | 1.4212 | 10 | 0.0602 | 0.0424 |
| Mean | 5.7418 | 4.2037 | Mean | 1.5765 | 1.6080 | Mean | 0.0467 | 0.0369 |

Table A4 – Case A pressure forecast PCE DMDc and DMDspc models for weekly, monthly, and yearly time scales

| Pressure | | | | | | | | |
|---|---|---|---|---|---|---|---|---|
| Weekly | | | Monthly | | | Yearly | | |
| Simulation Period | | | | | | | | |
| PCE | DMDc | DMDscp | MAE | DMDc | DMDscp | PCE | DMDc | DMDscp |
| 2 to 3 | 6.9E+00 | 1.3E+01 | 2 to 3 | 1.3E+10 | 1.7E+01 | 2 to 3 | 1.9E+00 | 2.2E+00 |
| 2 to 5 | 1.5E+01 | 2.3E+01 | 2 to 5 | 2.7E+10 | 4.0E+01 | 2 to 5 | 3.8E+00 | 2.0E+00 |
| 2 to 7 | 1.5E-01 | 4.1E+00 | 2 to 7 | 4.6E+09 | 3.7E+00 | 2 to 7 | 1.2E+00 | 5.5E+00 |
| 2 to 8 | 6.8E+00 | 1.3E+01 | 2 to 8 | 9.7E+09 | 1.6E+01 | 2 to 8 | 1.2E+00 | 2.9E+00 |
| 2 to 10 | 3.3E+00 | 3.5E+00 | 2 to 10 | 2.7E+10 | 4.0E+01 | 2 to 10 | 3.6E+00 | 2.0E+00 |
| Mean | 6.4E+00 | 1.1E+01 | Mean | 1.6E+10 | 2.3E+01 | Mean | 2.4E+00 | 2.9E+00 |
| 5 to 3 | 8.2E+00 | 7.4E+00 | 5 to 3 | 4.1E+04 | 4.4E+04 | 5 to 3 | 1.9E+01 | 1.7E+01 |
| 5 to 2 | 1.5E+01 | 1.4E+01 | 5 to 2 | 7.5E+04 | 7.7E+04 | 5 to 2 | 3.5E+01 | 3.1E+01 |
| 5 to 7 | 1.6E+01 | 1.4E+01 | 5 to 7 | 8.1E+04 | 8.3E+04 | 5 to 7 | 3.8E+01 | 3.4E+01 |
| 5 to 8 | 8.3E+00 | 7.5E+00 | 5 to 8 | 4.4E+04 | 4.8E+04 | 5 to 8 | 2.0E+01 | 1.8E+01 |
| 5 to 10 | 7.8E-02 | 2.7E-01 | 5 to 10 | 1.2E+03 | 5.7E+03 | 5 to 10 | 5.7E-01 | 3.3E-01 |
| Mean | 9.5E+00 | 8.5E+00 | Mean | 4.9E+04 | 5.2E+04 | Mean | 2.3E+01 | 2.0E+01 |
| 7 to 8 | 6.1E+00 | 5.7E+00 | 7 to 8 | 5.2E+14 | 7.9E+00 | 7 to 8 | 6.7E+00 | 7.4E+00 |
| 7 to 10 | 1.4E+01 | 1.3E+01 | 7 to 10 | 1.1E+15 | 2.1E+01 | 7 to 10 | 1.5E+01 | 1.7E+01 |
| 7 to 5 | 1.4E+01 | 1.3E+01 | 7 to 5 | 1.2E+15 | 2.1E+01 | 7 to 5 | 1.5E+01 | 1.7E+01 |
| 7 to 3 | 6.2E+00 | 5.8E+00 | 7 to 3 | 6.6E+14 | 7.9E+00 | 7 to 3 | 6.4E+00 | 7.1E+00 |
| 7 to 2 | 1.7E-01 | 4.6E-01 | 7 to 2 | 2.2E+14 | 3.1E+00 | 7 to 2 | 6.9E-01 | 9.0E-01 |
| Mean | 7.9E+00 | 7.6E+00 | Mean | 7.9E+14 | 1.3E+01 | Mean | 8.7E+00 | 9.8E+00 |
| 10 to 8 | 4.1E+01 | 1.3E+01 | 10 to 8 | 4.8E+04 | 3.8E+03 | 10 to 8 | 1.3E+01 | 1.1E+01 |
| 10 to 7 | 5.1E+01 | 3.2E+01 | 10 to 7 | 8.9E+04 | 6.5E+03 | 10 to 7 | 2.5E+01 | 2.1E+01 |
| 10 to 2 | 7.5E+01 | 2.3E+01 | 10 to 2 | 8.2E+04 | 6.2E+03 | 10 to 2 | 2.3E+01 | 1.9E+01 |
| 10 to 3 | 4.0E+01 | 1.3E+01 | 10 to 3 | 4.4E+04 | 3.6E+03 | 10 to 3 | 1.2E+01 | 9.9E+00 |
| 10 to 5 | 4.4E-01 | 3.3E-01 | 10 to 5 | 1.4E+03 | 5.5E+02 | 10 to 5 | 3.8E-01 | 5.0E-01 |
| Mean | 4.2E+01 | 1.6E+01 | Mean | 5.3E+04 | 4.1E+03 | Mean | 1.5E+01 | 1.2E+01 |
| Wells active | | | | | | | | |
| PCE | DMDc | DMDscp | MAE | DMDc | DMDscp | PCE | DMDc | DMDscp |
| 2 to 3 | 1.3E+00 | 1.5E+00 | 2 to 3 | 1.3E+01 | 9.4E-01 | 2 to 3 | 1.1E+00 | 1.3E+00 |
| 2 to 5 | 3.3E+00 | 3.6E+00 | 2 to 5 | 2.6E+01 | 2.5E+00 | 2 to 5 | 1.6E+00 | 1.1E+00 |
| 2 to 7 | 3.3E-01 | 2.8E-01 | 2 to 7 | 5.2E+00 | 2.6E-01 | 2 to 7 | 1.5E+00 | 3.3E+00 |
| 2 to 8 | 1.1E+00 | 1.4E+00 | 2 to 8 | 9.4E+00 | 9.0E-01 | 2 to 8 | 5.0E-01 | 2.0E+00 |
| 2 to 10 | 1.8E-02 | 3.6E-01 | 2 to 10 | 2.5E+01 | 2.5E+00 | 2 to 10 | 1.4E+00 | 1.3E+00 |
| Mean | 1.2E+00 | 1.4E+00 | Mean | 1.6E+01 | 1.4E+00 | Mean | 1.2E+00 | 1.8E+00 |
| 5 to 3 | 1.2E+00 | 9.9E-01 | 5 to 3 | 6.6E+00 | 9.0E+00 | 5 to 3 | 3.8E+00 | 2.3E+00 |
| 5 to 2 | 3.4E+00 | 9.1E-01 | 5 to 2 | 1.2E+01 | 1.6E+01 | 5 to 2 | 8.4E+00 | 5.6E+00 |
| 5 to 7 | 3.4E+00 | 9.0E-01 | 5 to 7 | 1.6E+01 | 1.9E+01 | 5 to 7 | 1.0E+01 | 7.4E+00 |
| 5 to 8 | 1.3E+00 | 9.5E-01 | 5 to 8 | 8.5E+00 | 1.1E+01 | 5 to 8 | 5.0E+00 | 3.4E+00 |
| 5 to 10 | 4.5E-02 | 4.5E-01 | 5 to 10 | 5.2E-01 | 2.0E+00 | 5 to 10 | 4.2E-01 | 3.8E-01 |
| Mean | 1.9E+00 | 8.4E-01 | Mean | 8.7E+00 | 1.1E+01 | Mean | 5.6E+00 | 3.8E+00 |
| 7 to 8 | 1.6E+00 | 8.3E-01 | 7 to 8 | 8.6E+00 | 1.9E+00 | 7 to 8 | 8.9E-01 | 1.5E+00 |
| 7 to 10 | 4.0E+00 | 3.5E+00 | 7 to 10 | 1.7E+01 | 5.0E+00 | 7 to 10 | 1.6E+00 | 2.4E+00 |
| 7 to 5 | 4.0E+00 | 3.5E+00 | 7 to 5 | 1.7E+01 | 5.0E+00 | 7 to 5 | 1.7E+00 | 2.3E+00 |
| 7 to 3 | 1.8E+00 | 9.9E-01 | 7 to 3 | 1.1E+01 | 2.1E+00 | 7 to 3 | 1.4E+00 | 1.0E+00 |
| 7 to 2 | 3.9E-01 | 7.7E-01 | 7 to 2 | 4.7E+00 | 3.1E-01 | 7 to 2 | 9.6E-01 | 1.2E+00 |
| Mean | 2.4E+00 | 1.9E+00 | Mean | 1.2E+01 | 2.8E+00 | Mean | 1.3E+00 | 1.7E+00 |
| 10 to 8 | 1.0E+00 | 4.9E+00 | 10 to 8 | 9.8E+00 | 2.1E+01 | 10 to 8 | 4.0E+00 | 1.7E+00 |
| 10 to 7 | 2.9E+00 | 8.7E+00 | 10 to 7 | 1.9E+01 | 3.9E+01 | 10 to 7 | 8.7E+00 | 4.1E+00 |
| 10 to 2 | 2.9E+00 | 8.2E+00 | 10 to 2 | 1.5E+01 | 3.6E+01 | 10 to 2 | 6.9E+00 | 2.6E+00 |
| 10 to 3 | 9.8E-01 | 4.5E+00 | 10 to 3 | 7.6E+00 | 1.9E+01 | 10 to 3 | 2.9E+00 | 8.4E-01 |
| 10 to 5 | 4.3E-02 | 2.9E-01 | 10 to 5 | 6.9E-01 | 3.6E+00 | 10 to 5 | 3.3E-01 | 5.9E-01 |
| Mean | 1.6E+00 | 5.3E+00 | Mean | 1.0E+01 | 2.4E+01 | Mean | 4.6E+00 | 2.0E+00 |

Table A5 – Case A $CO_2$ saturation forecast PCE DMDc and DMDspc models for weekly, monthly, and yearly time scales

| $CO_2$ Saturation | | | | | | | | |
|---|---|---|---|---|---|---|---|---|
| Weekly | | | Monthly | | | Yearly | | |
| Simulation period | | | | | | | | |
| MAE | DMDc | DMDspc | MAE | DMDc | DMDspc | MAE | DMDc | DMDspc |
| 2 to 3 | 1.68E+67 | 2.06E+67 | 2 to 3 | 4.64E+41 | 4.64E+41 | 2 to 3 | 8.27E-03 | 8.27E-03 |
| 2 to 5 | 4.21E+66 | 5.14E+66 | 2 to 5 | 1.16E+41 | 1.16E+41 | 2 to 5 | 3.21E-03 | 3.21E-03 |
| 2 to 7 | 5.10E+67 | 6.23E+67 | 2 to 7 | 1.40E+42 | 1.41E+42 | 2 to 7 | 2.87E-02 | 2.87E-02 |
| 2 to 8 | 5.78E+67 | 5.78E+67 | 2 to 8 | 7.02E+41 | 7.03E+41 | 2 to 8 | 1.84E-02 | 1.84E-02 |
| 2 to 10 | 6.38E+66 | 5.10E+67 | 2 to 10 | 1.75E+41 | 1.76E+41 | 2 to 10 | 5.50E-03 | 5.50E-03 |
| Mean | 1.96E+67 | 2.40E+67 | Mean | 5.72E+41 | 5.73E+41 | Mean | 1.28E-02 | 1.28E-02 |
| 5 to 3 | 1.08E+14 | 1.08E+14 | 5 to 3 | 4.76E+04 | 4.76E+04 | 5 to 3 | 1.10E-02 | 1.10E-02 |
| 5 to 2 | 8.65E+14 | 8.65E+14 | 5 to 2 | 9.53E+04 | 9.53E+04 | 5 to 2 | 2.69E-02 | 2.69E-02 |
| 5 to 7 | 1.37E+15 | 1.37E+15 | 5 to 7 | 1.64E+05 | 1.64E+05 | 5 to 7 | 1.54E-04 | 1.54E-04 |
| 5 to 8 | 6.86E+14 | 6.86E+14 | 5 to 8 | 8.18E+04 | 8.18E+04 | 5 to 8 | 2.75E-02 | 2.75E-02 |
| 5 to 10 | 1.72E+14 | 1.72E+14 | 5 to 10 | 2.05E+04 | 2.05E+04 | 5 to 10 | 5.17E-03 | 5.17E-03 |
| Mean | 6.41E+14 | 6.41E+14 | Mean | 8.18E+04 | 8.18E+04 | Mean | 1.42E-02 | 1.42E-02 |
| 7 to 8 | 7.60E-03 | 7.64E-03 | 7 to 8 | 7.60E-03 | 7.61E-03 | 7 to 8 | 7.54E-03 | 7.54E-03 |
| 7 to 10 | 3.11E-03 | 3.12E-03 | 7 to 10 | 3.11E-03 | 3.11E-03 | 7 to 10 | 3.08E-03 | 3.08E-03 |
| 7 to 5 | 3.21E-03 | 3.22E-03 | 7 to 5 | 3.20E-03 | 3.20E-03 | 7 to 5 | 3.07E-03 | 3.07E-03 |
| 7 to 3 | 8.07E-03 | 8.11E-03 | 7 to 3 | 8.06E-03 | 8.07E-03 | 7 to 3 | 7.69E-03 | 7.69E-03 |
| 7 to 2 | 4.20E-03 | 4.29E-03 | 7 to 2 | 4.18E-03 | 4.20E-03 | 7 to 2 | 3.15E-03 | 3.15E-03 |
| Mean | 5.24E-03 | 5.28E-03 | Mean | 5.23E-03 | 5.24E-03 | Mean | 4.90E-03 | 4.90E-03 |
| 10 to 8 | 5.54E+22 | 5.54E+22 | 10 to 8 | 9.85E-03 | 9.85E-03 | 10 to 8 | 9.64E-03 | 9.65E-03 |
| 10 to 7 | 1.11E+23 | 1.11E+23 | 10 to 7 | 2.49E-02 | 2.49E-02 | 10 to 7 | 2.45E-02 | 2.45E-02 |
| 10 to 2 | 7.45E+22 | 7.45E+22 | 10 to 2 | 2.67E-02 | 2.67E-02 | 10 to 2 | 2.49E-02 | 2.49E-02 |

| MAE | DMDc | DMDspc | MAE | DMDc | DMDspc | MAE | DMDc | DMDspc |
|---|---|---|---|---|---|---|---|---|
| 10 to 3 | 3.73E+22 | 3.73E+22 | 10 to 3 | 1.09E-02 | 1.09E-02 | 10 to 3 | 9.98E-03 | 9.99E-03 |
| 10 to 5 | 9.31E+21 | 9.32E+21 | 10 to 5 | 7.74E-04 | 7.75E-04 | 10 to 5 | 5.65E-04 | 5.66E-04 |
| Mean | 5.74E+22 | 5.75E+22 | Mean | 1.46E-02 | 1.46E-02 | Mean | 1.39E-02 | 1.39E-02 |
| Wells active | | | | | | | | |
| MAE | DMDc | DMDspc | MAE | DMDc | DMDspc | MAE | DMDc | DMDspc |
| 2 to 3 | 2.78E+02 | 3.40E+02 | 2 to 3 | 6.43E-03 | 6.45E-03 | 2 to 3 | 4.04E-03 | 4.04E-03 |
| 2 to 5 | 6.95E+01 | 8.49E+01 | 2 to 5 | 2.96E-03 | 2.97E-03 | 2 to 5 | 2.19E-03 | 2.19E-03 |
| 2 to 7 | 8.42E+02 | 1.03E+03 | 2 to 7 | 1.20E-02 | 1.20E-02 | 2 to 7 | 3.62E-03 | 3.62E-03 |
| 2 to 8 | 4.21E+02 | 5.15E+02 | 2 to 8 | 8.30E-03 | 8.32E-03 | 2 to 8 | 4.74E-03 | 4.74E-03 |
| 2 to 10 | 1.05E+02 | 1.29E+02 | 2 to 10 | 3.52E-03 | 3.53E-03 | 2 to 10 | 2.52E-03 | 2.52E-03 |
| Mean | 3.43E+02 | 4.19E+02 | Mean | 6.64E-03 | 6.66E-03 | Mean | 3.42E-03 | 3.42E-03 |
| 5 to 3 | 1.59E-05 | 1.64E-05 | 5 to 3 | 8.20E-03 | 8.20E-03 | 5 to 3 | 6.82E-03 | 6.82E-03 |
| 5 to 2 | 2.00E-02 | 2.00E-02 | 5 to 2 | 2.00E-02 | 2.00E-02 | 5 to 2 | 1.70E-02 | 1.70E-02 |
| 5 to 7 | 2.28E-02 | 2.28E-02 | 5 to 7 | 2.28E-02 | 2.28E-02 | 5 to 7 | 1.54E-04 | 1.54E-04 |
| 5 to 8 | 9.63E-03 | 9.63E-03 | 5 to 8 | 9.64E-03 | 9.64E-03 | 5 to 8 | 8.36E-03 | 8.36E-03 |
| 5 to 10 | 8.71E-04 | 8.71E-04 | 5 to 10 | 8.70E-04 | 8.71E-04 | 5 to 10 | 5.53E-04 | 5.53E-04 |
| Mean | 1.07E-02 | 1.07E-02 | Mean | 1.23E-02 | 1.23E-02 | Mean | 6.58E-03 | 6.58E-03 |
| 7 to 8 | 4.90E-03 | 5.13E-03 | 7 to 8 | 4.89E-03 | 4.94E-03 | 7 to 8 | 4.48E-03 | 4.48E-03 |
| 7 to 10 | 2.72E-03 | 2.78E-03 | 7 to 10 | 2.71E-03 | 2.72E-03 | 7 to 10 | 2.41E-03 | 2.41E-03 |
| 7 to 5 | 2.45E-03 | 2.51E-03 | 7 to 5 | 2.44E-03 | 2.45E-03 | 7 to 5 | 2.13E-03 | 2.13E-03 |
| 7 to 3 | 4.28E-03 | 4.50E-03 | 7 to 3 | 4.28E-03 | 4.33E-03 | 7 to 3 | 3.98E-03 | 3.98E-03 |
| 7 to 2 | 2.73E-03 | 3.15E-03 | 7 to 2 | 2.72E-03 | 2.81E-03 | 7 to 2 | 2.54E-03 | 2.54E-03 |
| Mean | 3.41E-03 | 3.61E-03 | Mean | 3.41E-03 | 3.45E-03 | Mean | 3.11E-03 | 3.11E-03 |
| 10 to 8 | 8.89E-03 | 8.89E-03 | 10 to 8 | 8.88E-03 | 8.89E-03 | 10 to 8 | 7.48E-03 | 7.48E-03 |
| 10 to 7 | 2.17E-02 | 2.17E-02 | 10 to 7 | 2.17E-02 | 2.17E-02 | 10 to 7 | 1.87E-02 | 1.87E-02 |
| 10 to 2 | 1.91E-02 | 1.91E-02 | 10 to 2 | 1.91E-02 | 1.91E-02 | 10 to 2 | 1.57E-02 | 1.57E-02 |
| 10 to 3 | 7.62E-03 | 7.63E-03 | 10 to 3 | 7.59E-03 | 7.59E-03 | 10 to 3 | 6.06E-03 | 6.06E-03 |
| 10 to 5 | 4.41E-04 | 4.41E-04 | 10 to 5 | 4.39E-04 | 4.41E-04 | 10 to 5 | 4.38E-04 | 4.39E-04 |
| Mean | 1.16E-02 | 1.16E-02 | Mean | 1.15E-02 | 1.15E-02 | Mean | 9.67E-03 | 9.67E-03 |

Table A6 – Case B Pressure forecast PCE DMDc and DMDspc models for weekly, monthly, and yearly time scales

| Pressure | | | | | | | | |
|---|---|---|---|---|---|---|---|---|
| Weekly | | | Monthly | | | Yearly | | |
| Simulation period | | | | | | | | |
| PCE | DMDc | DMDspc | PCE | DMDc | DMDspc | PCE | DMDc | DMDspc |
| 2 to 3 | 1.50E+01 | 1.46E+01 | 2 to 3 | 8.77E+07 | 5.42E+07 | 2 to 3 | 1.27E+01 | 1.30E+01 |
| 2 to 5 | 1.41E+01 | 1.44E+01 | 2 to 5 | 6.77E+07 | 2.68E+07 | 2 to 5 | 1.57E+01 | 1.70E+01 |
| 2 to 7 | 1.02E+00 | 2.35E+00 | 2 to 7 | 4.03E+07 | 2.57E+07 | 2 to 7 | 2.05E+00 | 1.26E+00 |
| 2 to 8 | 1.49E+01 | 1.45E+01 | 2 to 8 | 5.38E+07 | 4.11E+07 | 2 to 8 | 1.26E+01 | 1.27E+01 |
| 2 to 10 | 1.42E+01 | 1.44E+01 | 2 to 10 | 4.53E+07 | 3.42E+07 | 2 to 10 | 1.56E+01 | 1.69E+01 |
| Mean | 1.18E+01 | 1.20E+01 | Mean | 4.64E+07 | 3.37E+07 | Mean | 1.17E+01 | 1.22E+01 |
| 5 to 3 | 2.49E+01 | 3.27E+01 | 5 to 3 | 2.02E+01 | 1.55E+01 | 5 to 3 | 2.85E+01 | 1.75E+01 |
| 5 to 2 | 1.26E+01 | 1.49E+01 | 5 to 2 | 5.30E+01 | 4.91E+01 | 5 to 2 | 4.33E+01 | 3.45E+01 |
| 5 to 7 | 1.26E+01 | 1.47E+01 | 5 to 7 | 5.32E+01 | 4.90E+01 | 5 to 7 | 4.05E+01 | 3.19E+01 |

| | | | | | | | |
|---|---|---|---|---|---|---|---|
| 5 to 8 | 2.47E+01 | 3.24E+01 | 5 to 8 | 2.10E+01 | 1.38E+01 | 5 to 8 | 2.69E+01 | 1.61E+01 |
| 5 to 10 | 2.18E-01 | 2.46E+00 | 5 to 10 | 2.57E-01 | 2.63E+00 | 5 to 10 | 3.22E-01 | 1.70E+00 |
| Mean | 1.50E+01 | 1.94E+01 | Mean | 2.95E+01 | 2.60E+01 | Mean | 2.79E+01 | 2.03E+01 |
| 7 to 8 | 2.63E+01 | 3.43E+02 | 7 to 8 | 1.90E+01 | 1.93E+01 | 7 to 8 | 1.28E+01 | 1.16E+01 |
| 7 to 10 | 1.39E+02 | 2.59E+02 | 7 to 10 | 1.92E+01 | 1.91E+01 | 7 to 10 | 1.38E+01 | 1.49E+01 |
| 7 to 5 | 1.38E+02 | 2.59E+02 | 7 to 5 | 1.91E+01 | 1.90E+01 | 7 to 5 | 1.39E+01 | 1.50E+01 |
| 7 to 3 | 3.19E+01 | 3.45E+02 | 7 to 3 | 4.00E+00 | 3.00E+00 | 7 to 3 | 1.26E+01 | 1.14E+01 |
| 7 to 2 | 9.29E+00 | 3.01E+02 | 7 to 2 | 2.00E+00 | 2.06E+00 | 7 to 2 | 1.40E+00 | 3.29E+00 |
| Mean | 6.90E+01 | 3.01E+02 | Mean | 1.27E+01 | 1.25E+01 | Mean | 1.09E+01 | 1.12E+01 |
| 10 to 8 | 2.20E+09 | 3.55E+09 | 10 to 8 | 1.01E+03 | 3.09E+01 | 10 to 8 | 1.28E+02 | 1.52E+02 |
| 10 to 7 | 4.82E+09 | 7.49E+09 | 10 to 7 | 2.41E+03 | 4.02E+01 | 10 to 7 | 9.80E+01 | 9.71E+01 |
| 10 to 2 | 2.71E+09 | 4.10E+09 | 10 to 2 | 2.26E+03 | 4.11E+01 | 10 to 2 | 5.93E+01 | 5.46E+01 |
| 10 to 3 | 9.83E+08 | 1.40E+09 | 10 to 3 | 9.08E+02 | 3.15E+01 | 10 to 3 | 1.52E+02 | 1.78E+02 |
| 10 to 5 | 2.56E+08 | 1.50E+08 | 10 to 5 | 2.57E+01 | 7.08E-01 | 10 to 5 | 4.14E+00 | 1.17E+01 |
| Mean | 2.19E+09 | 3.34E+09 | Mean | 1.32E+03 | 2.89E+01 | Mean | 8.82E+01 | 9.88E+01 |
| | | | | Wells active | | | | |
| PCE | DMDc | DMDspc | PCE | DMDc | DMDspc | PCE | DMDc | DMDspc |
| 2 to 3 | 1.33E+00 | 2.12E+00 | 2 to 3 | 8.13E+00 | 4.87E+00 | 2 to 3 | 5.47E+00 | 8.26E+00 |
| 2 to 5 | 7.73E+00 | 7.93E+00 | 2 to 5 | 1.53E+01 | 1.30E+01 | 2 to 5 | 7.23E+00 | 1.02E+01 |
| 2 to 7 | 1.41E+00 | 2.35E+00 | 2 to 7 | 5.84E+00 | 3.73E+00 | 2 to 7 | 3.81E+00 | 9.89E-01 |
| 2 to 8 | 1.86E+00 | 2.71E+00 | 2 to 8 | 4.91E+00 | 2.78E+00 | 2 to 8 | 3.68E+00 | 6.23E+00 |
| 2 to 10 | 7.86E+00 | 8.04E+00 | 2 to 10 | 1.46E+01 | 1.25E+01 | 2 to 10 | 6.83E+00 | 9.79E+00 |
| Mean | 4.04E+00 | 4.63E+00 | Mean | 9.76E+00 | 7.37E+00 | Mean | 5.40E+00 | 7.10E+00 |
| 5 to 3 | 1.48E+00 | 2.35E+00 | 5 to 3 | 4.06E+00 | 9.33E+00 | 5 to 3 | 5.44E+00 | 2.12E+00 |
| 5 to 2 | 2.16E+00 | 3.28E+00 | 5 to 2 | 1.44E+01 | 2.75E+01 | 5 to 2 | 5.74E+00 | 1.05E+01 |
| 5 to 7 | 3.39E+00 | 3.81E+00 | 5 to 7 | 2.63E+00 | 6.13E-01 | 5 to 7 | 4.67E+00 | 8.64E+00 |
| 5 to 8 | 1.17E+00 | 1.86E+00 | 5 to 8 | 4.29E+00 | 1.04E+01 | 5 to 8 | 4.40E+00 | 1.62E+00 |
| 5 to 10 | 1.92E-01 | 5.51E-01 | 5 to 10 | 3.11E-01 | 2.63E+00 | 5 to 10 | 3.59E-01 | 1.24E+00 |
| Mean | 1.68E+00 | 2.37E+00 | Mean | 5.13E+00 | 1.01E+01 | Mean | 4.12E+00 | 4.81E+00 |
| 7 to 8 | 1.52E+00 | 7.80E+00 | 7 to 8 | 2.54E+00 | 1.94E+00 | 7 to 8 | 3.82E+00 | 6.76E+00 |
| 7 to 10 | 9.00E+00 | 1.18E+01 | 7 to 10 | 1.30E+01 | 1.30E+01 | 7 to 10 | 8.45E+00 | 1.08E+01 |
| 7 to 5 | 8.93E+00 | 1.18E+01 | 7 to 5 | 1.30E+01 | 1.30E+01 | 7 to 5 | 8.66E+00 | 1.10E+01 |
| 7 to 3 | 1.06E+00 | 7.60E+00 | 7 to 3 | 2.20E+00 | 1.56E+00 | 7 to 3 | 4.89E+00 | 7.89E+00 |
| 7 to 2 | 1.16E+00 | 6.53E+00 | 7 to 2 | 9.37E-01 | 1.57E+00 | 7 to 2 | 2.34E+00 | 4.84E+00 |
| Mean | 4.34E+00 | 9.11E+00 | Mean | 6.35E+00 | 6.21E+00 | Mean | 5.63E+00 | 8.26E+00 |
| 10 to 8 | 1.03E+02 | 1.79E+02 | 10 to 8 | 7.08E+00 | 1.22E+01 | 10 to 8 | 2.23E+00 | 2.25E+00 |
| 10 to 7 | 2.62E+02 | 4.16E+02 | 10 to 7 | 1.13E+01 | 3.16E+01 | 10 to 7 | 1.10E+01 | 2.15E+01 |
| 10 to 2 | 1.67E+02 | 2.35E+02 | 10 to 2 | 8.60E+00 | 3.04E+01 | 10 to 2 | 3.35E+00 | 1.16E+01 |
| 10 to 3 | 3.18E+01 | 4.55E+01 | 10 to 3 | 5.59E+00 | 1.11E+01 | 10 to 3 | 7.06E+00 | 7.74E+00 |
| 10 to 5 | 2.75E+01 | 2.11E+01 | 10 to 5 | 3.93E-01 | 3.88E-01 | 10 to 5 | 1.12E+00 | 2.39E+00 |
| Mean | 1.18E+02 | 1.79E+02 | Mean | 6.59E+00 | 1.71E+01 | Mean | 4.94E+00 | 9.10E+00 |

Table A7 – Case B $CO_2$ saturation forecast PCE DMDc and DMDspc models for weekly, monthly, and yearly time scales

| $CO_2$ Saturation |
|---|

|  | Weekly |  |  | Monthly |  |  | Yearly |  |
| --- | --- | --- | --- | --- | --- | --- | --- | --- |
|  |  |  |  | Simulation period |  |  |  |  |
| MAE | DMDc | DMDspc | MAE | DMDc | DMDspc | MAE | DMDc | DMDspc |
| 2 to 3 | 9.72E-03 | 9.72E-03 | 2 to 3 | 1.17E-02 | 1.06E-02 | 2 to 3 | 7.69E-03 | 7.69E-03 |
| 2 to 5 | 5.13E-03 | 5.13E-03 | 2 to 5 | 6.58E-03 | 6.12E-03 | 2 to 5 | 4.18E-03 | 4.18E-03 |
| 2 to 7 | 4.90E-03 | 4.90E-03 | 2 to 7 | 9.98E-03 | 8.30E-03 | 2 to 7 | 5.90E-03 | 5.90E-03 |
| 2 to 8 | 1.02E-02 | 1.02E-02 | 2 to 8 | 1.49E-02 | 1.27E-02 | 2 to 8 | 9.36E-03 | 9.36E-03 |
| 2 to 10 | 5.13E-03 | 5.13E-03 | 2 to 10 | 7.44E-03 | 6.45E-03 | 2 to 10 | 4.07E-03 | 4.07E-03 |
| Mean | 7.02E-03 | 7.02E-03 | Mean | 1.01E-02 | 8.84E-03 | Mean | 6.24E-03 | 6.24E-03 |
| 5 to 3 | 5.23E+24 | 5.23E+24 | 5 to 3 | 9.08E+17 | 9.08E+17 | 5 to 3 | 1.32E-02 | 1.32E-02 |
| 5 to 2 | 2.09E+25 | 2.09E+25 | 5 to 2 | 3.63E+18 | 3.63E+18 | 5 to 2 | 3.12E-02 | 3.12E-02 |
| 5 to 7 | 4.47E+25 | 4.47E+25 | 5 to 7 | 7.84E+18 | 7.84E+18 | 5 to 7 | 3.21E-02 | 3.21E-02 |
| 5 to 8 | 1.71E+25 | 1.71E+25 | 5 to 8 | 3.01E+18 | 3.01E+18 | 5 to 8 | 1.35E-02 | 1.35E-02 |
| 5 to 10 | 2.98E+24 | 2.97E+24 | 5 to 10 | 5.26E+17 | 5.26E+17 | 5 to 10 | 6.34E-04 | 6.34E-04 |
| Mean | 1.82E+25 | 1.82E+25 | Mean | 3.18E+18 | 2.65E+18 | Mean | 1.82E-02 | 1.82E-02 |
| 7 to 8 | 9.98E-03 | 9.98E-03 | 7 to 8 | 1.09E-02 | 9.96E-03 | 7 to 8 | 1.09E-02 | 1.09E-02 |
| 7 to 10 | 5.91E-03 | 5.90E-03 | 7 to 10 | 6.45E-03 | 6.01E-03 | 7 to 10 | 6.43E-03 | 6.43E-03 |
| 7 to 5 | 6.51E-03 | 6.51E-03 | 7 to 5 | 6.84E-03 | 6.21E-03 | 7 to 5 | 6.56E-03 | 6.56E-03 |
| 7 to 3 | 1.34E-02 | 1.34E-02 | 7 to 3 | 1.28E-02 | 1.10E-02 | 7 to 3 | 1.16E-02 | 1.16E-02 |
| 7 to 2 | 1.21E-02 | 1.21E-02 | 7 to 2 | 6.26E-03 | 4.58E-03 | 7 to 2 | 5.23E-03 | 5.23E-03 |
| Mean | 9.59E-03 | 9.58E-03 | Mean | 8.64E-03 | 7.56E-03 | Mean | 8.14E-03 | 8.14E-03 |
| 10 to 8 | 4.65E+56 | 2.53E+56 | 10 to 8 | 7.04E+02 | 7.03E+02 | 10 to 8 | 1.20E-02 | 1.20E-02 |
| 10 to 7 | 1.85E+57 | 1.01E+57 | 10 to 7 | 2.82E+03 | 2.81E+03 | 10 to 7 | 2.65E-02 | 2.65E-02 |
| 10 to 2 | 5.95E+56 | 3.24E+56 | 10 to 2 | 1.54E+03 | 1.54E+03 | 10 to 2 | 2.61E-02 | 2.61E-02 |
| 10 to 3 | 1.63E+56 | 8.89E+55 | 10 to 3 | 6.16E+02 | 6.16E+02 | 10 to 3 | 1.18E-02 | 1.18E-02 |
| 10 to 5 | 1.56E+56 | 8.50E+55 | 10 to 5 | 2.68E+02 | 2.68E+02 | 10 to 5 | 6.05E-04 | 6.05E-04 |
| Mean | 6.46E+56 | 3.52E+56 | Mean | 1.19E+03 | 1.19E+03 | Mean | 1.54E-02 | 1.54E-02 |
|  |  |  |  | Wells active |  |  |  |  |
| MAE | DMDc | DMDspc | MAE | DMDc | DMDspc | MAE | DMDc | DMDspc |
| 2 to 3 | 7.35E-03 | 7.35E-03 | 2 to 3 | 7.54E-03 | 7.62E-03 | 2 to 3 | 7.69E-03 | 7.69E-03 |
| 2 to 5 | 4.52E-03 | 4.53E-03 | 2 to 5 | 5.10E-03 | 5.12E-03 | 2 to 5 | 4.10E-03 | 4.10E-03 |
| 2 to 7 | 8.76E-03 | 8.77E-03 | 2 to 7 | 7.10E-03 | 7.18E-03 | 2 to 7 | 1.25E-02 | 1.25E-02 |
| 2 to 8 | 6.52E-03 | 6.52E-03 | 2 to 8 | 6.54E-03 | 6.67E-03 | 2 to 8 | 6.56E-03 | 6.56E-03 |
| 2 to 10 | 4.10E-03 | 4.11E-03 | 2 to 10 | 4.79E-03 | 4.81E-03 | 2 to 10 | 3.46E-03 | 3.46E-03 |
| Mean | 6.25E-03 | 6.25E-03 | Mean | 6.21E-03 | 6.28E-03 | Mean | 6.87E-03 | 6.87E-03 |
| 5 to 3 | 1.64E+01 | 1.64E+01 | 5 to 3 | 1.15E-02 | 1.15E-02 | 5 to 3 | 8.58E-03 | 8.58E-03 |
| 5 to 2 | 6.55E+01 | 6.55E+01 | 5 to 2 | 3.54E-02 | 3.54E-02 | 5 to 2 | 2.47E-02 | 2.47E-02 |
| 5 to 7 | 1.40E+02 | 1.40E+02 | 5 to 7 | 6.04E-02 | 6.04E-02 | 5 to 7 | 2.92E-02 | 2.92E-02 |
| 5 to 8 | 5.35E+01 | 5.35E+01 | 5 to 8 | 2.33E-02 | 2.33E-02 | 5 to 8 | 1.03E-02 | 1.03E-02 |
| 5 to 10 | 9.30E+00 | 9.30E+00 | 5 to 10 | 3.46E-03 | 3.46E-03 | 5 to 10 | 1.37E-03 | 1.37E-03 |
| Mean | 5.69E+01 | 5.69E+01 | Mean | 2.68E-02 | 2.68E-02 | Mean | 1.48E-02 | 1.48E-02 |
| 7 to 8 | 7.71E-03 | 7.75E-03 | 7 to 8 | 7.17E-03 | 7.24E-03 | 7 to 8 | 1.01E-02 | 1.01E-02 |
| 7 to 10 | 4.94E-03 | 4.95E-03 | 7 to 10 | 4.96E-03 | 4.98E-03 | 7 to 10 | 6.14E-03 | 6.14E-03 |
| 7 to 5 | 4.97E-03 | 4.98E-03 | 7 to 5 | 5.10E-03 | 5.13E-03 | 7 to 5 | 6.71E-03 | 6.71E-03 |
| 7 to 3 | 8.51E-03 | 8.54E-03 | 7 to 3 | 8.58E-03 | 8.65E-03 | 7 to 3 | 1.32E-02 | 1.32E-02 |
| 7 to 2 | 5.88E-03 | 5.94E-03 | 7 to 2 | 5.31E-03 | 5.40E-03 | 7 to 2 | 1.12E-02 | 1.12E-02 |
| Mean | 6.40E-03 | 6.43E-03 | Mean | 6.22E-03 | 6.28E-03 | Mean | 9.45E-03 | 9.45E-03 |
| 10 to 8 | 3.69E+02 | 2.01E+02 | 10 to 8 | 9.16E-03 | 9.16E-03 | 10 to 8 | 9.14E-03 | 9.14E-03 |
| 10 to 7 | 1.47E+03 | 8.00E+02 | 10 to 7 | 2.00E-02 | 2.00E-02 | 10 to 7 | 2.47E-02 | 2.47E-02 |
| 10 to 2 | 4.72E+02 | 2.57E+02 | 10 to 2 | 1.76E-02 | 1.76E-02 | 10 to 2 | 2.00E-02 | 2.00E-02 |
| 10 to 3 | 1.30E+02 | 7.05E+01 | 10 to 3 | 8.04E-03 | 8.04E-03 | 10 to 3 | 6.78E-03 | 6.78E-03 |
| 10 to 5 | 1.24E+02 | 6.74E+01 | 10 to 5 | 5.63E-04 | 5.63E-04 | 10 to 5 | 1.32E-03 | 1.32E-03 |
| Mean | 5.13E+02 | 2.79E+02 | Mean | 1.11E-02 | 1.11E-02 | Mean | 1.24E-02 | 1.24E-02 |

References

[1] *Energy Technology Perspectives 2020 - Special Report on Carbon Capture Utilisation and Storage*. International Energy Agency, 2020. doi: 10.1787/208b66f4-en.

[2] D. Deel, K. Mahajan, C. R. Mahoney, H. G. McIlvried, and R. D. Srivastava, "Risk assessment and management for long-term storage of CO 2 in geologic formations - United states department of energy R&D," *WMSCI 2006 - The 10th World Multi-Conference on Systemics, Cybernetics and Informatics, Jointly with the 12th International Conference on Information Systems Analysis and Synthesis, ISAS 2006 - Proc.*, vol. 6, no. 1, pp. 326–331, 2006.

[3] P. N. Price, T. E. McKone, and M. D. Sohn, "Carbon Sequestration Risks and Risk Management," *Lawrence Berkeley National Laboratory*, no. January 2007, 2007, [Online]. Available: https://escholarship.org/uc/item/0x18n8qm

[4] M. He, S. Luis, S. Rita, G. Ana, V. Euripedes, and N. Zhang, "Risk assessment of CO2 injection processes and storage in carboniferous formations: a review," *Journal of Rock Mechanics and Geotechnical Engineering*, vol. 3, no. 1, pp. 39–56, 2011, doi: 10.3724/sp.j.1235.2011.00039.

[5] G. J. Moridis, M. T. Reagan, T. Huang, and T. A. Blasingame, "Practical Aspects and Implications of Long-Term CO2 Sequestration in Saline Aquifers Using Vertical Wells," *SPE Latin American and Caribbean Petroleum Engineering Conference Proceedings*, vol. 2023-June, no. c, 2023, doi: 10.2118/213168-MS.

[6] A. K. Furre, O. Eiken, H. Alnes, J. N. Vevatne, and A. F. Kiær, "20 Years of Monitoring CO2-injection at Sleipner," *Energy Procedia*, vol. 114, no. August 1997, pp. 3916–3926, 2017, doi: 10.1016/j.egypro.2017.03.1523.

[7] R. Gholami, A. Raza, and S. Iglauer, "Leakage risk assessment of a CO2 storage site: A review," *Earth-Science Reviews*, vol. 223, no. June, p. 103849, 2021, doi: 10.1016/j.earscirev.2021.103849.

[8] D. Savage, P. R. Maul, S. Benbow, and R. C. Walke, "A generic FEP database for the assessment of long-term performance and safety of the geological storage of CO2," *Quintessa Report QRS-1060A-1*, no. June, p. 73, 2004.

[9] J. L. Lewicki, J. Birkholzer, and C. F. Tsang, "Natural and industrial analogues for leakage of CO2 from storage reservoirs: Identification of features, events, and processes and lessons learned," *Environmental Geology*, vol. 52, no. 3, pp. 457–467, 2007, doi: 10.1007/s00254-006-0479-7.

[10] A. Ramakrishnan, T. C. Zhang, and R. Y. Surampalli, "Monitoring, Verification and Accounting of CO 2 Stored in Deep Geological Formations," *Carbon Capture and Storage*, pp. 159–194, 2015, doi: 10.1061/9780784413678.ch06.

[11] M. A. Cardoso, L. J. Durlofsky, and P. Sarma, "Development and application of reduced-order modeling procedures for subsurface flow simulation," *International Journal for Numerical Methods in Engineering*, vol. 77, no. 9, pp. 1322–1350, Feb. 2009, doi: 10.1002/nme.2453.

[12] H. Florez and E. Gildin, "Global/local model order reduction in coupled flow and linear thermal-poroelasticity," *Computational Geosciences*, vol. 24, no. 2, pp. 709–735, 2020, doi: 10.1007/s10596-019-09834-7.

[13] W. Sun and L. J. Durlofsky, "Data-space approaches for uncertainty quantification of CO2 plume location in geological carbon storage," *Advances in Water Resources*, vol. 123, no. November 2018, pp. 234–255, 2019, doi: 10.1016/j.advwatres.2018.10.028.

[14] M. Tang, X. Ju, and L. J. Durlofsky, "Deep-learning-based coupled flow-geomechanics surrogate model for CO2 sequestration," *International Journal of Greenhouse Gas Control*, vol. 118, no. February, p. 103692, 2022, doi: 10.1016/j.ijggc.2022.103692.

[15] E. Artun, "Characterizing interwell connectivity in waterflooded reservoirs using data-driven and reduced-physics models: A comparative study," *Neural Computing and Applications*, vol. 28, no. 7, pp. 1729–1743, 2017, doi: 10.1007/s00521-015-2152-0.

[16] R. W. De Holanda, E. Gildin, J. L. Jensen, L. W. Lake, and C. Shah Kabir, "A state-of-the-art literature review on capacitance resistance models for reservoir characterization and performance forecasting," *Energies*, vol. 11, no. 12, 2018, doi: 10.3390/en11123368.

[17] A. Alghamdi, M. Hiba, M. Aly, and A. Awotunde, "A Critical Review of Capacitance-Resistance Models," in *SPE Russian Petroleum Technology Conference*, SPE, Oct. 2021. doi: 10.2118/206555-MS.

[18] D. W. Vasco and A. Datta-Gupta, "Asymptotic solutions for solute transport: A formalism for tracer tomography," *Water Resources Research*, vol. 35, no. 1, pp. 1–16, 1999, doi: 10.1029/98WR02742.

[19] K. Nakajima, A. Datta-Gupta, and M. J. King, "Application of the Fast Marching Method in Heterogeneous Dual PorosityReservoirs," *SPE Latin American and Caribbean Petroleum Engineering Conference Proceedings*, 2020, doi: 10.2118/199098-MS.

[20] A. Iino, A. Vyas, J. Huang, A. Datta-Gupta, Y. Fujita, and S. Sankaran, "Rapid compositional simulation and history matching of shale oil reservoirs using the fast marching method," *SPE/AAPG/SEG Unconventional Resources Technology Conference 2017*, 2017, doi: 10.15530/urtec-2017-2693139.

[21] T. Ertekin and Q. Sun, "Artificial intelligence applications in reservoir engineering: A status check," *Energies*, vol. 12, no. 15, 2019, doi: 10.3390/en12152897.

[22] H. Wang and S. Chen, "Insights into the Application of Machine Learning in Reservoir Engineering: Current Developments and Future Trends," *Energies*, vol. 16, no. 3, 2023, doi: 10.3390/en16031392.

[23] M. Tang, Y. Liu, and L. J. Durlofsky, "A deep-learning-based surrogate model for data assimilation in dynamic subsurface flow problems," *Journal of Computational Physics*, vol. 413, p. 109456, 2020, doi: 10.1016/j.jcp.2020.109456.

[24] F. He, W. Dong, and J. Wang, "Modeling and numerical investigation of transient two-phase flow with liquid phase change in porous media," *Nanomaterials*, vol. 11, no. 1, pp. 1–14, 2021, doi: 10.3390/nano11010183.

[25] I. Goodfellow, Y. Bengio, and A. Courville, *Deep Learning*. The MIT Press, 2017.

[26] N. Andrew, "Machine Learning Yearning," 2018, [Online]. Available: https://www.google.com/url?sa=t&rct=j&q=&esrc=s&source=web&cd=&cad=rja&uact=8&ved=2ahUKEwjKk5WtqNWIAxWTHNAFHcJaB-aMQFnoECBkQAQ&url=https%3A%2F%2Finfo.deeplearning.ai%2Fmachine-learning-yearning-book&usg=AOvVaw2ytesPttvL5Ns1wynh-kL4&opi=89978449

[27] A. Vaswani *et al.*, "Attention is All you Need," in *Advances in Neural Information Processing Systems*, I. Guyon, U. V. Luxburg, S. Bengio, H. Wallach, R. Fergus, S. Vishwanathan, and R. Garnett, Eds., New York, NY, USA: Curran Associates, Inc., Oct. 2017, pp. 4752–4758. doi: 10.1145/3583780.3615497.

[28] S. Minaee, Y. Boykov, F. Porikli, A. Plaza, N. Kehtarnavaz, and D. Terzopoulos, "Image Segmentation Using Deep Learning: A Survey," *IEEE Transactions on Pattern Analysis and Machine Intelligence*, vol. 44, no. 7, pp. 3523–3542, 2022, doi: 10.1109/TPAMI.2021.3059968.

[29] C. S. W. Ng, A. Jahanbani Ghahfarokhi, M. Nait Amar, and O. Torsæter, "Smart Proxy Modeling of a Fractured Reservoir Model for Production Optimization: Implementation of Metaheuristic Algorithm and Probabilistic Application," *Natural Resources Research*, vol. 30, no. 3, pp. 2431–2462, 2021, doi: 10.1007/s11053-021-09844-2.

[30] J. Bi *et al.*, "A Physics-Informed Spatial-Temporal Neural Network for Reservoir Simulation and Uncertainty Quantification," *SPE Journal*, vol. 28, no. 04, pp. 2026–2043, Apr. 2024, doi: 10.2118/218386-pa.

[31] J. F. M. Doren, R. Markovinović, and J. D. Jansen, "Reduced-order optimal control of water flooding using proper orthogonal decomposition," *Computational Geosciences*, vol. 10, no. 1, pp. 137–158, 2006, doi: 10.1007/s10596-005-9014-2.

[32] J. He and L. J. Durlofsky, "Reduced-order modeling for compositional simulation by use of trajectory piecewise linearization," *SPE Journal*, vol. 19, no. 5, pp. 858–872, 2014, doi: 10.2118/163634-pa.

[33] Y. Yang, M. Ghasemi, E. Gildin, Y. Efendiev, and V. Calo, "Fast multiscale reservoir simulations with POD-DEIM model reduction," *SPE Journal*, vol. 21, no. 6, pp. 2141–2154, 2016, doi: 10.2118/173271-PA.

[34] H. Florez, E. Gildin, and P. Morkos, *Realization and Model Reduction of Dynamical Systems*. Springer International Publishing, 2022. doi: 10.1007/978-3-030-95157-3.

[35] M. J. Dall'Aqua, E. J. R. Coutinho, E. Gildin, Z. Guo, H. Zalavadia, and S. Sankaran, "Input-Output Invariant Fast Proxy Models for Production Optimization," *SPE Latin American and Caribbean Petroleum Engineering Conference Proceedings*, vol. 2023-June, 2023, doi: 10.2118/213117-MS.

[36] Z. L. Jin and L. J. Durlofsky, "Reduced-order modeling of CO2 storage operations," *International Journal of Greenhouse Gas Control*, vol. 68, no. August 2017, pp. 49–67, 2018, doi: 10.1016/j.ijggc.2017.08.017.

[37] M. Th. van Genuchten, "A Closed-form Equation for Predicting the Hydraulic Conductivity of Unsaturated Soils," *Soil Science Society of America Journal*, vol. 44, no. 5, pp. 892–898, 1980, doi: 10.2136/sssaj1980.03615995004400050002x.

[38] D. Y. Peng and D. B. Robinson, "A New Two-Constant Equation of State," *Industrial and Engineering Chemistry Fundamentals*, vol. 15, no. 1, pp. 59–64, 1976, doi: 10.1021/i160057a011.

[39] N. Spycher, K. Pruess, and J. Ennis-King, "CO2-H2O mixtures in the geological sequestration of CO2. I. Assessment and calculation of mutual solubilities from 12 to 100°C and up to 600 bar," *Geochimica et Cosmochimica Acta*, vol. 67, no. 16, pp. 3015–3031, 2003, doi: 10.1016/S0016-7037(03)00273-4.

[40] Schlumberger, "ECLIPSE 300 2023 Manual." 2023. [Online]. Available: software.slb.com/eclipse

[41] P. J. Schmid, "Dynamic mode decomposition of numerical and experimental data," *Journal of Fluid Mechanics*, vol. 656, pp. 5–28, 2010, doi: 10.1017/S0022112010001217.

[42] P. J. Schmid, L. Li, M. P. Juniper, and O. Pust, "Applications of the dynamic mode decomposition," *Theoretical and Computational Fluid Dynamics*, vol. 25, no. 1–4, pp. 249–259, 2011, doi: 10.1007/s00162-010-0203-9.

[43] M. R. Jovanović, P. J. Schmid, and J. W. Nichols, "Sparsity-promoting dynamic mode decomposition," *Physics of Fluids*, vol. 26, no. 2, 2014, doi: 10.1063/1.4863670.

[44] J. Annoni, P. Seiler, and M. R. Jovanovic, "Sparsity-promoting dynamic mode decomposition for systems with inputs," *2016 IEEE 55th Conference on Decision and Control, CDC 2016*, no. Cdc, pp. 6506–6511, 2016, doi: 10.1109/CDC.2016.7799270.

[45] A. Tsolovikos, E. Bakolas, S. Suryanarayanan, and D. Goldstein, "Estimation and control of fluid flows using sparsity-promoting dynamic mode decomposition," *IEEE Control Systems Letters*, vol. 5, no. 4, pp. 1145–1150, 2021, doi: 10.1109/LCSYS.2020.3015776.

[46] P. J. Schmid, "Dynamic Mode Decomposition and Its Variants," *Annual Review of Fluid Mechanics*, vol. 54, pp. 225–254, 2021, doi: 10.1146/annurev-fluid-030121-015835.

[47] Q. A. Huhn, M. E. Tano, J. C. Ragusa, and Y. Choi, "Parametric dynamic mode decomposition for reduced order modeling," *Journal of Computational Physics*, vol. 475, p. 111852, 2023, doi: 10.1016/j.jcp.2022.111852.

[48] J. L. Proctor, S. L. Brunton, and J. N. Kutz, "Dynamic Mode Decomposition with Control," *SIAM Journal on Applied Dynamical Systems*, vol. 15, no. 1, pp. 142–161, Jan. 2016, doi: 10.1137/15M1013857.

[49] J. L. Proctor, S. L. Brunton, and J. N. Kutz, "Dynamic Mode Decomposition with Control," *SIAM J. Appl. Dyn. Syst.*, vol. 15, no. 1, pp. 142–161, Jan. 2016, doi: 10.1137/15M1013857.

[50] J. H. Tu, C. W. Rowley, D. M. Luchtenburg, S. L. Brunton, and J. N. Kutz, "On dynamic mode decomposition: Theory and applications," *Journal of Computational Dynamics*, vol. 1, no. 2, pp. 391–421, 2014, doi: 10.3934/jcd.2014.1.391.

[51] O. Andersen, K. A. Lie, and H. M. Nilsen, "An open-source toolchain for simulation and optimization of aquifer-wide CO2 storage," *Energy Procedia*, vol. 86, no. 1876, pp. 324–333, 2016, doi: 10.1016/j.egypro.2016.01.033.

[52] K.-A. Lie, *An Introduction to Reservoir Simulation Using MATLAB/GNU Octave*. 2019. doi: 10.1017/9781108591416.

[53] *Advanced Modeling with the MATLAB Reservoir Simulation Toolbox*. 2021. doi: 10.1017/9781009019781.

[54] A. Paszke *et al.*, "PyTorch: An imperative style, high-performance deep learning library," *Advances in Neural Information Processing Systems*, vol. 32, no. NeurIPS, 2019.

[55] C. Witzgall and R. Fletcher, *Practical Methods of Optimization*, vol. 53. 1989. doi: 10.2307/2008742.

[56] D. C. Liu and J. Nocedal, "On the limited memory BFGS method for large scale optimization," *Mathematical Programming*, vol. 45, no. 1–3, pp. 503–528, Aug. 1989, doi: 10.1007/BF01589116.